\documentclass[12pt,a4paper]{article}
\usepackage{amsmath,amsfonts,amsthm,graphicx}
\allowdisplaybreaks[3]
\numberwithin{equation}{section} 
\newtheorem{theorem}{Theorem}[section]
\newtheorem{lemma}[theorem]{Lemma}
\newtheorem{proposition}[theorem]{Proposition}
\newtheorem{conjecture}[theorem]{Conjecture}
\usepackage[body={15.5cm,22cm}]{geometry}
\newcounter{app}
\newcounter{sapp}[app]
\newcounter{ssapp}[sapp]
\def\theapp{\Alph{app}}
\newcommand{\app}[1]{
\refstepcounter{app}{\vspace{7mm}
\noindent\Large\bf Appendix
\theapp.
 \ #1 \par \vspace{5mm}}
\setcounter{equation}{0}
\def\theequation{\Alph{app}.\arabic{equation}}}
\def\thesapp{\Alph{app}.\arabic{sapp}}
\newcommand{\sapp}[1]{\par \refstepcounter{sapp}{\vspace{2mm} 
\noindent\large\bf \thesapp
\ #1 \par \vspace{2mm}}
\def\theequation{\Alph{app}.\arabic{equation}}}

\newcommand{\ds}{\displaystyle}
\newcommand{\Qs}{{\mathsf Q}}
\newcommand{\Ts}{{\mathsf T}}
\newcommand{\Ds}{{\mathsf D}}
\begin{document}
\title{
Solutions of the $T$-system and Baxter equations for supersymmetric spin chains
}
\author{Zengo Tsuboi 
\footnote{
E-mail address: 
ztsuboi
$\bullet$
yahoo.co.jp 
($\bullet$ should be at mark)}
\\
{\it Okayama Institute for Quantum Physics
\footnote{URL: 
http://www.pref.okayama.jp/kikaku/kouryoushi/english/kouryoushi.htm
},}
 \\
{\it Kyoyama 1-9-1, Okayama 700-0015, Japan}
}
\date{}
\maketitle
\begin{abstract} 
We propose Wronskian-like determinant formulae for the Baxter $\Qs$-functions and 
the eigenvalues of transfer matrices for spin chains related to the quantum affine superalgebra 
$U_{q}(\widehat{gl}(M|N))$. In contrast to the supersymmetric Bazhanov-Reshetikhin formula 
 (the quantum supersymmetric Jacobi-Trudi formula) proposed in 
[Z.\ Tsuboi, J.\ Phys.\ A: Math.\ Gen.\ 30 (1997) 7975], 
the size of the matrices of these Wronskian-like formulae is 
less than or equal to $M+N$. 
Base on these formulae, 
we give new expressions of the 
solutions of the $T$-system (fusion relations for transfer matrices) for 
supersymmetric spin chains 
proposed in the abovementioned paper. 
Baxter equations also follow 
from the Wronskian-like formulae. They are finite order linear difference equations  
with respect to the Baxter $\Qs$-functions. 
 Moreover, the Wronskian-like formulae also explicitly solve the functional relations for
B\"{a}cklund flows proposed in 
[V.\ Kazakov, A.\ Sorin, A.\ Zabrodin, Nucl.\ Phys.\ B790 (2008) 345 
[arXiv:hep-th/0703147]].
\end{abstract}
Journal reference: Nucl. Phys. B826 [PM] (2010) 399-455\\
DOI: 10.1016/j.nuclphysb.2009.08.009 \\
Report number: OIQP-09-04\\[6pt]
PACS 2008: 
02.20.Uw  02.30.Ik  75.10.Pq  11.25.Hf
\\
\setlength{\baselineskip}{14.99pt}
\newpage 
\section{Introduction}
\begin{figure}
  \begin{center}
    \setlength{\unitlength}{2pt}
    \begin{picture}(85,55) 
      \put(0,0){\line(0,1){50}}
      \put(20,0){\line(0,1){20}} 
      \put(20,20){\line(1,0){60}}
      {\thicklines \put(30,30){\line(0,1){20}}}
      {\thicklines \put(0,30){\line(1,0){30}}}
      {\thicklines \put(0,30){\line(0,1){20}}}
      {\thicklines \put(0,50){\line(1,0){30}}}
      \put(0,50){\line(1,0){80}}
       \put(-5,39){$a$}
       \put(14,52){$s$}
       \put(1.5,10){$\underleftrightarrow{\quad N\quad }$}
       \put(50,21){\rotatebox{90}{$\underleftrightarrow{\qquad \rotatebox{-90}{M} \qquad }$}}
    \end{picture}
  \end{center}
  \caption{The $a\times s $ rectangular Young diagram ($a,s \in {\mathbb Z}_{\ge 1}$) 
 to label the $\Ts$-function $\Ts^{(a)}_{s}(x)$ of the $T$-system 
 \eqref{ori-t-system1}-\eqref{ori-t-sys-bc1} 
has to be in the so-called $(M,N)$-hook.}
  \label{MN-hook}
\end{figure}
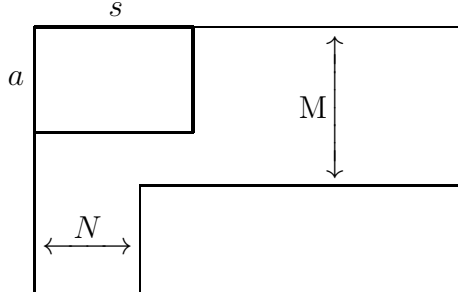

The $T$-system plays important roles in the study of quantum integrable systems
(see for example, earlier papers \cite{BP82,BR90}, which include the $T$-system 
 related to $U_{q}(\widehat{sl}(M))$). 
It is a system of fusion relations for a commuting family of 
transfer matrices of solvable lattice models. 
There is a large class of models whose $R$-matrices 
satisfy the graded Yang-Baxter equation \cite{KulSk82,Kul85}. 
In general, their symmetries are described by the quantum affine superalgebras \cite{Y99}. 
In a series of papers \cite{T97,T98,T98-2}, we proposed 
the $T$-system for solvable lattice models associated with 
$U_{q}(\widehat{sl}(M|N))$ or $U_{q}(\widehat{gl}(M|N))$. 
In an appropriate normalization, it has the following form
\footnote{A fusion relation corresponding to 
\eqref{ori-t-system1} for $(M,N)=(2,1)$ and $a=s=1$ was 
considered in \cite{Maassarani95}.}:
\begin{align}
\begin{split}
& \Ts^{(a)}_{s}(xq^{-1})\Ts^{(a)}_{s}(xq)=\Ts^{(a)}_{s-1}(x)\Ts^{(a)}_{s+1}(x)+
\Ts^{(a-1)}_{s}(x)\Ts^{(a+1)}_{s}(x)  \\[.1cm]
& \hspace{12pt} \text{for} \quad a \in \{1,2,\dots, M-1\} \quad \text{or} \quad 
s \in \{1,2,\dots, N-1\} 
 \quad \text{or} \quad (a,s)=(M,N), 
\end{split}
\label{ori-t-system1} \\[4pt]
& \Ts^{(M)}_{s}(xq^{-1})\Ts^{(M)}_{s}(xq)=
\Ts^{(M)}_{s-1}(x)\Ts^{(M)}_{s+1}(x)
\quad \text{for} \quad s \in {\mathbb Z}_{\ge N+1}, \\[6pt]
& \Ts^{(a)}_{N}(xq^{-1})\Ts^{(a)}_{N}(xq)=
\Ts^{(a-1)}_{N}(x)\Ts^{(a+1)}_{N}(x)
\quad \text{for} \quad a \in {\mathbb Z}_{\ge M+1}, \\[6pt]
& \Ts^{(M+b)}_{N}(x)=\varepsilon_{b} \Ts^{(M)}_{N+b}(x) 
\quad \text{for} \quad b \in {\mathbb Z}_{\ge 0},
\label{ori-t-sys-bc1}
\end{align}
where $a,s \in {\mathbb Z}_{\ge 1}$ is assumed; 
the factor $\varepsilon_{b} \in {\mathbb C}$ depends on the definition of the 
transfer matrices (see, (\ref{dual-mn}) for $(m,n)=(M,N)$). $x \in {\mathbb C}$ is 
a multiplicative spectral parameter whose origin goes 
back to the evaluation map from $U_{q}(\widehat{gl}(M|N))$ to $U_{q}(gl(M|N))$,  
or an automorphism of  $U_{q}(\widehat{gl}(M|N))$. 
$\Ts^{(a)}_{s}(x)$ is a transfer matrix (or its eigenvalues) 
 whose auxiliary space (the space where the (super)trace is taken)
 is an evaluation representation of $U_{q}(\widehat{gl}(M|N))$ based on 
a tensor representation of $U_{q}(gl(M|N))$ labeled by a 
 rectangular Young diagram with a hight of $a$ and a width of $s$ in the 
$(M,N)$-hook (see Figure \ref{MN-hook}). 
%
In particular, $\Ts^{(1)}_{1}(x)$ corresponds to the transfer matrix  of the 
Perk-Schultz-type model \cite{Perk:1981}.
In contrast to the bosonic algebra $U_{q}(\widehat{gl}(M))$ case, the index $a$ can take 
arbitrary non-negative integer values. 
For $N=0$, the ``duality relation" (\ref{ori-t-sys-bc1}) becomes trivial 
$\Ts^{(M)}_{b}(x)= \varepsilon_{b}^{-1}$, which means that the $M$-th antisymmetric 
tensor representation 
of $U_{q}(gl(M))$ is trivial; while this is not the case for $N>0$. 
The functions 
$\Ts^{(a)}_{0}(x),\Ts^{(0)}_{s}(x)$ ($a,s \in {\mathbb Z}_{\ge 0}$) 
appearing in the boundary 
depend on each model and the normalization. We sometimes chose 
a normalization where these functions are just $1$ (this corresponds to 
the normalization of the universal $R$-matrix).
As was pointed out in \cite{T97}, the above $T$-system 
is a reduction of the so-called Hirota bilinear difference equation \cite{Hirota81}. 
Due to the commutativity of the transfer matrices, the eigenvalues 
of the transfer matrices obey the same functional relations as the transfer matrices. 
In this paper, the eigenvalue of the transfer matrix will be called $\Ts$-function. 
These equations (\ref{ori-t-system1})-(\ref{ori-t-sys-bc1}) were transformed into the form of  
 the $Y$-system and used \cite{JKS98} to derive the thermodynamic Bethe ansatz 
equations. We also used \cite{Tsuboi06} 
these equations (\ref{ori-t-system1})-(\ref{ori-t-sys-bc1}) 
 to derive nonlinear integral equations with a finite number of unknown functions, 
which are equivalent to the thermodynamic Bethe ansatz equations. 
We remark that $T$-systems closely related to \eqref{ori-t-system1}-\eqref{ori-t-sys-bc1} 
also appeared recently in the study of the AdS/CFT correspondence 
in particle physics \cite{Beisert07,B07-2,GKV09}. 
 In particular, two copies of a $gl(2|2)$-like $T$-system coupled
\footnote{These two $gl(2|2)$-like $T$-systems decouple in the limit $L \to \infty $.} 
each other appeared as 
the ``left wing" and the ``right wing"  of the AdS/CFT $T$-system 
(in the form of the $Y$-system) \cite{GKV09,BFT09}. 

The so-called Bazhanov-Reshetikhin formula \cite{BR90}
 is a determinant expression of the eigenvalue of the transfer matrix 
for the fusion \cite{KRS81} model whose auxiliary space is labeled by a general Young 
diagram for a tensor representation of $U_{q}(\widehat{sl}(M))$ 
 (or $U_{q}(\widehat{gl}(M))$). 
This formula also has a tableaux sum expression. 
The Bazhanov-Reshetikhin formula allows a supersymmetric generalization for 
$U_{q}(\widehat{sl}(M|N))$ or $U_{q}(\widehat{gl}(M|N))$, 
which may be called 
``the quantum supersymmetric Jacobi-Trudi formula" 
or ``the supersymmetric Bazhanov-Reshetikhin formula" \cite{T97,T98,T98-2} 
(cf.\ (\ref{superJT1}) and (\ref{superJT2});\ see also a recent 
paper \cite{KV07}, and also \cite{KOS95} for $U_{q}(B^{(1)}_{r})$ case). 
This formula or its tableaux sum expression for the Young diagram of rectangular shape 
gives \cite{T97,T98,T98-2} the solution of the $T$-system \eqref{ori-t-system1}-\eqref{ori-t-sys-bc1}. 
The supersymmetric Bazhanov-Reshetikhin formula is a quantum affine superalgebra 
analogue (or the Yang-Baxterization) of the second 
Weyl formula for the transfer matrices since this formula reduces to 
the supersymmetric Jacobi-Trudi formula on the supercharacter of $gl(M|N)$ \cite{BB81} 
if the spectral parameter dependence is dropped (this corresponds to the limit $x \to 0$ in 
our normalization of the spectral parameter $x$).
 Then, it is natural to consider an analogue of the first Weyl formula. 
 The first Weyl formula for the superalgebra $gl(M|N)$ is often
called ``Sergeev-Pragacz formula" in mathematical literature 
\cite{Pragacz91} (see \eqref{Sergeev-Pragacz}). In addition, 
the Sergeev-Pragacz formula has a nice Wronskian-like determinant expression \cite{MV03}. 
In this paper, we propose  Wronskian-like determinant expressions 
of the $\Ts$-functions for $U_{q}(\widehat{gl}(M|N))$ 
(or $U_{q}(\widehat{sl}(M|N))$) for any $M,N$ 
(cf.\ (\ref{9-thvari-1})-(\ref{9-thvari-2})). 
These include formulae in our previous paper \cite{BT08} on $U_{q}(\widehat{sl}(2|1))$ 
and also similar formulae for the bosonic case $U_{q}(\widehat{gl}(M))$ \cite{KLWZ97,BLZ97,BHK02,Kojima08} 
(see also papers from a different approach \cite{BDKM06,DM08}).
In particular for the Young diagram of rectangular shape, 
these formulae give Wronskian-like determinant solutions of the $T$-system 
(\ref{ori-t-system1})-(\ref{ori-t-sys-bc1}) for $U_{q}(\widehat{gl}(M|N))$ 
(cf.\ Theorem \ref{solution-t-sys-th}). 
We also remark that these Wronskian-like determinants can be viewed as 
 generalization of the {\em ninth variation of Schur function in terms of the
 first Weyl formula} \cite{Mac92,NNSY00}. 

The so-called Baxter $\Qs$-operators were introduced by Baxter in his pioneering work on the 
eight-vertex model \cite{Bax72}, and have attracted interest from various point of views 
(see for example, \cite{BaxQ-papers,BLZ97,BHK02,BDKM06,Kojima08,BT08,DM08}). 
They belong to the same commuting family of operators as the transfer matrices. 
In this paper, the eigenvalue of the Baxter $\Qs$-operator will be called the Baxter $\Qs$-function. 
Zeros of the Baxter $\Qs$-function correspond to the roots of the Bethe ansatz equation. 
We have $2^{M+N}$ kind of the Baxter $\Qs$-functions. And two of them are ``trivial''  
in the sense that they can be normalized to just $1$. 
Thus we have $2^{M+N}-2$ kind of non-trivial Baxter $\Qs$-functions. 
But they are not independent as they obey the functional relations (\ref{QQ-rel1})-(\ref{QQ-rel2})
(cf.\ \cite{Pronko-Stroganov00,BHK02,GS03,BKSZ05,DDMST06,BDKM06,KSZ07,Zabrodin07,GV07,BT08}). 
As remarked many times 
\cite{Woynarovich83,T98,GS03,BKSZ05,BDKM06,KSZ07,GV07,BT08}, 
 there are equivalent, but different, forms of the
Bethe ansatz in the model.  In our case, 
there are $(M+N)!$ different Bethe ansatz. 
They are connected by these functional relations 
among the Baxter $\Qs$-functions (\ref{QQ-rel1})-(\ref{QQ-rel2}). 
In \cite{T98}, we discussed equivalence of the Bethe ansatz in relation to the Weyl (super) group 
for any $M,N \in {\mathbb Z}_{\ge 0}$.  
We also find Wronskian-like determinant expressions of the Baxter $\Qs$-functions, 
which solve the functional relations (\ref{QQ-rel1})-(\ref{QQ-rel2}) 
(cf.\ Theorem \ref{solution-Q}). These determinant expressions are similar to 
the ones for the $\Ts$-functions, 
but labelled by an {\em empty} Young diagram. 
In \cite{BLZ97}, the Baxter $\Qs$-operators are defined as trace of the universal $R$-matrix 
over $q$-oscillator representations of the quantum affine algebra $U_{q}(\widehat{sl}(2))$. 
This construction of the Baxter $\Qs$-operators was generalized in the subsequent papers  
\cite{BHK02,Kojima08,BT08}. 
In particular, importance of boundary twists or horizontal fields to 
regularize the trace over the infinite dimensional space was recognized in 
\cite{BLZ97} for the first time. 
Although we do not discuss operator realization of our formulae on the Baxter $\Qs$-functions, 
we design our formulae with the construction of the 
Baxter $\Qs$-operators in \cite{BLZ97} in mind. 

In \cite{KLWZ97}, it was shown for $gl(M)$-related elliptic models 
 that $\Ts$-functions for quantum integrable models  
 can be treated by methods of classical theory of solitons and 
discrete integrable equations. Moreover 
 a part of discussions in \cite{KLWZ97} was generalized,
 in the recent papers \cite{KSZ07,Zabrodin07}, 
 to $gl(M|N)$-related rational models, where $(M+1)(N+1)$ kind of Baxter $\Qs$-functions were treated. 
In particular, B\"{a}cklund transformations among solutions of the $gl(m|n)$-type $T$-systems  
($0 \le m \le M, 0 \le n \le N $) were proposed in \cite{KSZ07,Zabrodin07} 
(cf.\ eqs.\ (\ref{bac1})-(\ref{bac4})), 
where the problem on $gl(M|N)$ was connected to the one on $gl(0|0)$. 
The $U_{q}(gl(m|n))$-type $T$-systems ($0 \le m < M, 0 \le n < N $; 
cf.\ eqs.\ (\ref{t-system1})-(\ref{dual-mn})) in the intermediate step 
in the B\"{a}cklund flow have the same form as the original $T$-system 
(\ref{ori-t-system1})-(\ref{ori-t-sys-bc1}) with $(M,N) \to (m,n)$ except for 
the functions 
$\Ts^{(a)}_{0}(x),\Ts^{(0)}_{s}(x)$ ($a,s \in {\mathbb Z}_{\ge 0}$) 
appearing in the boundary.
The functions $\Ts^{(a)}_{0}(x),\Ts^{(0)}_{s}(x)$ 
for these intermediate functional relations are proportional to 
 the Baxter $\Qs$-functions (cf.\ eqs.\ (\ref{t-sys-bc1}), (\ref{t-sys-bc2})). 
 We find that our Wronskian-like determinant formulae 
 explicitly solve all these functional relations 
(cf.\ Theorems \ref{solution-t-sys-th}, \ref{solution-back}). 

In the representation theoretical context, the $\Ts$-functions can be interpreted 
as the so-called $q$-characters \cite{FR99}.
Originally, the $q$-character is defined \cite{FR99} as a partial trace of 
the universal $R$-matrix for some representation of the quantum affine algebra. 
Consequently, the eigenvalue formula of the transfer matrix by 
the Bethe ansatz has a similar 
\footnote{Compare \cite{Reshetikhin83} with \cite{FR99}.} 
form as the $q$-character if 
the Baxter $\Qs$-functions are defined in the normalization of 
the universal $R$-matrix
 (this corresponds to case where the vacuum part is formally put
\footnote{This does not always mean that the quantum space of the model 
has ``zero-spin". 
Rather, this means that the quantum space is arbitrary.}
 to 1). 
 In this sense, the $\Ts$-functions by the analytic Bethe ansatz in 
\cite{T97,T98,T98-2} are prototypes of the 
 $q$-(super)characters for the quantum affine superalgebra $U_{q}(\widehat{sl}(M|N))$ 
or $U_{q}(\widehat{gl}(M|N))$. 
 Thus our new  Wronskian-like expression of the $\Ts$-function 
is yet another form of the $q$-(super)character.
In this expression, the Weyl group invariance of the $q$-(super) character becomes manifest. 

The size of the Wronskian-like determinant for the $\Ts$-function  
is less than (for atypical representations) 
or equal to $M+N$ (for typical representations). 
On the other hand, the size of the determinant for 
 the supersymmetric Bazhanov-Reshetikhin formula 
depends on the representation and has no upper bound. 
In addition, the number of the terms in the tableaux sum expression of the 
$\Ts$-function is the dimension of the auxiliary space. 
Thus our new Wronskian-like determinant formulae will be particularly 
useful to analyze eigenvalues of the transfer matrices 
for large dimensional representations in the auxiliary space. 

In section 2, we reformulate formulae on the analytic  Bethe ansatz 
in our previous papers \cite{T97,T98,T98-2}. 
In section 3, we propose the Wronskian-like determinant 
expressions of the $\Ts$-functions and the Baxter $\Qs$-functions, and 
give solutions of the $T$-system for $U_{q}(\widehat{gl}(M|N))$. 
These are our main results in this paper. 
We will also mention relation among these Wronskian-like formulae and 
the formulae in section 2. 
In section 4, we will briefly comment on how to obtain the 
$\Ts$-functions for typical representations. 
The factorization property of the $\Ts$-functions straightforwardly follows 
from the Wronskian-like determinant formulae. 
In \cite{T98-2}, we showed that the $\Ts$-function for the typical representation 
can be obtained as a deformation of the tableaux sum expression of the $\Ts$-function. 
We rederive this result for the typical representation  
 from our new Wronskian-like determinant formula. 
In section 5, we see that Baxter equations for $U_{q}(\widehat{gl}(M|N))$ straightforwardly 
follow from the Wronskian-like determinant formulae. 
They are finite order difference equations and linear with respect to 
the Baxter $\Qs$-functions. 
Section 6 is devoted to discussions. 
In Appendix \ref{representation}, we briefly mention representations 
of the superalgebra $gl(M|N)$ and their characters. 
It will be good to see how the $\Ts$-function in this paper reduces to the 
character formula (in particular, the determinant formula in \cite{MV03}) 
of $gl(M|N)$ (or its subalgebras) in the limit $x \to 0$. 
In Appendix \ref{vaumm}, we will discuss,  
 the normalization of the Baxter $\Qs$-functions. 
We will prove our main theorems in Appendix \ref{proof-th}. 
In Appendix \ref{Conserved}, we will derive conserved quantities, and obtain 
determinant formulae of generalized Baxter equations for $U_{q}(\widehat{gl}(M|N))$.  
In this paper, we assume that $q$ is not a root of unity. 
\section{Analytic Bethe ansatz for the Perk-Schultz-type models and 
their fusion hierarchies revisited} 
The Perk-Schultz model \cite{Perk:1981} is a trigonometric vertex model related to 
 the $(M+N)$ dimensional evaluation representation of $U_{q}(\widehat{gl}(M|N))$. 
It is just a representative of an infinite set of fusion models. 
The fusion models related to $U_{q}(\widehat{gl}(M|N))$ were studies in \cite{DA90,DM92}. 
In this section, we briefly reformulate formulae for the $\Ts$-functions of the 
fusion hierarchies of the Perk-Schultz-type models from the 
analytic Bethe ansatz in our previous papers \cite{T97,T98,T98-2}, keeping in mind 
recent developments \cite{KSZ07,Zabrodin07,BT08}. 

Let us introduce the following sets 
\begin{equation}
\begin{split}
&{\mathfrak I}:=\{ 1,2,\dots,M+N \}= {\mathfrak B} \sqcup {\mathfrak F}, \\[10pt]
&   {\mathfrak B}:=\{ 1,2,\dots,M\} , \qquad 
        {\mathfrak F}:=\{ M+1,M+2,\dots,M+N \}, 
\end{split}
  \label{set}
\end{equation}
and the grading parameters 
\begin{eqnarray}
p_{a}=1 \quad \text{for} \quad a  \in {\mathfrak B} \quad \text{and} \quad 
p_{a}=-1 \quad \text{for} \quad a  \in {\mathfrak F}. 
\end{eqnarray}
Consider any one of the permutations of the components of
 the tuple $(1,2,\dots,M+N )$, and write it
 as a $(M+N)$-tuple $I_{M+N}=(i_{1},i_{2},\dots, i_{M+N})$. 
Of course, $I_{M+N}$ coincides with 
${\mathfrak I}$ as a set. 
Let us take the first $a$-components ($0\le a \le M+N$) of 
the tuple $I_{M+N}$ and write it as $a$-tuple $I_{a}=(i_{1},i_{2},\dots, i_{a})$. 
We also take the rest of $(M+N-a)$ components of $I_{M+N}$, and write it as 
$(M+N-a)$-tuple  $\overline{I}_{a}=(i_{a+1},i_{a+2},\dots, i_{M+N})$. 
Similarly, consider any one of the permutations of the components of the 
tuple $(1,2,\dots,M)$ 
(resp.\ $(M+1,M+2,\dots,M+N)$), and write it
 as a $M$-tuple $B_{M}=(b_{1},b_{2},\dots, b_{M})$ 
(resp.\ $N$-tuple $F_{N}=(f_{1},f_{2},\dots, f_{N})$). 
Let us take the first $m$-components ($0 \le m \le M$) of 
the tuple $B_{M}$ (resp.\ the first $n$-components ($0 \le n \le N$)
 of the tuple $F_{N}$) and write it as a $m$-tuple 
$B_{m}=(b_{1},b_{2},\dots, b_{m})$ (resp.\ a $n$-tuple $F_{n}=(f_{1},f_{2},\dots, f_{n})$). 
We will use a symbol $ B_{m}\times F_{n}:=(b_{1},\dots,b_{m},f_{1},\dots,f_{n})$. 
We will denote $I_{0},\overline{I}_{M+N},B_{0},F_{0}$ as $\emptyset$. 
We also use the symbol $I$ to denote any one of the tuples $I_{0},I_{1},\dots, I_{M+N}$. 
In case we need not mind order of the components of these tuples $I_{a},B_{m},F_{n}$, 
we will treat these as just sets, but will use the same symbols. 
We have $(M+N)!$ different choices of the tuple $I_{M+N}$. 
There are $\binom{M+N}{a}$ different choices of $I_{a}$ (as sets) for each $a$, 
and in total 
$2^{M+N}$ for all $a\in \{0,1,\dots, M+N \}$. 
As sets, they form a poset with respect to the inclusion relations. 
Each $\{I_{a} \}_{a=0}^{M+N}$ forms a chain as a family of sets:
\begin{eqnarray}
\emptyset =
I_{0}\subset I_{1}\subset \cdots  \subset 
I_{M+N}, 
\qquad {\mathrm {Card}}(I_{a} \setminus I_{a-1})=1, \qquad a \in {\mathfrak I}, \label{Isets}
\end{eqnarray}
where ${\mathrm {Card}}(J)$ is the number of the elements of the set $J$. 
We also have an opposite  family of sets: 
$I_{M+N}=\overline{I}_{0}\supset\overline{I}_{1}\supset \cdots \supset 
\overline{I}_{M+N}=\emptyset$. 
%

We will consider functions $\{\Qs_{I_{a}}(x)\}_{a=0}^{M+N}$ 
of a multiplicative spectral parameter $x \in {\mathbb C}$ labeled by $\{I_{a}\}_{a=0}^{M+N}$. 
These are the so-called Baxter $\Qs$-functions
\footnote{The Baxter $\Qs$-operators $({\mathbb A}_{1}(x),{\mathbb A}_{2}(x),
{\mathbb A}_{3}(x),\overline{\mathbb A}_{1}(x),
\overline{\mathbb A}_{2}(x),\overline{\mathbb A}_{3}(x))$
for $U_{q}(\widehat{sl}(2|1))$ in \cite{BT08} 
corresponds to the functions $(\Qs_{\{2,3\}}(x),\Qs_{\{1,3\}}(x),\Qs_{\{1,2\}}(x),
\Qs_{\{1\}}(x),\Qs_{\{2\}}(x),\Qs_{\{3\}}(x))$ for $(M,N)=(2,1)$ with $z_{1}z_{2}z_{3}^{-1}=1$  
in this paper. $q$ in \cite{BT08} corresponds to $q^{-1}$ here.}, which will be discussed 
in what follows. 
We will use a notation  
\begin{eqnarray}
\overline{\Qs}_{I_{a}}(x):=\Qs_{\overline{I}_{a}}(x). \label{jouken}
\end{eqnarray}
We will treat the tuple $I_{a}$ just a set for 
$\Qs_{I_{a}}(x)$ and $\overline{\Qs}_{I_{a}}(x)$, where $0 \le a \le M+N$. 
For example, $\Qs_{(1,2,3)}(x),\Qs_{(2,3,1)}(x),\Qs_{(3,1,2)}(x),
\Qs_{(1,3,2)}(x),\Qs_{(3,2,1)}(x),\Qs_{(2,1,3)}(x)$ 
are the same function and denoted as $ \Qs_{\{1,2,3\}}(x)$. 
There are $2^{M+N}$ Baxter $\Qs$-functions, in total. 
In this paper, we normalize the Baxter $\Qs$-functions as 
\begin{align}
\Qs_{J}(0)=1 \qquad \text{for any} \qquad J \subset {\mathfrak I}.
\label{Q-norm}
\end{align}

Before we proceed discussions, we must comment on a subset of the $2^{M+N}$ Baxter $\Qs$-functions. 
Let us consider the case where the components of the tuple $I_{M+N}$ 
are arranged in the following way:  
for any $a \in \{0,1,\dots, M+N\}$, $I_{a}$ has the form 
$\{1,2,\dots,m\}\sqcup \{M+1,M+2,\dots,M+n\}$ as a set, 
where 
$m={\rm Card}(I_{a}\cap {\mathfrak B})$, $n={\rm Card}(I_{a}\cap {\mathfrak F})$ and 
$a=m+n$.  
There are $\frac{(M+N)!}{M!N!}$ different 
choice of such $I_{M+N}$. And in total,   
there are  $(M+1)(N+1)$-$\Qs$-functions 
$\{\Qs_{\{1,2,\dots,m\}\sqcup \{M+1,M+2,\dots,M+n\}}(x)\}_{0 \le m \le M, 0 \le n \le N}$ 
labeled by such $\{I_{a}\}_{a=0}^{M+N}$. 
These $(M+1)(N+1)$-$\Qs$-functions (for rational case) 
essentially correspond to the ones considered in \cite{KSZ07,Zabrodin07}. 

The eigenvalue formula of the transfer matrix of the Perk-Schultz-type model 
by the Bethe ansatz has the following form
\footnote{Here we are considering the case where the 
auxiliary space is an $(M+N)$ dimensional evaluation representation of
 $U_{q}(\widehat{gl}(M|N))$. 
The quantum space of the original Perk-Schultz model is also the 
$(M+N)$ dimensional representation 
on each site. In our normalization of the Baxter $\Qs$-functions, 
the vacuum part is included in the Baxter $\Qs$-functions. Thus the formula 
has the same form even if the model has a 
 more general quantum space \cite{Kul85,T98-2,KSZ07,BR08}.
}
 \cite{Sc83}:
\begin{eqnarray}
{\mathcal F}_{(1)}^{I_{M+N}}(x)=
\sum_{a=1}^{M+N}p_{i_{a}}{\mathcal   X}_{I_{a}}(x), \label{tab-fund}
\end{eqnarray}
where the functions $\{{\mathcal   X}_{I_{a}}(x)\}_{a=1}^{M+N}$ are defined in the 
following way. 
Let us take a  
 $K$-tuple ${\mathtt I}=(\gamma_{1},\gamma_{2},\dots,\gamma_{K})$ 
whose components are mutually distinct elements of ${\mathfrak I}$, 
where $1 \le K \le M+N$. 
We also introduce 
$(K-1)$-tuples 
$ \hat{\mathtt I}=(\gamma_{1},\gamma_{2},\dots,\gamma_{K-1})$ and 
$ \check{\mathtt I}=(\gamma_{2},\gamma_{3},\dots,\gamma_{K})$.  
Then we define functions labeled by ${\mathtt I}$ as 
\begin{align}
{\mathcal   X}_{{\mathtt I}}(x)&:=
z_{\gamma_{K}}
\frac{\Qs_{\hat{\mathtt I}}
(xq^{-\sum_{j \in \hat{\mathtt I}}p_{j}-2p_{\gamma_{K}}
+\frac{M-N}{2}})
\Qs_{{\mathtt I}}
(xq^{-\sum_{j \in {\mathtt I}}p_{j}+2p_{\gamma_{K}}
+\frac{M-N}{2}})
}{
\Qs_{\hat{\mathtt I}}
(xq^{-\sum_{j \in \hat{\mathtt I}}p_{j}
+\frac{M-N}{2}})
\Qs_{{\mathtt I}}
(xq^{-\sum_{j \in {\mathtt I}}p_{j}+\frac{M-N}{2}})
} 
\label{boxes} 
\end{align}
and
\begin{align}
\overline{\mathcal   X}_{{\mathtt I}}(x)&:=
z_{\gamma_{1}}
\frac{\overline{\Qs}_{\check{\mathtt I}}
(xq^{\sum_{j \in \check{\mathtt I}}p_{j}+2p_{\gamma_{1}}
-\frac{M-N}{2}})
\overline{\Qs}_{{\mathtt I}}
(xq^{\sum_{j \in {\mathtt I}}p_{j}-2p_{\gamma_{1}}
-\frac{M-N}{2}})
}{
\overline{\Qs}_{\check{\mathtt I}}
(xq^{\sum_{j \in \check{\mathtt I}}p_{j}
-\frac{M-N}{2}})
\overline{\Qs}_{{\mathtt I}}
(xq^{\sum_{j \in {\mathtt I}}p_{j}-\frac{M-N}{2}})
} 
. \label{boxes2}
\end{align}
We note that 
the relation 
$\overline{{\mathcal   X}}_{\overline{I}_{a-1}}(x)={\mathcal   X}_{I_{a}}(x)$ holds  
due the relation $\sum_{j\in I_{a}}p_{j}+\sum_{j\in \overline{I}_{a}}p_{j}=M-N
$ and 
\eqref{jouken}.
  The parameters $z_{a} \in {\mathbb C}$ correspond to boundary twists or horizontal fields; 
namely we are considering a transfer matrix with a twisted boundary condition. 
The functions $\Qs_{I}(x)$ for $I=\emptyset $ or $I=I_{M+N} $ 
are special. 
In this paper, we will 
normalize these as 
\begin{eqnarray}
\Qs_{\emptyset}(x)=1 \label{tri1}
\end{eqnarray}
or 
\begin{eqnarray}
\overline{\Qs}_{\emptyset}(x)=1. \label{tri2}
\end{eqnarray}
In the normalization of the universal $R$-matrix, the 
conformal field theoretical models or the $q$-(super)character, 
we impose both \eqref{tri1} and \eqref{tri2}, at least for the case $M \ne N$ 
(cf.\ \cite{BLZ97,BHK02,Kojima08,BT08}; see Appendix \ref{vaumm}). 
For the Perk-Schultz model, we may use the normalization \eqref{tri2} and 
\begin{align}
\overline{\Qs}_{I_{M+N}}(x)=\prod_{j=1}^{L}\left(1-\frac{x}{w_{j}}\right),
\label{quantum-wron-ps}
\end{align} 
where  $L$ is the number of the lattice site; 
$w_{j}$ is an inhomogeneity on the spectral parameter on the $j$-th site 
of the quantum space (the homogeneous case corresponds to $w_{j}=1$). 
\eqref{quantum-wron-ps} corresponds to the so-called {\em quantum Wronskian condition}. 
In this case, $\{\Qs_{I_{a}}(x)\}_{a=1}^{M+N-1}$ are polynomials of 
$x$ for finite $L$: 
\begin{align}
\Qs_{I_{a}}(x)=\prod_{k=1}^{n_{I_{a}}}\left(1-\frac{x}{x_{k}^{I_{a}}}\right), 
\label{baxterQ-root}
\end{align}
where $\{x_{k}^{I_{a}}\}$ are roots
\footnote{Here we treat the tuple $I_{a}$ just a set for $x_{k}^{I_{a}}$.}
 of the Bethe ansatz equation: 
\begin{align}
& -1=\frac{p_{i_{a+1}}z_{i_{a+1}}}{p_{i_{a}}z_{i_{a}}}
\frac{\Qs_{I_{a-1}}(x_{k}^{I_{a}}q^{p_{i_{a}}})
\Qs_{I_{a}}(x_{k}^{I_{a}}q^{-2p_{i_{a+1}}})
\Qs_{I_{a+1}}(x_{k}^{I_{a}}q^{p_{i_{a+1}}})}
{\Qs_{I_{a-1}}(x_{k}^{I_{a}}q^{-p_{i_{a}}})
\Qs_{I_{a}}(x_{k}^{I_{a}}q^{2p_{i_{a}}})
\Qs_{I_{a+1}}(x_{k}^{I_{a}}q^{-p_{i_{a+1}}})} 
\label{BAE}
\\[3pt]
& \hspace{130pt}
\text{for} \quad k\in \{1,2,\dots, n_{I_{a}}\} \quad \text{and} \quad a \in \{1,2,\dots, M+N-1 \} .
\nonumber 
\end{align}
The condition that the poles of \eqref{tab-fund}  
from the zeros of $\Qs_{I_{a}}(x)$ ($1\le a \le M+N-1$) chancel each other:  
\begin{align}
{\rm Res}_{x=x_{k}^{I_{a}}q^{\sum_{j \in I_{a}}p_{j}-\frac{M-N}{2}}}
(p_{i_{a}}{\mathcal   X}_{I_{a}}(x)+p_{i_{a+1}}{\mathcal   X}_{I_{a+1}}(x))=0 
\label{chancels}
\end{align}
produces the Bethe ansatz equation \eqref{BAE}.
One can obtain a Bethe ansatz system, which has one-to-one correspondence with 
the above mentioned one for the Perk-Schultz model by the 
 normalization \eqref{tri1} and 
\begin{align}
\Qs_{I_{M+N}}(x)=\prod_{j=1}^{L}\left(1-\frac{x}{w_{j}}\right). 
\label{quantum-wron-ps2}
\end{align} 
The rational case of this system  corresponds to 
 the one considered in \cite{KSZ07,Zabrodin07}.
In the context of the theory of the $q$-character \cite{FR99}, (\ref{tab-fund}) 
corresponds to the $q$-(super)character
\footnote{
Difference between the supercharacter and the character is not essential here as 
the supercharacter becomes the character by the transformation: 
$z_{a}\to -z_{a}$ for $a \in {\mathfrak F}$. 
In the theory of the $q$-characters, one usually includes 
the parameters $\{z_{a}\}$ into the Baxter $\Qs$-functions and 
writes the $q$-characters in terms of the variables of 
the form $Y_{I_{a}}(x)=\Qs_{I_{a}}(xq)/\Qs_{I_{a}}(xq^{-1})$.  
}
 for the $M+N$ dimensional evaluation representation of 
$U_{q}(\widehat{gl}(M|N))$, and ${\mathcal X}_{I_{1}}(x)$
 is called the ``highest weight monomial''. 

Let us write the symmetric group over the components of 
the tuple $I$ (or elements of the set $I$) as $S(I)$.  
We assume that $\sigma \in S(I)$ acts on $a \in {\mathfrak I}\setminus I$ 
as $\sigma(a)=a$. 
The action on ${\mathcal F}^{I_{M+N}}_{(1)}(x)$ is defined as 
$\sigma[{\mathcal F}^{I_{M+N}}_{(1)}(x)]:={\mathcal F}^{\sigma(I_{M+N})}_{(1)}(x)
={\mathcal F}^{(\sigma(i_{1}),\sigma(i_{2}),\dots, \sigma(i_{M+N}))}_{(1)}(x)$. 
If one interprets the action of $\sigma $ on ${\mathcal X}_{I_{a}}(x)$, $\Qs_{I_{a}}(x)$,
$z_{a}$ and $p_{a}$ as 
$\sigma[{\mathcal X}_{I_{a}}(x)]={\mathcal X}_{\sigma(I_{a})}(x)$, 
$\sigma [\Qs_{I_{a}}(x)]=\Qs_{\sigma (I_{a})}(x)$, 
$\sigma(z_{a})=z_{\sigma(a)},\sigma(p_{a})=p_{\sigma(a)}$, then 
one sees that these are compatible with the definitions \eqref{tab-fund} and \eqref{boxes}. 
Note that 
$\sigma [\Qs_{J}(x)]=\Qs_{J}(x)$ for any 
$\sigma \in S(I)$ such that $I \subset J$ or $I \subset {\mathfrak I}\setminus J$. 
The direct product of 
 the symmetric groups $S({\mathfrak B}) \times S({\mathfrak F})$ 
corresponds to the Weyl group of $gl(M|N)$ if the parameters $\{z_{a}\}$ are 
interpreted as in Appendix \ref{representation}. 
The $\Qs$-functions $\Qs_{I}(x)$ with the same ``level'' ${\rm Card}(I)$ are in 
the same ``$S({\mathfrak I})$-orbit''. 

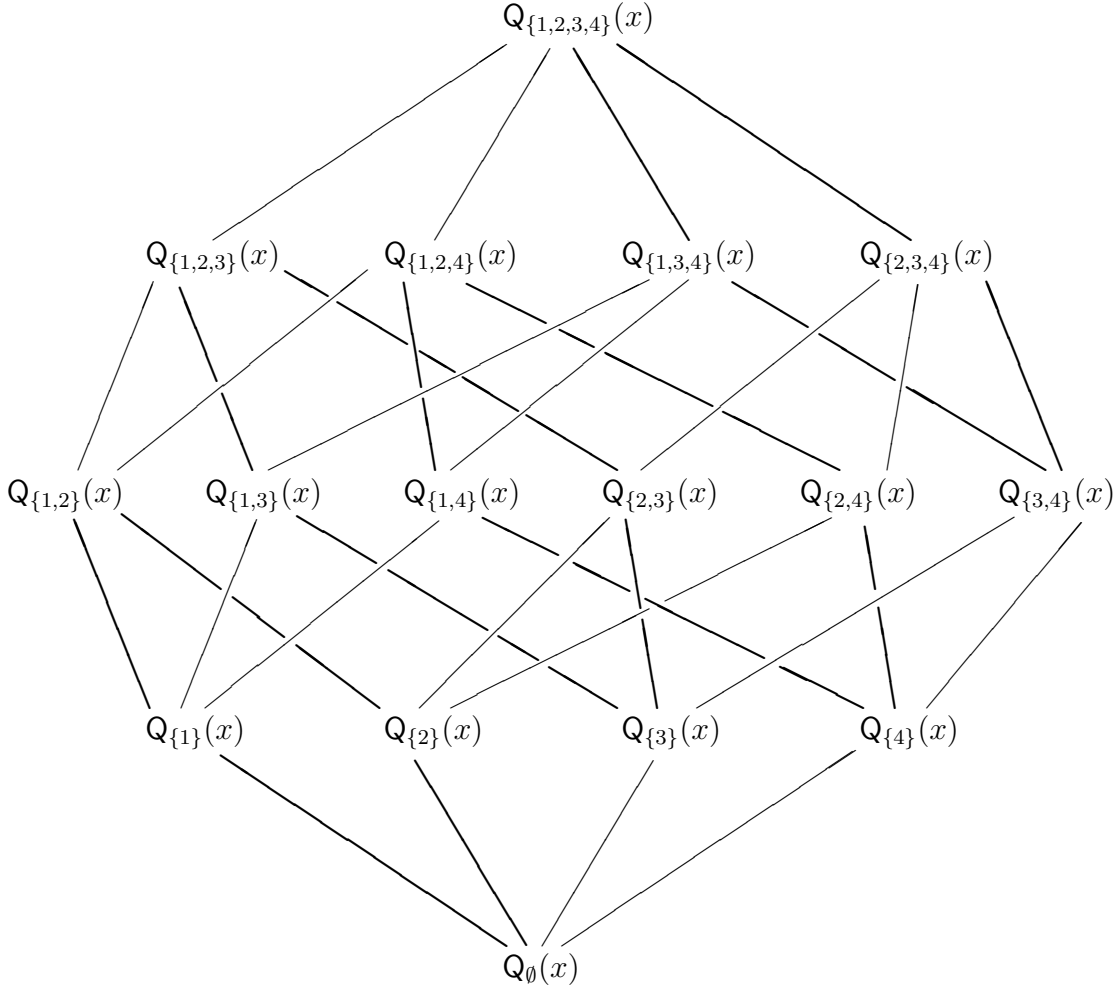
\begin{figure}
  \begin{center}
    \setlength{\unitlength}{3pt}
    \begin{picture}(140,125) 
      {\thicklines 
      \put(63,4){\line(-3,2){36}}
      \put(65.5,4){\line(-3,5){14}}
      }
      \put(67.5,4){\line(3,5){14}}
      \put(70,4){\line(3,2){37}}
      {\thicklines
      \put(18,34){\line(-2,5){9.5}}
      }
      \put(22,34){\line(2,5){9.5}}
      \put(25,34){\line(5,4){29.5}}
       {\thicklines 
       \put(47,34){\line(-4,3){10.5}} 
       \put(35,43){\line(-4,3){6.5}}
       \put(27,49){\line(-4,3){12.5}}
      }
      \put(52,34){\line(1,1){24}}
      \put(56,34){\line(2,1){25}}
      \put(83,47.5){\line(2,1){21}}
      {\thicklines 
      \put(77,34){\line(-5,3){8.6}}
      \put(67,40){\line(-5,3){5}}
      \put(60.6,43.84){\line(-5,3){12.5}}
      \put(46.6,52.25){\line(-5,3){10}}
      %
      \put(82,34){\line(-1,6){1.85}}
      \put(79.9,46.6){\line(-1,6){1.85}}
      }
      \put(87,34){\line(5,3){40}}
      {\thicklines 
      \put(108,34){\line(-2,1){10.5}}
      \put(95.5,40.25){\line(-2,1){15}}
      \put(78.5,48.75){\line(-2,1){7}}
      \put(69.5,53.25){\line(-2,1){9.5}}
      %
      \put(112,34){\line(-1,6){2.1}}
      \put(109.5,49){\line(-1,6){1.3}}
      }
      \put(116,34){\line(5,6){19.5}}
      \put(9,64){\line(2,5){9.5}}
      \put(14,64){\line(5,4){31.5}} 
      {\thicklines
      \put(31,64){\line(-2,5){3.85}}
      \put(26.5,75.25){\line(-2,5){4.7}}
      }
      \put(33,64){\line(2,1){48}}
      {\thicklines 
      \put(54,64){\line(-1,6){1.5}}
      \put(52.25,74.5){\line(-1,6){2.2}}
      }
      \put(56,64){\line(5,4){30}} 
      {\thicklines 
      \put(77,64){\line(-5,3){11.5}}
      \put(64.5,71.5){\line(-5,3){7}}
      \put(56.5,76.3){\line(-5,3){4.4}}
      \put(50.5,79.9){\line(-5,3){8.8}}
      \put(40.5,85.9){\line(-5,3){5.4}}
      }
      \put(80,64){\line(5,4){30}} 
      {\thicklines 
      \put(105,64){\line(-2,1){14.7}}
      \put(89,72){\line(-2,1){13.5}}
      \put(74,79.5){\line(-2,1){4.3}}
      \put(68,82.5){\line(-2,1){10}}
      }
      \put(111,64){\line(1,6){3.9}}
       {\thicklines 
      \put(131,64){\line(-5,3){17.5}}
      \put(112,75.4){\line(-5,3){9.5}}
      \put(101,82){\line(-5,3){9.5}}
      \put(133,64){\line(-2,5){9.5}}
      }
      \put(26,93){\line(3,2){37}}
      \put(54,93){\line(3,5){14.5}}
      {\thicklines
      \put(86,93){\line(-3,5){14.5}}
      }
      {\thicklines
      \put(114,93){\line(-3,2){37}}
      }
      \put(62.5,120){$\Qs_{\{1,2,3,4\}}(x)$}
      \put(17.5,90){$\Qs_{\{1,2,3\}}(x)$}
      \put(47.5,90){$\Qs_{\{1,2,4\}}(x)$}
      \put(77.5,90){$\Qs_{\{1,3,4\}}(x)$}
      \put(107.5,90){$\Qs_{\{2,3,4\}}(x)$}
      \put(0,60){$\Qs_{\{1,2\}}(x)$}
      \put(25,60){$\Qs_{\{1,3\}}(x)$}
      \put(50,60){$\Qs_{\{1,4\}}(x)$}
      \put(75,60){$\Qs_{\{2,3\}}(x)$}
      \put(100,60){$\Qs_{\{2,4\}}(x)$}
      \put(125,60){$\Qs_{\{3,4\}}(x)$}
      \put(17.5,30){$\Qs_{\{1\}}(x)$}
      \put(47.5,30){$\Qs_{\{2\}}(x)$}
      \put(77.5,30){$\Qs_{\{3\}}(x)$}
      \put(107.5,30){$\Qs_{\{4\}}(x)$}
      \put(62.5,0){$\Qs_{\emptyset}(x)$}
    \end{picture}
  \end{center}
  \caption{The hierarchy of the $2^{4}$ Baxter $\Qs$-functions for $U_{q}(\widehat{gl}(2|2))$. 
 The functions $\Qs_{\emptyset}(x)$ and $\Qs_{\{1,2,3,4\}}(x)$ are special on the point that 
 they do not contain roots of the Bethe ansatz equation, and often normalized as just $1$. 
  The index sets $I$ of the Baxter $\Qs$-functions $\Qs_{I}(x)$ 
form a Hasse diagram. $\Qs_{I_{a}}(x)$ and $\Qs_{I_{a-1}}(x)$ are connected by a thick line 
if $I_{a}\setminus I_{a-1} \subset {\mathfrak B}=\{1,2\}$, 
a thin line if  $I_{a}\setminus I_{a-1} \subset {\mathfrak F}=\{3,4\}$. 
As a graph, this contains many cycles. 
Minimal non-trivial cycles are 4-cycles, which  
correspond to functional relations among the Baxter $\Qs$-functions on 
the vertexes. 
4-cycles containing only thick lines correspond to \eqref{QQ-rel1} for $p_{i}=p_{j}=1$; 
only thin lines correspond to \eqref{QQ-rel1} for $p_{i}=p_{j}=-1$; 
both thick lines and thin lines correspond to \eqref{QQ-rel2}. 
In addition, the function ${\mathcal X}_{I_{a}}(x)$
 (\eqref{boxes} for ${\mathtt I}=I_{a}$) lives on the edge which connects 
two vertexes for 
$\Qs_{I_{a}}(x)$ and $\Qs_{I_{a-1}}(x)$. Thus the function ${\mathcal F}_{(1)}^{I_{a}}(x)$ 
 (\eqref{DVF-tab1} for $\lambda \subset \mu =(1)$, ${\mathtt I}=I_{a}$, $M=N=2$) 
lives on a path from $\Qs_{I_{0}}(x)=\Qs_{\emptyset}(x)$ to $\Qs_{I_{a}}(x)$; 
the function $\overline{\mathcal F}_{(1)}^{\overline{I}_{a}}(x)$
 (cf. \eqref{DVF-tab2} for $\lambda \subset \mu =(1)$, 
${\mathtt I}=\overline{I}_{a}$, $M=N=2$)  
lives on a path from $\Qs_{I_{a}}(x)$ to $\Qs_{I_{4}}(x)=\Qs_{\{1,2,3,4\}}(x)$, 
where these paths must not contain more than two functions $\Qs_{I}(x),\Qs_{J}(x)$ 
of the same level ${\rm Card}(I)={\rm Card}(J)$. In particular, 
they do not depend on the paths.}
  \label{Q-fun-fig}
\end{figure}

As already remarked many times in various context 
\cite{Woynarovich83,T98,GS03,BKSZ05,BDKM06,KSZ07,GV07,BT08}, 
(\ref{tab-fund}) does not depend on the choice of the tuple $I_{M+N}$ since 
there are relations among the Baxter $\Qs$-functions (cf.\ Figure \ref{Q-fun-fig}). 
%
Let us consider a permutation $\tau \in S(I_{M+N})=S({\mathfrak I})$ 
such that $\tau(i_{a})=i_{a+1}, \tau(i_{a+1})=i_{a}$ and $\tau(i_{b})=i_{b}$ 
for $b \ne a,a+1$,  
for a fixed $a \in \{1,2,\dots, M+N-1\}$. 
Note that $\tau[\Qs_{I_{b}}(x)]=\Qs_{I_{b}}(x)$ if $b\ne a$. 
Thus  $\tau[{\mathcal   X}_{I_{b}}(x)]={\mathcal   X}_{I_{b}}(x)$ if 
$b \ne a,a+1$. Let us write $i_{a}=i,i_{a+1}=j,I_{a-1}=I$. 
The condition 
$\tau[ {\mathcal F}_{(1)}^{I_{M+N}}(x)]={\mathcal F}_{(1)}^{I_{M+N}}(x)$ 
is equivalent to 
\begin{eqnarray}
p_{i}{\mathcal   X}_{I_{a}}(x)+
p_{j}{\mathcal   X}_{I_{a+1}}(x)=
p_{j}{\mathcal   X}_{\tau(I_{a})}(x)+p_{i}{\mathcal   X}_{\tau(I_{a+1})}(x).
\label{eqiv}
\end{eqnarray}
This implies the following functional relation (cf.\ \cite{Pronko-Stroganov00,BHK02,DDMST06,BDKM06,GV07,BT08}) 
\footnote{Functional relations similar to 
\eqref{QQ-rel1} appeared in \cite{Pronko-Stroganov00} 
for rational vertex models related to $A_{2}$; 
in \cite{DDMST06} in relation to the ODE/IM correspondence for 
$A_{n},B_{n},C_{n},D_{n}$. 
 \eqref{QQ-rel1} appeared in \cite{BHK02} 
for ${\mathcal W}_{3}$ CFT related to $U_{q}(\widehat{sl}(3))$. 
As for the superalgebra related models, functional relations 
similar to \eqref{QQ-rel1} appeared in \cite{BDKM06} 
for rational non-compact $sl(2|1)$ spin chains; 
in \cite{GV07}, in the context of the 
AdS/CFT correspondence, for rational models related to $su(M|N)$. 
In \cite{BT08}, we proposed \eqref{QQ-rel1} 
 for both CFT and trigonometric vertex models associated with 
$U_{q}(\widehat{sl}(2|1))$. 
We had also reported 
these functional relations \eqref{QQ-rel1}-\eqref{QQ-rel2} 
and some Wronskian-like formulae on $T$- and $\Qs$-{\em operators} 
for $U_{q}(\widehat{sl}(2|1))$ case, 
before the preprint version of \cite{BT08} appeared,  
on many conferences which include the following two: 
``Workshop and Summer School: 
From Statistical Mechanics to Conformal and Quantum Field Theory'', 
the university of Melbourne, January, 2007 
[http://www.smft2007.ms.unimelb.edu.au/program/LectureSeries.html];  
La 79eme Rencontre entre physiciens theoriciens et mathematiciens 
``Supersymmetry and Integrability", IRMA Strasbourg, June, 2007 [http://www-irma.u-strasbg.fr/article383.html].}: 
\begin{align}
& (z_{i}-z_{j})\Qs_{I}(x)\Qs_{I\sqcup \{i,j\}}(x)
=z_{i}\Qs_{I\sqcup \{i\}}(xq^{p_{i}})
\Qs_{I\sqcup \{j\}}(xq^{-p_{i}})-
z_{j}\Qs_{I\sqcup \{i\}}(xq^{-p_{i}})
\Qs_{I\sqcup \{j\}}(xq^{p_{i}}) \nonumber \\
& \hspace{170pt} \text{for} \qquad p_{i}=p_{j}, \label{QQ-rel1}
\end{align}
and also the following (cf.\ \cite{GS03,BKSZ05,BDKM06,KSZ07,Zabrodin07,BT08}): 
\begin{align}
& (z_{i}-z_{j})
\Qs_{I\sqcup \{i\}}(x)\Qs_{I \sqcup \{j \}}(x)=
z_{i}\Qs_{I}(xq^{-p_{i}})
\Qs_{I\sqcup \{i,j\}}(xq^{p_{i}})-
z_{j}\Qs_{I}(xq^{p_{i}})
\Qs_{I\sqcup \{i,j\}}(xq^{-p_{i}}) 
\nonumber \\
& \hspace{170pt} \text{for} \qquad p_{i}=-p_{j}. \label{QQ-rel2}
\end{align}
In fact, one can easily show the relation (\ref{eqiv}) based on the functional relations 
(\ref{QQ-rel1}) and (\ref{QQ-rel2}). 
Functional relations corresponding to \eqref{QQ-rel2} were discussed in \cite{KSZ07,Zabrodin07} 
from a point view of a classical theory of the soliton, and also to 
\eqref{QQ-rel1} in \cite{GV07}. 
In addition, 
\eqref{QQ-rel1} was proved based on decompositions of 
$q$-oscillator representations of 
 $U_{q}(\widehat{sl}(2))$ \cite{BLZ97}, 
$U_{q}(\widehat{sl}(3))$ \cite{BHK02}
 and $U_{q}(\widehat{sl}(2|1))$ \cite{BT08}. 
\eqref{QQ-rel2} was also proved similarly for $U_{q}(\widehat{sl}(2|1))$ \cite{BT08}. 
For $U_{q}(\widehat{gl}(2))$ case, \eqref{QQ-rel1} reduces to 
the quantum Wronskian condition. 
One may also unify the above two relations.
\begin{multline}
z_{i}\Qs_{I\sqcup \{i\}}(xq^{\frac{p_{i}+p_{j}}{2}})
\Qs_{I \sqcup \{j \}}(xq^{-\frac{p_{i}+p_{j}}{2}})-
z_{j}\Qs_{I\sqcup \{i\}}(xq^{-\frac{p_{i}+p_{j}}{2}})
\Qs_{I \sqcup \{j \}}(xq^{\frac{p_{i}+p_{j}}{2}})  \\ 
=z_{i}\Qs_{I}(xq^{-\frac{p_{i}-p_{j}}{2}})
\Qs_{I\sqcup \{i,j\}}(xq^{\frac{p_{i}-p_{j}}{2}})-
z_{j}\Qs_{I}(xq^{\frac{p_{i}-p_{j}}{2}})
\Qs_{I\sqcup \{i,j\}}(xq^{-\frac{p_{i}-p_{j}}{2}}).
\end{multline}
Note that the relations (\ref{QQ-rel1}) and (\ref{QQ-rel2}) can be rewritten as 
\begin{align}
& (z_{i}-z_{j})\overline{\Qs}_{I}(x)\overline{\Qs}_{I\sqcup \{i,j\}}(x)
=z_{i}\overline{\Qs}_{I\sqcup \{i\}}(xq^{-p_{i}})
\overline{\Qs}_{I\sqcup \{j\}}(xq^{p_{i}})-
z_{j}\overline{\Qs}_{I\sqcup \{i\}}(xq^{p_{i}})
\overline{\Qs}_{I\sqcup \{j\}}(xq^{-p_{i}}) \nonumber \\
& \hspace{170pt} \text{for} \qquad p_{i}=p_{j},  \label{QQ-rel3}
\\[5pt]
& (z_{i}-z_{j})
\overline{\Qs}_{I\sqcup \{i\}}(x)\overline{\Qs}_{I \sqcup \{j \}}(x)=
z_{i}\overline{\Qs}_{I}(xq^{p_{i}})
\overline{\Qs}_{I\sqcup \{i,j\}}(xq^{-p_{i}})-
z_{j}\overline{\Qs}_{I}(xq^{-p_{i}})
\overline{\Qs}_{I\sqcup \{i,j\}}(xq^{p_{i}}) 
\nonumber \\
& \hspace{170pt} \text{for} \qquad p_{i}=-p_{j}. \label{QQ-rel4}
\end{align}
The form of the functional relations \eqref{QQ-rel1}-\eqref{QQ-rel2} are invariant under 
the gauge transformation 
$\Qs_{I}(x) \to g_{1}(xq^{\sum_{k \in I}p_{k}})g_{2}(xq^{-\sum_{k \in I}p_{k}})\Qs_{I}(x)$ 
for all $I \subset {\mathfrak I}$, 
where $g_{1}(x),g_{2}(x)$ are arbitrary functions of $x$. 
Thus one can always chose the normalization \eqref{tri1} based on this freedom. 
The same discussion can be applied for \eqref{QQ-rel3}-\eqref{QQ-rel4} and \eqref{tri2}. 
In particular for $I=\emptyset$ case, (\ref{QQ-rel2}) with the normalization \eqref{tri1} reduces to 
\begin{align}
  (z_{i}-z_{j})
\Qs_{i}(x)\Qs_{j}(x)=
z_{i}
\Qs_{ \{i,j\}}(xq)-
z_{j}
\Qs_{\{i,j\}}(xq^{-1}) 
\qquad 
\text{for} \quad i \in {\mathfrak B}, \quad j \in {\mathfrak F},\label{QQ-bf1} 
\end{align}
and (\ref{QQ-rel4}) with \eqref{tri2} reduces to
\begin{align}
& (z_{i}-z_{j})
\overline{\Qs}_{i}(x)\overline{\Qs}_{j}(x)=
z_{i}
\overline{\Qs}_{ \{i,j\}}(xq^{-1})-
z_{j}
\overline{\Qs}_{\{i,j\}}(xq) 
\qquad 
\text{for} \quad i \in {\mathfrak B}, \quad j \in {\mathfrak F}, \label{QQ-bf2}
\end{align}
where we used abbreviations $\Qs_{i}(x):=\Qs_{\{i\}}(x)$ and 
$\overline{\Qs}_{i}(x):=\overline{\Qs}_{\{i\}}(x)$. 
Thus $\Qs_{\{i,j\}}(x)$ with \eqref{tri1} 
(resp.\ $\overline{\Qs}_{\{i,j\}}(x)$ with \eqref{tri2}) can be expressed as a 
kind of {\em convolution} of 
 $\Qs_{i}(x)$ and $\Qs_{j}(x)$ (resp.\  $\overline{\Qs}_{i}(x)$ and $\overline{\Qs}_{j}(x)$): 
 $\Qs_{\{i,j\}}(x)=\frac{z_{i}-z_{j}}{z_{i}{\mathbf D}-z_{j}{\mathbf D}^{-1}}\Qs_{i}(x)\Qs_{j}(x)$ 
(resp.\  $\overline{\Qs}_{\{i,j\}}(x)=
\frac{z_{i}-z_{j}}{z_{i}{\mathbf D}^{-1}-z_{j}{\mathbf D}}
\overline{\Qs}_{i}(x)\overline{\Qs}_{j}(x)$), where 
 ${\mathbf D}$ is a difference operator: 
 ${\mathbf D}f(x)=f(xq)$ for any function $f(x)$ of $x$. 
Explicitly, we have 
\begin{eqnarray}
\Qs_{\{i,j\}}(x)=\frac{z_{i}-z_{j}}{z_{i}}\sum_{k=0}^{\infty} \left(\frac{z_{j}}{z_{i}}\right)^{k}
 \Qs_{i}(xq^{-2k-1})\Qs_{j}(xq^{-2k-1}) \label{expan-aij}
\end{eqnarray}
or 
\begin{eqnarray}
\Qs_{\{i,j\}}(x)=\frac{z_{j}-z_{i}}{z_{j}}\sum_{k=0}^{\infty} \left(\frac{z_{i}}{z_{j}}\right)^{k}
 \Qs_{i}(xq^{2k+1})\Qs_{j}(xq^{2k+1})
\end{eqnarray}
depending on their convergence regions. 
Or instead, one may expand the Baxter $\Qs$-function as
 $\Qs_{I}(x)=\sum_{k=0}^{\infty}a_{I}^{(k)}x^{k}$ 
($I=\{i\},\{j\},\{i,j\}$, $a_{I}^{(0)}=1$, $a_{i}^{(k)}:=a_{\{i\}}^{(k)}$)
and plug this into (\ref{QQ-bf1}). 
 Then we obtain
\begin{eqnarray}
\Qs_{\{i,j\}}(x)=\sum_{\alpha=0}^{\infty}\sum_{\beta=0}^{\infty} 
 \frac{(z_{i}-z_{j})a_{i}^{(\alpha)}a_{j}^{(\beta)}}
{q^{\alpha+\beta}z_{i}-q^{-\alpha-\beta}z_{j}}x^{\alpha+\beta}.
\end{eqnarray}

Substitute $x=x_{k}^{I \sqcup \{i\}}q^{-p_{i}}$ and 
 $x=x_{k}^{I \sqcup \{i\}}q^{p_{i}}$ into \eqref{QQ-rel1} 
with \eqref{baxterQ-root}, we obtain 
\begin{align}
& (z_{i}-z_{j})\Qs_{I}(x_{k}^{I \sqcup \{i\}}q^{-p_{i}})
\Qs_{I\sqcup \{i,j\}}(x_{k}^{I \sqcup \{i\}}q^{-p_{i}})
=-z_{j}\Qs_{I\sqcup \{i\}}(x_{k}^{I \sqcup \{i\}}q^{-2p_{i}})
\Qs_{I\sqcup \{j\}}(x_{k}^{I \sqcup \{i\}}), \\[.2cm]
& (z_{i}-z_{j})\Qs_{I}(x_{k}^{I \sqcup \{i\}}q^{p_{i}})
\Qs_{I\sqcup \{i,j\}}(x_{k}^{I \sqcup \{i\}}q^{p_{i}})
=z_{i}\Qs_{I\sqcup \{i\}}(x_{k}^{I \sqcup \{i\}}q^{2p_{i}})
\Qs_{I\sqcup \{j\}}(x_{k}^{I \sqcup \{i\}}).
\end{align}
Let us divide these two equations by one another:
\begin{align}
 -1=\frac{z_{j}}{z_{i}}
\frac{\Qs_{I}(x_{k}^{I\sqcup \{i \}}q^{p_{i}})
\Qs_{I\sqcup \{i \} }(x_{k}^{I \sqcup \{i \}}q^{-2p_{i}})
\Qs_{I \sqcup \{i,j \}}(x_{k}^{I \sqcup \{i \}}q^{p_{i}})}
{\Qs_{I}(x_{k}^{I\sqcup \{i \}}q^{-p_{i}})
\Qs_{I\sqcup \{i \} }(x_{k}^{I \sqcup \{i \}}q^{2p_{i}})
\Qs_{I \sqcup \{i,j \}}(x_{k}^{I \sqcup \{i \}}q^{-p_{i}})} 
\quad \text{for} \quad p_{i}=p_{j}.
\label{BAE-b}
\end{align}
Substitute  $x=x_{k}^{I \sqcup \{i\}}$ into (\ref{QQ-rel2}) 
with \eqref{baxterQ-root}, we obtain
\begin{align}
 1=\frac{z_{j}}{z_{i}}
\frac{\Qs_{I}(x_{k}^{I\sqcup \{i \}}q^{p_{i}})
\Qs_{I \sqcup \{i,j \}}(x_{k}^{I \sqcup \{i \}}q^{-p_{i}})}
{\Qs_{I}(x_{k}^{I\sqcup \{i \}}q^{-p_{i}})
\Qs_{I \sqcup \{i,j \}}(x_{k}^{I \sqcup \{i \}}q^{p_{i}})} 
\quad \text{for} \quad p_{i}=-p_{j}.
\label{BAE-f}
\end{align}
 (\ref{BAE-b}) and (\ref{BAE-f}) for $I=I_{a-1},i=i_{a},j=i_{a+1}$ 
coincides with the Bethe ansatz equation (\ref{BAE}). 

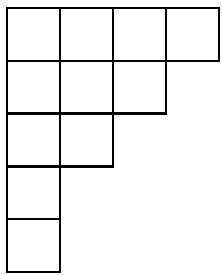
\begin{figure}
  \begin{center}
    \setlength{\unitlength}{2pt}
    \begin{picture}(40,55) 
      \put(0,0){\line(0,1){50}}
      \put(10,0){\line(0,1){50}}
      \put(20,20){\line(0,1){30}}
      \put(30,30){\line(0,1){20}}
      \put(40,40){\line(0,1){10}}
      \put(0,0){\line(1,0){10}}      
      \put(0,10){\line(1,0){10}}
      \put(0,20){\line(1,0){20}}
      \put(0,30){\line(1,0){30}}
      \put(0,40){\line(1,0){40}}
      \put(0,50){\line(1,0){40}}
    \end{picture}
  \end{center}
  \caption{The Young diagram with shape $\mu=(4,3,2,1^2)$.}
  \label{Young1}
\end{figure}

Let us introduce notations for the partitions and Young
diagrams (see \cite{M95} for details). 
A {\em partition} is a non increasing sequence of positive 
 integers $\mu=(\mu_{1},\mu_{2},\dots) $: 
$\mu_{1} \ge \mu_{2} \ge \dots \ge 0$. 
We often write this in the form $\mu=(r^{m_{r}},(r-1)^{m_{r-1}},\dots,2^{m_{2}},1^{m_{1}})$, 
where $r=\mu_{1}$, and 
 $m_{k}={\rm Card}\{j|\mu_{j}=k \}$ is called the {\em multiplicity} of $k$ in $\mu$. 
Two partitions are regarded equivalent if all their non-zero elements of positive 
multiplicity coincide. Elements of the multiplicity zero should be abbreviated. 
For example, $(4,3,2,1,1,0,0)=(4,3,2,1,1)=(5^{0},4,3,2,1^2)=(4,3,2,1^2)$. 
We will also express the partitions $\mu$ as {\em Young diagrams}, and will use 
the same symbol $\mu$ for the Young diagrams. 
The Young diagram $\mu$, corresponding to a partition $\mu$, has 
$\mu_{k}$ boxes in the $k$-th 
row in the plane (see Figure \ref{Young1}).
%
\begin{figure}
  \begin{center}
    \setlength{\unitlength}{2pt}
    \begin{picture}(50,45) 
      \put(0,0){\line(0,1){40}}
      \put(10,0){\line(0,1){40}}
      \put(20,10){\line(0,1){30}}
      \put(30,20){\line(0,1){20}}
      \put(40,30){\line(0,1){10}}  
      \put(50,30){\line(0,1){10}}  
      \put(0,0){\line(1,0){10}}   
      \put(0,10){\line(1,0){20}}
      \put(0,20){\line(1,0){30}}
      \put(0,30){\line(1,0){50}}
      \put(0,40){\line(1,0){50}}
    \end{picture}
  \end{center}
  \caption{The Young diagram for the partition $\mu^{\prime}=(5,3,2,1)$, 
conjugated to $\mu=(4,3,2,1^2)$ in Figure \ref{Young1}.}
  \label{Young2}
\end{figure}
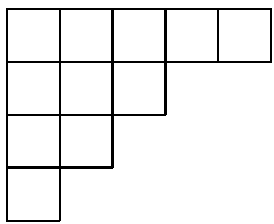
%
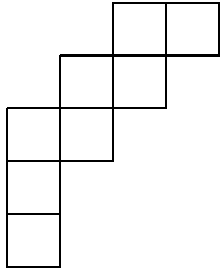
\begin{figure}
  \begin{center}
    \setlength{\unitlength}{2pt}
    \begin{picture}(40,55) 
      \put(0,0){\line(0,1){30}}
      \put(10,0){\line(0,1){40}}
      \put(20,20){\line(0,1){30}}
      \put(30,30){\line(0,1){20}}
      \put(40,40){\line(0,1){10}}
      \put(0,0){\line(1,0){10}}      
      \put(0,10){\line(1,0){10}}
      \put(0,20){\line(1,0){20}}
      \put(0,30){\line(1,0){30}}
      \put(10,40){\line(1,0){30}}
      \put(20,50){\line(1,0){20}}
    \end{picture}
  \end{center}
  \caption{The skew Young diagram $\lambda \subset \mu$ with 
  $\lambda=(2,1)$ and $\mu=(4,3,2,1^2)$.}
  \label{Young3}
\end{figure}
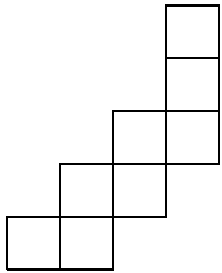
\begin{figure}
  \begin{center}
    \setlength{\unitlength}{2pt}
    \begin{picture}(40,55) 
      \put(40,20){\line(0,1){30}}
      \put(30,10){\line(0,1){40}}
      \put(20,0){\line(0,1){30}}
      \put(10,0){\line(0,1){20}}
      \put(0,0){\line(0,1){10}}
      \put(30,50){\line(1,0){10}}      
      \put(30,40){\line(1,0){10}}
      \put(20,30){\line(1,0){20}}
      \put(10,20){\line(1,0){30}}
      \put(0,10){\line(1,0){30}}
      \put(0,0){\line(1,0){20}}
    \end{picture}
  \end{center}
  \caption{
  The 180 degrees rotation of the skew-Young diagram $\lambda \subset \mu$  with 
$\lambda=(2,1)$ and $\mu=(4,3,2,1^2)$ in Figure \ref{Young3} is the 
skew Young diagram with shape $\widetilde{\lambda \subset \mu}=(3^2,2,1)\subset (4^{3},3,2)$.}
  \label{Young-rot}
\end{figure}
Each box in the Young diagram is specified by a coordinate 
$(i,j)\in {\mathbb Z}^{2}$, 
where the row index $i$ increases as one goes downwards, and the column 
index $j$ increases as one goes from left to right. 
The top left corner of $\mu$ has coordinates $(1,1)$. 
The partition $\mu^{\prime}=(\mu_{1}^{\prime},\mu_{2}^{\prime},\dots)$ 
is called {\em conjugate} of $\mu$, and  
is defined as $\mu_{j}^{\prime}={\rm Card}\{k| \mu_{k} \ge j\}$ (see Figure \ref{Young2}). 
The Young diagram $\mu^{\prime}$ is obtained by the transposition of rows and columns 
of the Young diagram $\mu$. 
Let $\lambda =(\lambda_{1},\lambda_{2},\dots)$ and 
$\mu =(\mu_{1},\mu_{2},\dots)$ be two partitions such that
$\mu_{i} \ge \lambda_{i}: i=1,2,\dots$ and 
$\lambda_{\mu_{1}^{\prime}}=\lambda^{\prime}_{\mu_{1}}=0$. 
We will express a skew-Young diagram defined by these two partitions as 
 $\lambda \subset \mu$. This is the domain given by the subtraction 
 $\mu-\lambda$  (see Figure \ref{Young3}). 
If $\lambda $ is an empty set $\emptyset$, then 
$\lambda \subset \mu$ coincides with $\mu$. 
Each box on the skew-Young diagram 
$\lambda \subset \mu$ is specified by its coordinate on $\mu$. 
We will also use the $180$ degrees rotation of the skew-Young diagram 
$\lambda \subset \mu $, and 
denote it as $\widetilde{\lambda \subset \mu}$ (see Figure \ref{Young-rot}):
\begin{multline}
\widetilde{\lambda \subset \mu}:=
(\underbrace{\mu_{1}-\mu_{\mu_{1}^{\prime}},\mu_{1}-\mu_{\mu_{1}^{\prime}-1}, 
\mu_{1}-\mu_{\mu_{1}^{\prime}-2},\dots,\mu_{1}-\mu_{2},0}_{\mu_{1}^{\prime}} )
\\
\subset 
(\mu_{1}^{\mu_{1}^{\prime}-\lambda_{1}^{\prime}},\mu_{1}-\lambda_{\lambda_{1}^{\prime}},
\mu_{1}-\lambda_{\lambda_{1}^{\prime}-1},
\mu_{1}-\lambda_{\lambda_{1}^{\prime}-2},
\dots, 
\mu_{1}-\lambda_{2},\mu_{1}-\lambda_{1}).
\label{180deg}
\end{multline}
Note that that $ \widetilde{\mu}=\mu $  if 
$\mu $ is 
 a Young diagram of rectangular shape. 

Let us take a $K$-tuple ${\mathtt I}=(\gamma_{1},\gamma_{2},\dots,\gamma_{K})$ 
whose components are mutually distinct elements of ${\mathfrak I}$, 
where $1 \le K \le M+N$. 
We also introduce 
$k$-tuple 
$ {\mathtt D}_{k}({\mathtt I})=(\gamma_{1},\gamma_{2},\dots,\gamma_{k})$ and 
$(K-k+1)$-tuple 
$ \overline{\mathtt D}_{k}({\mathtt I})=(\gamma_{k},\gamma_{k+1},\dots,\gamma_{K})$, 
where $1 \le k \le K$.  
Next, let us define a space of {\em admissible tableaux} 
$\mathsf{Tab}_{{\mathtt I}}(\lambda\subset \mu)$ 
for the $K$-tuple ${\mathtt I}$  
on a (skew) Young diagram $\lambda\subset \mu$. 
%
We assign an integer $t_{ij}$ in each box $(i,j)$ of the diagram. 
An admissible tableau 
$t\in\mathsf{Tab}_{{\mathtt I}}(\lambda\subset \mu)$ 
 is a set of integers $t=\{t_{jk}\}_{(j,k)\in \lambda\subset \mu}$, 
where all $t_{jk} \in \{1, 2, \dots, K \}$  
satisfy the conditions  
\begin{itemize} 
\item[(i)] $t_{jk}\le t_{j+1,k},t_{j,k+1}$ \\
\item[(ii)]  $t_{jk} < t_{j,k+1}$ if $\gamma_{t_{jk}}\in {\mathfrak F}$ or
  $\gamma_{t_{j,k+1}} \in {\mathfrak F}$  \\ 
\item[(iii)]  $t_{jk} < t_{j+1,k}$ if $\gamma_{t_{jk}}\in {\mathfrak B}$ or
  $\gamma_{t_{j+1,k}} \in {\mathfrak B}$.
\end{itemize}

The fusion models are described by the Young diagrams. 
Here we supposed that the fusion is performed in the auxiliary space of 
the transfer matrix. 
In particular, 
the $\Ts$-functions of the fusion models from the analytic Bethe ansatz 
can be written as summation over Young 
tableaux with spectral parameters \cite{T97,T98,T98-2} 
(see also \cite{BR90} for $M>0,N=0$ case, and \cite{KOS95} for $U_{q}(B_{r}^{(1)})$). 
Here we introduce slightly generalized versions of such $\Ts$-functions
\footnote{The Young diagram $\mu$ in 
${\mathcal  F}_{\mu}^{(1,2,\dots,M+N)}(x)$ is related to the representation of 
$gl(M|N)$ as \eqref{kacdynkin}.} 
${\mathcal  F}_{\lambda\subset \mu}^{{\mathtt I}}(x)$ and 
${\overline {\mathcal  F}}_{\lambda \subset \mu}^{{\mathtt I}}(x)$. 
We remark that the following discussion on the $\Ts$-functions is essentially 
independent of the ``vacuum part'', and thus quantum space, 
where the transfer matrices act, is arbitrary. 
For any skew Young diagram $\lambda\subset \mu$,  
define functions
\footnote{Eq.\ (C.8) in \cite{BT08} corresponds to 
${\mathcal  F}_{\lambda \subset \mu}^{(1,2,3)}(x)$
 for $K=3$ and $(M,N)=(2,1)$ with $z_{1}z_{2}z_{3}^{-1}=1$.}:  
\begin{align}
{\mathcal  F}_{\lambda\subset \mu}^{{\mathtt I}}(x)=
\sum_{t\in\mathsf{Tab}_{{\mathtt I}}(\lambda\subset \mu)}
\prod_{(j,k) \in \lambda\subset \mu}
p_{\gamma_{t_{j,k}}}
{\mathcal  X}_{{\mathtt D}_{t_{j,k}}({\mathtt I})}
(xq^{\mu_{1}-\mu_{1}^{\prime}+2j-2k+\frac{m-n}{2}-\frac{M-N}{2}}),
\label{DVF-tab1} 
\end{align}
\begin{align}
{\overline {\mathcal  F}}_{\lambda \subset \mu}^{{\mathtt I}}(x)
=
\sum_{t \in {\mathsf{  Tab}}_{{\mathtt I}}(\lambda \subset \mu)}
\prod_{(j,k) \in \lambda \subset \mu}
p_{\gamma_{t_{j,k}}}
\overline{{\mathcal  X}}_{\overline{\mathtt D}_{t_{j,k}}({\mathtt I})}
(xq^{\mu_{1}-\mu_{1}^{\prime}+2j-2k-\frac{m-n}{2}+\frac{M-N}{2}}), 
\label{DVF-tab2}
\end{align}
where the summations are taken over all admissible tableaux, and 
the products are taken over all boxes of the 
Young diagram $\lambda\subset \mu$;   
$m:={\rm card} ({\mathtt I} \cap {\mathfrak B})$, 
$n:={\rm card} ({\mathtt I} \cap {\mathfrak F})$. 
We also set ${\mathcal  F}_{\emptyset}^{I}(x)=
{\overline {\mathcal  F}}_{\emptyset}^{I}(x)=1$, and 
${\mathcal  F}_{\mu}^{\emptyset}(x)=
{\overline {\mathcal  F}}_{\mu}^{\emptyset}(x)=0$ and 
for the non-empty Young diagram $\mu $.
Note that the admissible tableaux $\mathsf{  Tab}_{{\mathtt I}}(\lambda \subset \mu)$ 
 becomes an empty set if the Young diagram $ \lambda \subset \mu$ 
contains a rectangular sub-diagram of a height of $m+1$ and a width of $n+1$,
 and consequently
(\ref{DVF-tab1}) and (\ref{DVF-tab2}) vanish for such Young diagram
\footnote{See \cite{DM92,T97,T98} for ${\mathtt I}=I_{M+N}$ case.}. 
After inspection of \eqref{boxes}, \eqref{boxes2}, 
\eqref{DVF-tab1} and \eqref{DVF-tab2}, 
one observes the following relation: 
\begin{align}
{\mathcal  F}_{\lambda\subset \mu}^{{\mathtt I}}(x)=
\overline{\mathcal  F}_{\widetilde{\lambda\subset \mu}}^{\widetilde{\mathtt I}}(x)|
_{\overline{\Qs}_{J}(xq^{{\mathtt s}}) \to \Qs_{J}(q^{-{\mathtt s}})} 
\quad \text{for all} \quad J \subset {\mathfrak I}, 
\label{reverseDVF}
\end{align}
where ${\mathtt s}$ is any shift of the $\Qs$-functions in
 $\overline{\mathcal  F}_{\widetilde{\lambda\subset \mu}}^{\widetilde{\mathtt I}}(x)$, 
 and 
$\widetilde{\mathtt I}=(\gamma_{K},\dots,\gamma_{2},\gamma_{1})$ is a reversed tuple of 
${\mathtt I}$. 

Let us take an integer $K$ ($0\le K \le M+N$). 
The tableau sum formula \eqref{DVF-tab1} has determinant expressions.
\begin{align}
{\mathcal F}_{\lambda \subset \mu}^{I_{K}}(x)&= \det_{1 \le i,j \le \mu_{1}}
    ({\mathcal F}^{I_{K}}_{(1^{\mu_{i}^{\prime}-\lambda_{j}^{\prime}-i+j})}
(xq^{\mu_{1}-\mu_{1}^{\prime}+\mu_{i}^{\prime}+\lambda_{j}^{\prime}-i-j+1}))	
 \label{superJT1}
	\\[7pt] 
& =\det_{1 \le i,j \le \mu_{1}^{\prime}}
    ({\mathcal F}^{I_{K}}_{(\mu_{j}-\lambda_{i}+i-j)}
    (xq^{\mu_{1}-\mu_{1}^{\prime}-\mu_{j}-\lambda_{i}+i+j-1})),
     \label{superJT2}
\end{align}
where  ${\mathcal F}_{(1^{0})}^{I_{K}}(x)={\mathcal F}_{(0)}^{I_{K}}(x)=1$ and 
${\mathcal F}_{(1^{a})}^{I_{K}}(x)={\mathcal F}_{(a)}^{I_{K}}(x) = 0$ for $a <0$. 
We also have the same type of formulae for \eqref{DVF-tab2}. 
These determinant expressions for $K=M+N$ correspond to 
the supersymmetric Bazhanov-Reshetikhin formulae \cite{T97,T98} 
(see also a recent paper \cite{KV07}; and for $U_{q}(B_{r}^{(1)})$ case, see \cite{KOS95}), which are 
supersymmetric extension of the Bazhanov-Reshetikhin formulae \cite{BR90}. 
The determinants \eqref{superJT1} and \eqref{superJT2} for 
$U_{q}(\widehat{gl}(M|N))$, $M,N>0$ superficially look 
 same as the ones for $M=0$ or $N=0$ case. 
However, one must note that their properties for $M,N>0$ case 
are different from the ones for $M=0$ or $N=0$ case 
since the conditions of the admissible tableaux are different, and that the 
matrix elements of the determinants are different. 

The above combinatorial tableau sum expressions of the $\Ts$-functions 
\eqref{DVF-tab1} and \eqref{DVF-tab2} are 
convenient for computer calculations. Next, we 
 consider other expressions of the $\Ts$-functions. 
One can rewrite the above tableau sum expressions for the Young diagrams of 
one row or one column in the form of the non-commutative 
 generating series \cite{T97,T98} (see also, \cite{KLWZ97,KSZ07,Zabrodin07}). 
\begin{align}
{\mathbf W}_{I_{K}}(x,{\mathbf X})&=
(1-{\mathcal X}_{I_{1}}(x){\mathbf X})^{-p_{i_{1}}}(1-{\mathcal X}_{I_{2}}(x){\mathbf X})^{-p_{i_{2}}}
\cdots (1-{\mathcal X}_{I_{K}}(x){\mathbf X})^{-p_{i_{K}}} \nonumber \\
&= \sum_{\alpha=0}^{\infty} {\mathcal F}_{(\alpha)}^{I_{K}}
(xq^{1-\alpha -\frac{m-n}{2}+\frac{M-N}{2}}){\mathbf X}^{\alpha }, 
\label{gene1}
\\[6pt]
\overline{\mathbf W}_{\overline{I}_{K}}(x,{\mathbf X}) & \nonumber \\
& \hspace{-25pt}  =
(1-\overline{\mathcal X}_{\overline{I}_{K}}(x){\mathbf X})^{-p_{i_{K+1}}}
(1-\overline{\mathcal X}_{\overline{I}_{K+1}}(x){\mathbf X})^{-p_{i_{K+2}}}
\cdots (1-\overline{\mathcal X}_{\overline{I}_{M+N-1}}(x){\mathbf X})^{-p_{i_{M+N}}}  \nonumber \\
&\hspace{-25pt} 
=(1-{\mathcal X}_{I_{K+1}}(x){\mathbf X})^{-p_{i_{K+1}}}(1-{\mathcal X}_{I_{K+2}}(x){\mathbf X})^{-p_{i_{K+2}}}
\cdots (1-{\mathcal X}_{I_{M+N}}(x){\mathbf X})^{-p_{i_{M+N}}}  \nonumber \\
&\hspace{-25pt} 
=\sum_{\alpha=0}^{\infty}\overline{\mathcal F}_{(\alpha)}^{\overline{I}_{K}}
(xq^{1-\alpha +\frac{\overline{m}-\overline{n}}{2}-\frac{M-N}{2}}){\mathbf X}^{\alpha },
\label{gene2}
\end{align}
where ${\mathbf X}$ is a shift operator ${\mathbf X}f(x)=f(xq^{-2}){\mathbf X}$ for any function of $x$; 
$m:={\rm card} (I_{K} \cap {\mathfrak B})$, 
$n:={\rm card} (I_{K} \cap {\mathfrak F})$, 
$\overline{m}:={\rm card} (\overline{I}_{K}\cap {\mathfrak B})=M-m$, 
$\overline{n}:={\rm card} (\overline{I}_{K}\cap {\mathfrak F})=N-n$. 
We also have the inverse of the above generating series. 
\begin{align}
{\mathbf W}_{I_{K}}(x,{\mathbf X})^{-1}&=
(1-{\mathcal X}_{I_{K}}(x){\mathbf X})^{p_{i_{K}}}
\cdots (1-{\mathcal X}_{I_{2}}(x){\mathbf X})^{p_{i_{2}}} (1-{\mathcal X}_{I_{1}}(x){\mathbf X})^{p_{i_{1}}}  \nonumber \\
&= \sum_{\alpha=0}^{\infty} (-1)^{\alpha} {\mathcal F}_{(1^{\alpha})}^{I_{K}}
(xq^{1-\alpha -\frac{m-n}{2}+\frac{M-N}{2}}){\mathbf X}^{\alpha }, 
\label{gene3}
\\[6pt]
\overline{\mathbf W}_{\overline{I}_{K}}(x,{\mathbf X})^{-1}& \nonumber \\
& \hspace{-25pt}=
(1-\overline{\mathcal X}_{\overline{I}_{M+N-1}}(x){\mathbf X})^{p_{i_{M+N}}} 
\cdots  (1-\overline{\mathcal X}_{\overline{I}_{K+1}}(x){\mathbf X})^{p_{i_{K+2}}} 
(1-\overline{\mathcal X}_{\overline{I}_{K}}(x){\mathbf X})^{p_{i_{K+1}}} \nonumber \\
&\hspace{-25pt} =\sum_{\alpha=0}^{\infty}(-1)^{\alpha}
\overline{\mathcal F}_{(1^{\alpha})}^{\overline{I}_{K}}
(xq^{1-\alpha +\frac{\overline{m}-\overline{n}}{2}-\frac{M-N}{2}}){\mathbf X}^{\alpha }.
\label{gene4}
\end{align}
They obey the following recurrence relations (cf.\ \cite{KSZ07,Zabrodin07}): 
\begin{align}
& {\mathbf W}_{I_{K}}(x,{\mathbf X})(1-{\mathcal X}_{I_{K+1}}(x){\mathbf X})^{-p_{i_{K+1}}}={\mathbf W}_{I_{K+1}}(x,{\mathbf X}), 
\label{rec1}
\\[5pt] 
& 
(1-\overline{\mathcal X}_{\overline{I}_{K}}(x){\mathbf X})^{-p_{i_{K+1}}}
\overline{\mathbf W}_{\overline{I}_{K+1}}(x,{\mathbf X})
=\overline{\mathbf W}_{\overline{I}_{K}}(x,{\mathbf X}).
\label{rec2}
\end{align}
These generating series \eqref{gene1} and \eqref{gene3} for $K=M+N$
(resp.\ \eqref{gene2} and \eqref{gene4} for $K=0$), which generate 
eigenvalue formulae of ``real" transfer matrices appeared in \cite{T97,T98}. 
In particular, 
${\mathbf W}_{I_{K}}(x,{\mathbf X})\overline{\mathbf W}_{\overline{I}_{K}}(x,{\mathbf X})
={\mathbf W}_{I_{M+N}}(x,{\mathbf X})=\overline{\mathbf W}_{\overline{I}_{0}}(x,{\mathbf X})$, 
 corresponds to eq.\ (3.11) in \cite{T98}. In addition, 
from this relation, one can derive the following relation for $s\in {\mathbb Z}_{\ge 0}$:  
\begin{align}
{\mathcal F}_{(s)}^{I_{M+N}}(x)=\overline{\mathcal F}_{(s)}^{\overline{I}_{0}}(x)
=
\sum_{\alpha=0}^{s} 
{\mathcal F}_{(\alpha)}^{I_{K}}(xq^{s-\alpha -\frac{m-n}{2}+\frac{M-N}{2}})
\overline{\mathcal F}_{(s-\alpha)}^{\overline{I}_{K}}
(xq^{-\alpha +\frac{\overline{m}-\overline{n}}{2}-\frac{M-N}{2}}).
\end{align}
This is a generalization of eq.\ (E.6) in \cite{T97}. 
On the other hand, 
``Intermediate''  ($0<K<M+N $) generating series 
\eqref{gene1} and \eqref{gene3} 
 were considered
\footnote{To be precise, the ones in \cite{KSZ07,Zabrodin07} were written in terms of 
 functions corresponding to 
\eqref{TF-rel03} for the rational case.} 
in \cite{KSZ07,Zabrodin07}. 
Some ``intermediate'' generating series were also written in Appendix E in  
\cite{T97}. 
We remark that the non-commutative generating series of the $T$-functions 
similar to the ones in \cite{T97,T98} were  
also used recently in the study of the AdS/CFT correspondence 
in particle physics \cite{BKSZ05,Beisert07,B07-2,GV07,GKV09}. 

We define the action of $\sigma \in S({\mathfrak I})$ on the 
$\Ts$-functions as 
$\sigma[{\mathcal F}^{I}_{\lambda \subset \mu}(x)]=
{\mathcal F}^{\sigma(I)}_{\lambda \subset \mu}(x)$ and 
$\sigma[\overline{\mathcal F}^{I}_{\lambda \subset \mu}(x)]=
\overline{\mathcal F}^{\sigma(I)}_{\lambda \subset \mu}(x)$. 
As was pointed out in \cite{KSZ07,Zabrodin07}, compatibility 
of this type of recurrence relations \eqref{rec1}-\eqref{rec2}, 
which corresponds to a 
discrete ``zero curvature condition", 
reproduces the functional relations among the Baxter $\Qs$-functions 
\eqref{QQ-rel1}-\eqref{QQ-rel4}. 
This corresponds
\footnote{The compatibility condition in \cite{KSZ07,Zabrodin07} corresponds to 
the one for the rational case with $p_{i_{a}}=-p_{i_{a+1}}$. 
The compatibility condition for the rational case with $p_{i_{a}}=p_{i_{a+1}}$ was 
discussed in \cite{GV07}.}
 to 
${\mathbf W}_{I_{a+1}}(x,{\mathbf X})={\mathbf W}_{\tau(I_{a+1})}(x,{\mathbf X})$ in the notation of 
\eqref{eqiv}. 
Therefore $\tau[{\mathcal F}^{I_{a+1}}_{(s)}(x)]=
{\mathcal F}^{I_{a+1}}_{(s)}(x)$ 
under the functional relations 
\eqref{QQ-rel1}-\eqref{QQ-rel2}, 
and thus $\tau[{\mathcal F}^{I_{a+1}}_{\lambda \subset \mu}(x)]=
{\mathcal F}^{I_{a+1}}_{\lambda \subset \mu}(x)$ through \eqref{superJT2}. 
Similarly, $\tau[\overline{\mathcal F}^{I_{a+1}}_{\lambda \subset \mu}(x)]=
\overline{\mathcal F}^{I_{a+1}}_{\lambda \subset \mu}(x)$ under the functional relations 
\eqref{QQ-rel3}-\eqref{QQ-rel4}. 
Thus $\sigma[{\mathcal F}^{I}_{\lambda \subset \mu}(x)]=
{\mathcal F}^{I}_{\lambda \subset \mu}(x)$ 
and  $\sigma[\overline{\mathcal F}^{I}_{\lambda \subset \mu}(x)]=
\overline{\mathcal F}^{I}_{\lambda \subset \mu}(x)$ 
for any $\sigma \in S(I)$ 
or $\sigma \in S({\mathfrak I} \setminus I)$. 
This means that 
 the $\Ts$-functions ${\mathcal F}^{I}_{\lambda \subset \mu}(x)$ and 
$\overline{\mathcal F}^{I}_{\lambda \subset \mu}(x)$ 
are independent of the order of the components of the 
 tuple $I$ (the order of the gradings) under the functional relations 
\eqref{QQ-rel1}-\eqref{QQ-rel4}. 
In particular, 
$\sigma[{\mathcal F}^{I_{M+N}}_{\lambda \subset \mu}(x)]=
{\mathcal F}^{I_{M+N}}_{\lambda \subset \mu}(x)$ for $\sigma \in 
S({\mathfrak B}) \times S({\mathfrak F})$ 
under \eqref{QQ-rel1} corresponds to the Weyl group invariance of 
the $\Ts$-functions, and 
for $\sigma \in S({\mathfrak I})/[S({\mathfrak B}) \times S({\mathfrak F})]$ 
under \eqref{QQ-rel2} corresponds to the {\em super}-Weyl group 
invariance (cf.\ \cite{T98}). 
Note that $S({\mathfrak I})/[S({\mathfrak B}) \times S({\mathfrak F})]$ 
is not a simple permutation on the parameters $\{z_{a} \}$; 
it affects the rule of the admissible tableau $\mathsf{  Tab}_{I_{M+N}}(\lambda \subset \mu)$. 

We can also consider
\footnote{See eq.\ (3.14) in \cite{BG90} for ``classical" case:
$x=0$, $\lambda=\emptyset $, $I_{K}=(1,2,\dots,M+N)$.}
 a non-commutative generating series of the $\Ts$-functions 
for the general skew-Young diagrams $\lambda \subset \mu$: 
\begin{multline}
 {\det }_{1 \le i,j \le \mu_{1}^{\prime}}({\mathbf X}_{i}^{\lambda_{j}+i-j})
\prod_{j=1}^{\mu_{1}^{\prime}}
{\mathbf W}_{I_{K}}(x_{j}q^{2j-2+\mu_{1}-\mu_{1}^{\prime}
+\frac{m-n}{2}-\frac{M-N}{2}},{\mathbf X}_{j})  \\
=\sum_{\sigma \in S_{\mu_{1}^{\prime}}} {\rm sgn}(\sigma)
\sum_{m_{1}=0}^{\infty}\sum_{m_{2}=0}^{\infty} \cdots \sum_{m_{\mu_{1}^{\prime}}=0}^{\infty}
\prod_{j=1}^{\mu_{1}^{\prime}} {\mathcal F}^{I_{K}}_{(m_{j})}
(x_{j}q^{-2\lambda_{\sigma(j)}+2\sigma(j)-m_{j}+\mu_{1}-\mu_{1}^{\prime}-1})
\\
\times 
{\mathbf X}_{1}^{m_{1}+\lambda_{\sigma(1)}-\sigma(1)+1}
{\mathbf X}_{2}^{m_{2}+\lambda_{\sigma(2)}-\sigma(2)+2}
\cdots 
{\mathbf X}_{\mu_{1}^{\prime}}^{m_{\mu_{1}^{\prime}}+\lambda_{\sigma(\mu_{1}^{\prime})}-
 \sigma(\mu_{1}^{\prime})+\mu_{1}^{\prime}}
\label{general},
\end{multline} 
where $S_{\mu_{1}^{\prime}}$ is the symmetric group of order $\mu_{1}^{\prime}$ 
and ${\rm sgn}(\sigma)$ is the signature of a permutation $\sigma$;
$\{{\mathbf X}_{j}\}$ are commutative (${\mathbf X}_{i}{\mathbf X}_{j}={\mathbf X}_{j}{\mathbf X}_{i}$) shift operators: 
${\mathbf X}_{j}f(x_{1},x_{2},\dots, x_{j},\dots)=
f(x_{1},x_{2},\dots, x_{j}q^{-2},\dots){\mathbf X}_{j}$ for any function 
$f(x_{1},x_{2},\dots )$ of $x_{1},x_{2},\dots $. 
Let us take the coefficient of ${\mathbf X}_{1}^{\mu_{1}}{\mathbf X}_{2}^{\mu_{2}}
\cdots {\mathbf X}_{\mu_{1}^{\prime}}^{\mu_{\mu_{1}^{\prime}}}$ in the right hand side of 
\eqref{general}, and write it as 
${\mathcal F}_{\lambda \subset \mu}^{I_{K}}(x_{1},x_{2},\dots, x_{\mu_{1}^{\prime}})$. 
We find 
${\mathcal F}_{\lambda \subset \mu}^{I_{K}}(x,x,\dots,x) =
{\mathcal F}_{\lambda \subset \mu}^{I_{K}}(x)$ (cf.\ \eqref{superJT2}). 
Thus ${\mathcal F}_{\lambda \subset \mu}^{I_{K}}
(x_{1},x_{2},\dots, x_{\mu_{1}^{\prime}})$ is a refinement of 
${\mathcal F}_{\lambda \subset \mu}^{I_{K}}(x)$. 
We have also derived a similar formula for $\overline{\mathcal F}_{\lambda \subset \mu}^{\overline{I}_{K}}(x)$.  

In general, the function \eqref{DVF-tab1} for ${\mathtt I}=I_{K}$ 
(resp.\ \eqref{DVF-tab2} for ${\mathtt I}=\overline{I}_{K}$ ) has poles from zeros 
of the Baxter $\Qs$-function $\Qs_{I_{K}}(x)$ 
(resp.\ $\overline{\Qs}_{\overline{I}_{K}}(x)$). 
For example, the function ${\mathcal F}_{(1)}^{I_{K}}(x)$ under 
\eqref{tri1}, \eqref{baxterQ-root} and \eqref{quantum-wron-ps2} 
has poles
\footnote{For $K=M+N$ case, the poles for $a=M+N$ (poles from the vacuum part) are located at 
$x=w_{k}q^{\sum_{j \in I_{M+N}}p_{j}-\frac{M-N}{2}}=
w_{k}q^{\frac{M-N}{2}}$.}
 at $x=x_{k}^{I_{a}}q^{\sum_{j \in I_{a}}p_{j}-\frac{M-N}{2}}$ 
for $1 \le a \le K$. And these poles for $1 \le a \le K-1$ chancel each other as in 
\eqref{chancels} under the Bethe ansatz equation \eqref{BAE}. 
But the poles for $a=K$ still survive. 
To kill
\footnote{There are cases where poles from the vacuum part still survive (see \eqref{norsl21}  
in Appendix \ref{vaumm}). 
This depends on the normalization of the Baxter $\Qs$-functions. 
In any case, we expect that poles from the roots of the Bethe ansatz equation 
disappear.}
 this kind of pole, we introduce 
the following transformations for any (non-skew) Young diagram $\mu$.  
\begin{align}
 {\mathsf F}_{\mu}^{I_{K}}(x)
 =
\Qs_{I_{K}}
(xq^{-\frac{m-n}{2}-\mu_{1}+\mu_{1}^{\prime}})
{\mathcal  F}_{\widetilde{\mu}}^{I_{K}}(x) 
\label{TF-rel03}
\end{align}
under the normalization \eqref{tri1} and the functional relations
 \eqref{QQ-rel1}-\eqref{QQ-rel2}, and 
\begin{align}
{\overline {\mathsf F}}_{\mu}^{\overline{I}_{K}}(x)
 =
\overline{\Qs}_{\overline{I}_{K}}
(xq^{\frac{\overline{m}-\overline{n}}{2}+\mu_{1}-\mu_{1}^{\prime}})
\overline{\mathcal  F}_{\mu}^{\overline{I}_{K}}(x)
\label{TF-rel04}
\end{align}
under the normalization \eqref{tri2} and the functional relations
 \eqref{QQ-rel3}-\eqref{QQ-rel4}. 
Here  $\widetilde{\mu}$ is 
 the 180 degrees rotated Young diagram of $\mu$ (see, \eqref{180deg} for $\lambda=\emptyset$). 
 
Let us write these functions for the Young diagram of a rectangular shape $\mu=(s^{a})$ 
as ${\mathsf F}_{s}^{(a),I_{K}}(x)={\mathsf F}_{(s^a)}^{I_{K}}(x)
=\Qs_{I_{K}}(xq^{-\frac{m-n}{2}-s+a}){\mathcal  F}_{(s^a)}^{I_{K}}(x)$ and 
$\overline{\mathsf F}_{s}^{(a),\overline{I}_{K}}(x)
=\overline{\mathsf F}_{(s^a)}^{\overline{I}_{K}}(x)
=\overline{\Qs}_{\overline{I}_{K}}
(xq^{\frac{\overline{m}-\overline{n}}{2}+s-a})
\overline{\mathcal  F}_{(s^a)}^{\overline{I}_{K}}(x)
$. 
In particular for $a=0$ or $s=0$ case, we interpret them 
\footnote{$\mu_{1}-\mu_{1}^{\prime}=s-a$ for $\mu=(s^a)$. 
And then we formally put $\mu_{1}-\mu_{1}^{\prime}=s$ for $a=0$ and 
$\mu_{1}-\mu_{1}^{\prime}=-a$ for $s=0$ in \eqref{TF-rel03} and \eqref{TF-rel04}. 
We included non-vanishing terms on $s<0$ for the case $a=0$ as in \cite{KSZ07}.} as:
\begin{align}
& 
{\mathsf F}^{(a),I_{K}}_{0}(x)=\Qs_{I_{K}}(xq^{-\frac{m-n}{2}+a}) 
\qquad   
\text{ for} \quad a \in {\mathbb Z}_{\ge 0},
\label{t-sys-bc01} 
\\[5pt]
& 
 {\mathsf F}^{(0),I_{K}}_{s}(x)=\Qs_{I_{K}}(xq^{-\frac{m-n}{2}-s}) 
\qquad    
\text{ for} \quad s \in {\mathbb Z},
 \label{t-sys-bc02}
\end{align}
\begin{align}
& 
\overline{{\mathsf F}}^{(a),\overline{I}_{K}}_{0}(x)=\overline{\Qs}_{\overline{I}_{K}}(xq^{\frac{\overline{m}-\overline{n}}{2}-a})
 \qquad   
\text{ for} \quad a \in {\mathbb Z}_{\ge 0} ,
\label{t-sys-bc03} 
\\[5pt]
&
 \overline{{\mathsf F}}^{(0),\overline{I}_{K}}_{s}(x)=\overline{\Qs}_{\overline{I}_{K}}(xq^{\frac{\overline{m}-\overline{n}}{2}+s}) 
 \qquad   
\text{ for} \quad s \in {\mathbb Z}.
 \label{t-sys-bc04}
\end{align}
For $a,s \in {\mathbb Z}$, 
${\mathsf F}_{s}^{(a),I_{K}}(x)=0$ if $a <0$, or $a>0 $ and $s<0$, or $a>m$ and $s>n$; 
$\overline{\mathsf F}_{s}^{(a),\overline{I}_{K}}(x)=0$
 if $a <0$, or $a>0 $ and $s<0$, or $a>\overline{m}$ and $s>\overline{n}$.

There are ``duality relations'' among the functions
\footnote{In the context of the analytic Bethe ansatz, the duality relation 
for $(m,n)=(M,N)$
appeared in \cite{T97} first. This relation with the factor 
$\left(\prod_{\gamma \in I_{K}}p_{\gamma}z_{\gamma}^{p_{\gamma}}
\right)^{a}$ for $(m,n)=(M,N)$ 
 appeared in \cite{Tsuboi06} (the parameters 
$\{z_{a}\}$ were used as chemical potentials).
This factor was also written as a superdeterminant in \cite{Zabrodin07}.}. 
\begin{align}
{\mathsf F}_{a+n}^{(m),I_{K}}(x)&=
\left(\prod_{\gamma \in I_{K}}p_{\gamma}z_{\gamma}^{p_{\gamma}}
\right)^{a}
{\mathsf F}_{n}^{(a+m),I_{K}}(x),
 \label{dual-mn0}
\\[6pt]
\overline{{\mathsf F}}_{a+\overline{n}}^{(\overline{m}),\overline{I}_{K}}(x)&=
\left(\prod_{\gamma \in \overline{I}_{K}}p_{\gamma}z_{\gamma}^{p_{\gamma}}
\right)^{a}
\overline{{\mathsf F}}_{\overline{n}}^{(a+\overline{m}),\overline{I}_{K}}(x)
 \quad \text{for} \quad a \in {\mathbb Z}_{\ge 0}. 
 \label{dual-mn02}
\end{align}

The $\Ts$-functions of fusion models obey functional relations called the $T$-system. 
For $a,s \in {\mathbb Z}_{\ge 1}$, ${\mathsf F}^{(a),I_{K}}_{s}(u)$ 
 satisfies 
the $U_{q}(gl(m|n))$-type $T$-system:
\begin{align}
\begin{split}
&  
{\mathsf F}^{(a),I_{K}}_{s}(xq^{-1}){\mathsf F}^{(a),I_{K}}_{s}(xq)={\mathsf F}^{(a),I_{K}}_{s-1}(x){\mathsf F}^{(a),I_{K}}_{s+1}(x) 
 +
{\mathsf F}^{(a-1),I_{K}}_{s}(x){\mathsf F}^{(a+1),I_{K}}_{s}(x)  \\[2pt]
& \hspace{80pt} \text{for} \quad 1 \le a \le m-1 \quad {\rm or} \quad 1\le s \le n-1
 \quad {\rm or} \quad (a,s)=(m,n), 
\end{split}
\label{t-system01} 
 \\[4pt]
&  
{\mathsf F}^{(m),I_{K}}_{s}(xq^{-1}){\mathsf F}^{(m),I_{K}}_{s}(xq)=
{\mathsf F}^{(m),I_{K}}_{s-1}(x){\mathsf F}^{(m),I_{K}}_{s+1}(x)
\quad \text{for} \quad s \in {\mathbb Z}_{\ge n+1}, \\[6pt]
&  
{\mathsf F}^{(a),I_{K}}_{n}(xq^{-1}){\mathsf F}^{(a),I_{K}}_{n}(xq)=
{\mathsf F}^{(a-1),I_{K}}_{n}(x){\mathsf F}^{(a+1),I_{K}}_{n}(x)
\quad \text{for} \quad a \in {\mathbb Z}_{\ge m+1},  
\end{align}
and  $\overline{{\mathsf F}}^{(a),\overline{I}_{K}}_{s}(u)$ 
 satisfies the 
$U_{q}(gl(\overline{m}|\overline{n}))$-type $T$-system:
\begin{align}
\begin{split}
&
\overline{{\mathsf F}}^{(a),\overline{I}_{K}}_{s}(xq^{-1})
\overline{{\mathsf F}}^{(a),\overline{I}_{K}}_{s}(xq)=
\overline{{\mathsf F}}^{(a),\overline{I}_{K}}_{s-1}(x)
\overline{{\mathsf F}}^{(a),\overline{I}_{K}}_{s+1}(x)  
 +
\overline{{\mathsf F}}^{(a-1),\overline{I}_{K}}_{s}(x)
\overline{{\mathsf F}}^{(a+1),\overline{I}_{K}}_{s}(x) 
  \\[2pt]
& \hspace{80pt} \text{for} \quad
 1 \le a \le \overline{m}-1 \quad {\rm or} \quad 1\le s \le \overline{n}-1
 \quad {\rm or} \quad (a,s)=(\overline{m},\overline{n}),
\end{split}
 \label{t-system02}
\\[4pt]
&  
\overline{{\mathsf F}}^{(\overline{m}),\overline{I}_{K}}_{s}(xq^{-1})
\overline{{\mathsf F}}^{(\overline{m}),\overline{I}_{K}}_{s}(xq)=
\overline{{\mathsf F}}^{(\overline{m}),\overline{I}_{K}}_{s-1}(x)
\overline{{\mathsf F}}^{(\overline{m}),\overline{I}_{K}}_{s+1}(x)
\quad \text{for} \quad s \in {\mathbb Z}_{\ge \overline{n}+1}, \\[6pt]
&
\overline{{\mathsf F}}^{(a),\overline{I}_{K}}_{\overline{n}}(xq^{-1})\overline{{\mathsf F}}^{(a),\overline{I}_{K}}_{\overline{n}}(xq)=
\overline{{\mathsf F}}^{(a-1),\overline{I}_{K}}_{\overline{n}}(x)\overline{{\mathsf F}}^{(a+1),\overline{I}_{K}}_{\overline{n}}(x)
\quad \text{for} \quad a \in {\mathbb Z}_{\ge \overline{m}+1}. 
\label{t-system04}
\end{align}
This type of ``boundary condition'' (\ref{t-sys-bc01}), (\ref{t-sys-bc02}), 
(\ref{t-sys-bc03}), (\ref{t-sys-bc04}) was discussed    
 in \cite{KSZ07,Zabrodin07} (for $N=0$ case, see \cite{KLWZ97}). 
For $K=M+N$ (resp.\ for $K=0$), the right hand side of 
(\ref{t-sys-bc01}), (\ref{t-sys-bc02}), (resp.\ (\ref{t-sys-bc03}), (\ref{t-sys-bc04}))
become model dependent scalar functions 
(see examples on the inhomogeneous Perk-Schultz models \eqref{quantum-wron-ps}, 
\eqref{quantum-wron-ps2}). 
And the above $T$-system corresponds to the original $T$-system 
(\ref{ori-t-system1})-(\ref{ori-t-sys-bc1}) 
for $U_{q}(\widehat{gl}(M|N))$ \cite{T97,T98}.   
These relations \eqref{t-system01}-\eqref{t-system04} 
 follow from the determinant formula \eqref{superJT1} or \eqref{superJT2}  
 and the Jacobi identity \eqref{jacobi}. 

The function ${\mathsf F}^{(a),I_{K}}_{s}(x)$ also satisfies
\footnote{This should follow from similar discussions given in \cite{KSZ07} for rational 
case. As we already remarked, the Baxter $\Qs$-functions in \cite{KSZ07} correspond to  
the rational case of $(M+1)(N+1)$ subset of $2^{M+N}$ Baxter $\Qs$-functions. Thus 
we have to consider actions of the Weyl group $S({\mathfrak B})\times S({\mathfrak F})$ 
 to get the whole set of the Baxter $\Qs$-functions, and also the whole set of
 $F^{(a),I}_{s}(x)$ and $\overline{F}^{(a),I}_{s}(x)$.} 
the following functional relations, which correspond to the B\"{a}cklund transformations
\footnote{For these relations for $gl(M)$ related elliptic models were proposed in \cite{KLWZ97}. 
They were generalized to $gl(M|N)$ related rational models 
in \cite{KSZ07} (without twist parameters case); 
\cite{Zabrodin07} (with twist parameters case). 
The normalization of the spectral parameter of the 
functions in \cite{KSZ07,Zabrodin07} are closer to the following 
(functions with prime):  
${\mathcal  F}_{\lambda\subset \mu}^{\prime I_{K}}(x)=
{\mathcal  F}_{\lambda\subset \mu}^{I_{K}}(xq^{-\frac{m-n}{2}})$, 
$\overline{\mathcal  F}_{\lambda\subset \mu}^{\prime \overline{I}_{K}}(x)=
\overline{\mathcal  F}_{\lambda\subset \mu}^{\overline{I}_{K}}
(xq^{\frac{\overline{m}-\overline{n}}{2}-M+N})$,
${\mathsf F}^{\prime (a),I_{K}}_{s}(x)={\mathsf F}^{(a),I_{K}}_{s}(xq^{-\frac{m-n}{2}})$, 
$\overline{\mathsf F}^{\prime (a),I_{K}}_{s}(x)=
\overline{\mathsf F}^{(a),I_{K}}_{s}(xq^{\frac{m-n}{2}-M+N})$, 
$\Qs^{\prime}_{I}(x)=\Qs_{I}(xq^{-\sum_{j \in I}p_{j}})$, 
$\overline{\Qs}^{\prime}_{I}(x)=\overline{\Qs}_{I}(xq^{\sum_{j \in I}p_{j}-M+N})$,
${\mathcal X}^{\prime}_{I}(x)={\mathcal X}_{I}(xq^{-\frac{M-N}{2}})$, 
$\overline{\mathcal X}^{\prime}_{I}(x)=\overline{\mathcal X}_{I}(xq^{-\frac{M-N}{2}})$.
In fact, \eqref{bac01}-\eqref{bac04} have the same type of shift of the spectral parameter 
as in \cite{KSZ07,Zabrodin07} 
if they are written in terms of ${\mathsf F}^{\prime (a),I_{K}}_{s}(x)$.} 
 \cite{KSZ07,Zabrodin07}. \\
The functional relations for $p_{i_{K}}=1$ have the following form: 
\begin{align}
\begin{split}
& {\mathsf F}^{(a+1),I_{K}}_{s}(x){\mathsf F}^{(a),I_{K-1}}_{s}(xq^{\frac{3}{2}})-
{\mathsf F}^{(a),I_{K}}_{s}(xq){\mathsf F}^{(a+1),I_{K-1}}_{s}(xq^{\frac{1}{2}})
\\
& \hspace{190pt} =
z_{i_{K}}{\mathsf F}^{(a+1),I_{K}}_{s-1}(xq){\mathsf F}^{(a),I_{K-1}}_{s+1}(xq^{\frac{1}{2}}), 
\end{split}
\label{bac01} 
\\[10pt] 
\begin{split}
&  {\mathsf F}^{(a),I_{K}}_{s+1}(xq){\mathsf F}^{(a),I_{K-1}}_{s}(xq^{\frac{1}{2}})-
{\mathsf F}^{(a),I_{K}}_{s}(x){\mathsf F}^{(a),I_{K-1}}_{s+1}(xq^{\frac{3}{2}})
 \\ 
& \hspace{190pt} =
z_{i_{K}}{\mathsf F}^{(a+1),I_{K}}_{s}(xq){\mathsf F}^{(a-1),I_{K-1}}_{s+1}(xq^{\frac{1}{2}}), 
\end{split}
\end{align}
and the functional relations for $p_{i_{K}}=-1$ have the following form:
\begin{align}
\begin{split}
&  {\mathsf F}^{(a+1),I_{K-1}}_{s}(x){\mathsf F}^{(a),I_{K}}_{s}(xq^{\frac{3}{2}})-
{\mathsf F}^{(a),I_{K-1}}_{s}(xq){\mathsf F}^{(a+1),I_{K}}_{s}(xq^{\frac{1}{2}})
 \\ & \hspace{190pt} 
=
z_{i_{K}}{\mathsf F}^{(a+1),I_{K-1}}_{s-1}(xq){\mathsf F}^{(a),I_{K}}_{s+1}(xq^{\frac{1}{2}}), 
\end{split}
\\[10pt] 
\begin{split}
& {\mathsf F}^{(a),I_{K-1}}_{s+1}(xq){\mathsf F}^{(a),I_{K}}_{s}(xq^{\frac{1}{2}})-
{\mathsf F}^{(a),I_{K-1}}_{s}(x){\mathsf F}^{(a),I_{K}}_{s+1}(xq^{\frac{3}{2}})
 \\ 
& \hspace{190pt} =
z_{i_{K}}{\mathsf F}^{(a+1),I_{K-1}}_{s}(xq){\mathsf F}^{(a-1),I_{K}}_{s+1}(xq^{\frac{1}{2}}).
\label{bac04}
\end{split}
\end{align}
There are also relations for $\overline{{\mathsf F}}^{(a),\overline{I}_{K}}_{s}(x)$. 
The functional relations for $p_{i_{K}}=1$ have the following form: 
\begin{align}
\begin{split}
&  \overline{{\mathsf F}}^{(a+1),\overline{I}_{K-1}}_{s}(x)\overline{{\mathsf F}}^{(a),\overline{I}_{K}}_{s}(xq^{-\frac{3}{2}})-
\overline{{\mathsf F}}^{(a),\overline{I}_{K-1}}_{s}(xq^{-1})\overline{{\mathsf F}}^{(a+1),\overline{I}_{K}}_{s}(xq^{-\frac{1}{2}})
 \\
& \hspace{180pt} =
z_{i_{K}}\overline{{\mathsf F}}^{(a+1),\overline{I}_{K-1}}_{s-1}(xq^{-1})\overline{{\mathsf F}}^{(a),\overline{I}_{K}}_{s+1}(xq^{-\frac{1}{2}}), 
\end{split}
\label{bac05}
\\[10pt] 
\begin{split}
&  \overline{{\mathsf F}}^{(a),\overline{I}_{K-1}}_{s+1}(xq^{-1})\overline{{\mathsf F}}^{(a),\overline{I}_{K}}_{s}(xq^{-\frac{1}{2}})-
\overline{{\mathsf F}}^{(a),\overline{I}_{K-1}}_{s}(x)\overline{{\mathsf F}}^{(a),\overline{I}_{K}}_{s+1}(xq^{-\frac{3}{2}})
 \\ 
& \hspace{170pt} =
z_{i_{K}}\overline{{\mathsf F}}^{(a+1),\overline{I}_{K-1}}_{s}(xq^{-1})\overline{{\mathsf F}}^{(a-1),\overline{I}_{K}}_{s+1}(xq^{-\frac{1}{2}}), 
\end{split}
\end{align}
and the functional relations for $p_{i_{K}}=-1$ have the following form:
\begin{align}
\begin{split}
&  \overline{{\mathsf F}}^{(a+1),\overline{I}_{K}}_{s}(x)\overline{{\mathsf F}}^{(a),\overline{I}_{K-1}}_{s}(xq^{-\frac{3}{2}})-
\overline{{\mathsf F}}^{(a),\overline{I}_{K}}_{s}(xq^{-1})\overline{{\mathsf F}}^{(a+1),\overline{I}_{K-1}}_{s}(xq^{-\frac{1}{2}})
 \\ 
& \hspace{180pt} =
z_{i_{K}}\overline{{\mathsf F}}^{(a+1),\overline{I}_{K}}_{s-1}(xq^{-1})\overline{{\mathsf F}}^{(a),\overline{I}_{K-1}}_{s+1}(xq^{-\frac{1}{2}}), 
\end{split}
\\[10pt] 
\begin{split}
&  \overline{{\mathsf F}}^{(a),\overline{I}_{K}}_{s+1}(xq^{-1})\overline{{\mathsf F}}^{(a),\overline{I}_{K-1}}_{s}(xq^{-\frac{1}{2}})-
\overline{{\mathsf F}}^{(a),\overline{I}_{K}}_{s}(x)\overline{{\mathsf F}}^{(a),\overline{I}_{K-1}}_{s+1}(xq^{-\frac{3}{2}})
 \\ 
& \hspace{180pt} =
z_{i_{K}}\overline{{\mathsf F}}^{(a+1),\overline{I}_{K}}_{s}(xq^{-1})\overline{{\mathsf F}}^{(a-1),\overline{I}_{K-1}}_{s+1}(xq^{-\frac{1}{2}}).
\end{split}
\label{bac08}
\end{align}
The functional relations \eqref{bac05}-\eqref{bac08} 
follow from \eqref{bac01}-\eqref{bac04} through \eqref{reverseDVF}. 
The functional relations \eqref{bac01}-\eqref{bac04} 
(resp.\ \eqref{bac05}-\eqref{bac08}) connect
 the solution of the $U_{q}(gl(m|n))$-type $T$-system 
(resp.\ $U_{q}(gl(\overline{m}|\overline{n}))$-type $T$-system) and 
that of the $U_{q}(gl(m-1|n))$-type $T$-system or the 
$U_{q}(gl(m|n-1))$-type $T$-system 
(resp.\ the $U_{q}(gl(\overline{m}+1|\overline{n}))$-type $T$-system  
or the $U_{q}(gl(\overline{m}|\overline{n}+1))$-type $T$-system). 
Thus an original problem on $U_{q}(gl(M|N))$ 
is connected to the one on $U_{q}(gl(0|0))$. 

Before closing this section, we would like to mention the $\Ts$-functions for
 conjugate representations (cf.\ Appendix B in \cite{T97}; 
eq.\ (3.11) in \cite{BT08}). 
We conjecture that they are obtained by a manipulation
\footnote{The $\Qs$-function itself is also a function of 
$\{z_{a}\}$ and $q$: $\Qs_{I}(x)=\Qs_{I}(x,q,z_{1},\dots,z_{M+N})$. 
But we need not touch these in our normalization of the spectral parameter of the $\Qs$-function. 
Namely, ${\mathfrak C}[\Qs_{I}(xq^{{\mathtt s}},q,z_{1},\dots,z_{M+N})]=
\Qs_{I}(xq^{-{\mathtt s}},q,z_{1},\dots,z_{M+N})$. 
If one applies ${\mathfrak C}$ to \eqref{t-sys-bc01}-\eqref{bac08}, one will obtain the $T$-system and the B\"{a}cklund transformations for the conjugate representations. 
There are cases where 
${\mathsf T}$-functions for both fundamental and 
its conjugate representations are necessary to analyze   
 physically interesting models (see for example, \cite{EFS05}).}
 ${\mathfrak C}$ for the $\Ts$-functions: 
\begin{align}
{\mathfrak C}: 
z_{a} \to z_{a}^{-1}, \quad \Qs_{J}(xq^{{\mathtt s}}) \to 
\Qs_{J}(xq^{-{\mathtt s}}) 
\label{conj}
\end{align}
for any shift ${\mathtt s}$ of the spectral parameter of the $\Qs$-functions in 
the $\Ts$-functions and for any
 $a \in {\mathfrak I}$ and $J \subset {\mathfrak I}$. 
 For example, \eqref{tab-fund} is transformed to 
 \begin{eqnarray}
{\mathfrak C}[{\mathcal F}_{(1)}^{I_{M+N}}(x)]=
\sum_{a=1}^{M+N}p_{i_{a}}{\mathfrak C}[{\mathcal   X}_{I_{a}}(x)], 
\end{eqnarray}
where 
\begin{align}
{\mathfrak C}[{\mathcal   X}_{I_{a}}(x)]&=
z_{i_{a}}^{-1}
\frac{ \Qs_{I_{a-1}}
(xq^{\sum_{j \in I_{a-1}} p_{j}+2p_{i_{a}}-\frac{M-N}{2} })
\Qs_{I_{a}}
(xq^{\sum_{j \in I_{a}}p_{j}-2p_{i_{a}}
-\frac{M-N}{2}})
}{
\Qs_{I_{a-1}}
(xq^{\sum_{j \in I_{a-1}}p_{j}
-\frac{M-N}{2}})
\Qs_{I_{a}}
(xq^{\sum_{j \in I_{a}}p_{j}-\frac{M-N}{2}})
} .
\label{boxes3} 
\end{align}
Note that the functional relations for the Baxter $\Qs$-functions 
\eqref{QQ-rel1}-\eqref{QQ-bf2} are invariant 
under \eqref{conj}.
\section{Wronskian-like formulae for the $T$- and $\Qs$-functions}
In this section, we propose Wronskian-like formulae for 
the $T$- and $\Qs$-functions (cf.\ \eqref{9-thvari-1}, \eqref{9-thvari-2}), 
and claim that they satisfy the functional relations.  
Theorems \ref{solution-Q}, \ref{solution-t-sys-th}, \ref{solution-back} 
are our main results of this paper. 
We will also mention relation among 
the Wronskian-like formulae and the formulae in the previous 
section (cf.\ \eqref{TF-rel1}, \eqref{TF-rel2}). 
In Proposition \ref{infi-pro}, we will rewrite
 the Wronskian-like formulae \eqref{9-thvari-1} and \eqref{9-thvari-2} 
as summations over a direct product of symmetric groups, 
which suggest $\Ts$-functions for infinite dimensional representations. 
There are Wronskian-like formulae for 
the $T$-functions for the bosonic case $N=0$  
 \cite{KLWZ97,BLZ97,BHK02,Kojima08,DM08}. 
However, our new formulae here are {\em not} straightforward generalization of the ones for $N=0$. 
The size of the matrices for the Wronskian-like formulae for the $T$-functions 
depends on the representation for $M,N>0$ case. The upper bound of the size is $M+N$. 
Note that the size is constantly $M$ for $N=0$ case. 
There are papers related to rational models for $sl(2|1)$ \cite{BDKM06} and 
trigonometric models for $U_{q}(\widehat{sl}(2|1))$ \cite{BT08}. 

We introduce the following infinite matrices
\begin{align}
& {\mathcal A}(x)
:=
\left(
\begin{array}{cc}
\left( {\mathsf Z}_{k,l} \right)_{1 \le k \le M,
M+1 \le l \le M+N} & 
\left( {\mathsf X}_{k,l} \right)_{1 \le k \le M,
l \in {\mathbb Z}}
\\[4pt]
\left( {\mathsf Y}_{k,l} \right)_{k \in {\mathbb Z}, 
M+1 \le l \le M+N} & ({\mathsf W}_{k,l})_{k,l \in {\mathbb Z}},
\end{array}
\right), \label{Acal-mat}
\\[13pt]
& \overline{\mathcal A}(x)
:=
\left(
\begin{array}{cc}
\left( \overline{\mathsf Z}_{k,l} \right)_{1 \le k \le M,
M+1 \le l \le M+N} & 
\left( \overline{\mathsf X}_{k,l} \right)_{1 \le k \le M,
l \in {\mathbb Z}}
\\[4pt]
\left( \overline{\mathsf Y}_{k,l} \right)_{k \in {\mathbb Z}, 
M+1 \le l \le M+N} & (\overline{\mathsf W}_{k,l})_{k,l \in {\mathbb Z}}
\end{array}
\right) 
\end{align}
whose matrix elements are given as follows: 
\begin{align}
\begin{split}
& {\mathsf Z}_{k,l}:=\frac{\Qs_{\{k,l \}}(x)}{z_{k}-z_{l}} , 
\quad {\mathsf X}_{k,l}:=z_{k}^{l-1}\Qs_{k}(xq^{2l-1}), 
\quad {\mathsf Y}_{k,l}:=(-z_{l})^{k-1}\Qs_{l}(xq^{-2k+1}), 
 \\[8pt]
& {\mathsf W}_{k,l}:=0, \label{Acal-mat-el}
 \\[12pt]
& \overline{\mathsf Z}_{k,l}:=\frac{\overline{\Qs}_{\{k,l \}}(x)}{z_{k}-z_{l}} , 
\quad \overline{\mathsf X}_{k,l}:=z_{k}^{l-1}\overline{\Qs}_{k}(xq^{-2l+1}), 
\quad \overline{\mathsf Y}_{k,l}:=(-z_{l})^{k-1}\overline{\Qs}_{l}(xq^{2k-1}), 
 \\[8pt]
& \overline{\mathsf W}_{k,l}:=0.
\end{split}
\end{align}
For $B_{m}=(b_{1},b_{2},\dots,b_{m})$, $ F_{n}=(f_{1},f_{2},\dots,f_{n})$ and 
$s_{1},s_{2},\dots,s_{\beta}, r_{1},r_{2},\dots,r_{\alpha} \in  {\mathbb Z} $ 
($m +\alpha =n +\beta $), 
we define minor determinants of ${\mathcal A}(x)$ and 
$\overline{\mathcal A}(x)$. 
\begin{align}
\begin{split}
& \Delta^{(b_{1},b_{2},\dots,b_{m} ), (r_{1},r_{2},\dots,r_{\alpha})}
_{(f_{1},f_{2},\dots,f_{n} ), (s_{1},s_{2},\dots,s_{\beta})}
(x)
:=
\det \left(
\begin{array}{cc}
\left( {\mathsf Z}_{b_{k},f_{l}} \right)_{1 \le k \le m ,
1 \le l \le n} & 
\left( {\mathsf X}_{b_{k},s_{l}} \right)_{1 \le k \le m,
1 \le l \le \beta }
\\[4pt]
\left( {\mathsf Y}_{r_{k},f_{l}} \right)_{1 \le k \le \alpha, 
1 \le l \le n } & ({\mathsf W}_{r_{k},s_{l}})_{1\le k \le \alpha, 1 \le l \le \beta }
\end{array}
\right) \\[14pt]
& 
=\det
\left(
\begin{array}{cc}
\left( \frac{\Qs_{\{b_{k},f_{l}\}}(x)}{z_{b_{k}}-z_{f_{l}}} \right)_{1 \le k \le m,
1 \le l \le n} & 
\left(z_{b_{k}}^{s_{l}-1}
\Qs_{b_{k}}(xq^{2s_{l}-1}) \right)_{1 \le k \le m, 1 \le l \le \beta, }
\\[11pt]
\left((-z_{f_{l}})^{r_{k}-1}
\Qs_{f_{l}}(xq^{-2r_{k}+1}) \right)_{1 \le k \le \alpha, 
1 \le l \le n} & (0)_{\alpha \times \beta}
\end{array}
\right),
\end{split} 
\label{minor1}
\\[6pt]
\begin{split}
& \overline{\Delta}^{(b_{1},b_{2},\dots,b_{m} ), (r_{1},r_{2},\dots,r_{\alpha})}
_{(f_{1},f_{2},\dots,f_{n} ), (s_{1},s_{2},\dots,s_{\beta})}
(x)
:=
\det \left(
\begin{array}{cc}
\left( \overline{\mathsf Z}_{b_{k},f_{l}} \right)_{1 \le k \le m ,
1 \le l \le n} & 
\left( \overline{\mathsf X}_{b_{k},s_{l}} \right)_{1 \le k \le m,
1 \le l \le \beta }
\\[4pt]
\left( \overline{\mathsf Y}_{r_{k},f_{l}} \right)_{1 \le k \le \alpha, 
1 \le l \le n } & (\overline{\mathsf W}_{r_{k},s_{l}})_{1\le k \le \alpha, 1 \le l \le \beta }
\end{array}
\right)
\\[7pt]
& 
=\det
\left(
\begin{array}{cc}
\left( \frac{\overline{\Qs}_{\{b_{k},f_{l}\}}(x)}{z_{b_{k}}-z_{f_{l}}} \right)_{1 \le k \le m,
1 \le l \le n} & 
\left(z_{b_{k}}^{s_{l}-1}
\overline{\Qs}_{b_{k}}(xq^{-2s_{l}+1}) \right)_{1 \le k \le m, 1 \le l \le \beta, }
\\[11pt]
\left((-z_{f_{l}})^{r_{k}-1}
\overline{\Qs}_{f_{l}}(xq^{2r_{k}-1}) \right)_{1 \le k \le \alpha, 
1 \le l \le n} & (0)_{\alpha \times \beta}
\end{array}
\right),
\end{split}
\label{minor2}
\end{align}
where $ (0)_{\alpha \times \beta}$ is 
a $\alpha \times \beta$ zero matrix. Note that $\{r_{k},s_{l} \}$ are supposed to be integers, 
but this determinant are still well defined even if they are complex numbers. 
And this fact will be used in section 4.
For $m,n\in {\mathbb Z}_{\ge 0}$ and Young diagram $\mu$, we introduce a 
number, called $(m,n)$-index \cite{MV03}.
\begin{align}
\xi_{m,n}(\mu):={\rm min}\{j \in {\mathbb Z}_{\ge 0}|\mu_{j}+m-j \le n-1\}.
\label{mn-index}
\end{align}
In particular, $1 \le \xi_{m,n}(\mu) \le m+1$, $\xi_{m,0}(\mu)=m+1$ and $\xi_{0,n}(\mu)=1$
 for $\mu_{m+1} \le n$, and 
$\xi_{m,n}(\mu)=m+1$ for $\mu_{m+1} \le n \le \mu_{m}$. 
We often abbreviate $\xi_{m,n}(\mu)$ as $\xi_{m,n}$. 

For $B_{m}=(b_{1},b_{2},\dots, b_{m})$, 
$F_{n}=(f_{1},f_{2},\dots, f_{n}) $, 
$x \in {\mathbb C}$, and Young diagram $\mu$, 
we introduce the following functions. 
\begin{multline}
 {\mathcal T}_{\mu}^{(b_{1},b_{2},\dots,b_{m}),(f_{1},f_{2},\dots,f_{n})}(x)
:= (-1)^{(m+n+1)(\xi_{m,n}(\mu ) +1)} 
 \\ 
 \times 
\Delta^{(b_{1},b_{2},\dots,b_{m}), (r_{1},r_{2},\dots,r_{n-m+\xi_{m,n}-1})}
_{(f_{1},f_{2},\dots,f_{n} ),(s_{1},s_{2},\dots,s_{\xi_{m,n}-1} )}
(xq^{-\frac{3(m-n)}{2} +\mu_{1}^{\prime}-\mu_{1}}),
\label{unnor-t1}
\end{multline}
\begin{multline}
 {\mathcal {\overline T}}_{\mu}^{(b_{1},b_{2},\dots,b_{m}),(f_{1},f_{2},\dots,f_{n})}(x)
:= (-1)^{(m+n+1)(\xi_{m,n}(\mu ) +1)} 
\\
   \times 
\overline{\Delta}^{(b_{1},b_{2},\dots,b_{m}), (r_{1},r_{2},\dots,r_{n-m+\xi_{m,n}-1})}
_{(f_{1},f_{2},\dots,f_{n} ),(s_{1},s_{2},\dots,s_{\xi_{m,n}-1} )}
(xq^{\frac{3(m-n)}{2} -\mu_{1}^{\prime}+\mu_{1}}),
\label{unnor-t2}
\end{multline}
where $s_{l}=\mu_{\xi_{m,n} -l}+m-n-\xi_{m,n}(\mu)+l+1$,
$r_{k}=\mu_{n-m+\xi_{m,n}-k}^{\prime}+k-\xi_{m,n}(\mu)+1$. 
By the definition of the $(m,n)$-index (\ref{mn-index}), we have (cf.\ Lemma 3.2 in \cite{MV03}) 
\begin{align}
& \mu_{l} + m-n-\xi_{m,n}(\mu )+1 \ge 0 \quad \text{for} \quad 1 \le l \le \xi_{m,n}(\mu)-1 , 
\\[6pt]
& \mu_{k}^{\prime}-\xi_{m,n}(\mu )+1 \ge 0 \quad \text{for} \quad 
 1 \le k \le n-m+\xi_{m,n}(\mu )-1.
\end{align}
Here the increasing sequence of positive integers 
$(r_{1},r_{2},\dots,r_{n-m+\xi_{m,n}-1})$ is called the {\em Maya diagram}
 for the 
Young diagram 
$(\mu_{1}^{\prime}-\xi_{m,n}+1,\mu_{2}^{\prime}-\xi_{m,n}+1,\dots, 
\mu_{n-m+\xi_{m,n}-1}^{\prime}-\xi_{m,n}+1)$; 
the increasing sequence of positive integers  $(s_{1},s_{2},\dots,s_{\xi_{m,n}-1})$ is  
the Maya diagram for the Young diagram 
$(\mu_{1} + m-n-\xi_{m,n}+1,\mu_{2} + m-n-\xi_{m,n}+1,\dots, \mu_{\xi_{m,n}-1} + m-n-\xi_{m,n}+1)$. 
Maya diagram is convenient to see relation to the fermion Fock space. 
Thus the above expressions will be useful when we realize the Baxter $\Qs$-operators in terms of 
the fermion operators. 
We will use many times 
 the above formula \eqref{unnor-t1} for the Young diagram with a rectangular shape $\mu=(s^a)$. 
We have to consider the following four cases as the $(m,n)$-index \eqref{mn-index}
 depends on $m,n,a,s$. \\[12pt]
\noindent 
For $a \le m-n$, we have $\xi_{m,n}((s^a))=m-n+1$ and 
\begin{align}
 {\mathcal T}_{(s^a)}^{B_{m},F_{n}}(x)
= 
\Delta^{B_{m}, \emptyset}
_{F_{n},(1,2,\dots,m-n-a,m-n-a+s+1,\dots,m-n+s)}
(xq^{-\frac{3(m-n)}{2} +a-s}).
\end{align}
For $a-s \le m-n \le a $, we have $\xi_{m,n}((s^a))=a+1$ and 
\begin{align}
 {\mathcal T}_{(s^a)}^{B_{m},F_{n}}(x)
= (-1)^{(m+n+1)a} 
\Delta^{B_{m}, (1,2,\dots, n-m+a)}
_{F_{n},(m-n-a+s+1,\dots,m-n+s)}
(xq^{-\frac{3(m-n)}{2} +a-s}).
\end{align}
For $-s \le m-n \le a-s $, we have $\xi_{m,n}((s^a))=m-n+s+1$ and 
\begin{align}
 {\mathcal T}_{(s^a)}^{B_{m},F_{n}}(x)
= (-1)^{(m+n+1)s} 
\Delta^{B_{m},(n-m-s+a+1,\dots,n-m+a)}
_{F_{n},(1,2,\dots, m-n+s)}
(xq^{-\frac{3(m-n)}{2} +a-s}).
\end{align}
For $m-n \le -s$, we have $\xi_{m,n}((s^a))=1$ and 
\begin{align}
 {\mathcal T}_{(s^a)}^{B_{m},F_{n}}(x)
= 
\Delta^{B_{m},(1,2,\dots,n-m-s,n-m-s+a+1,\dots,n-m+a)}
_{F_{n}, \emptyset}
(xq^{-\frac{3(m-n)}{2} +a-s}).
\end{align}
\\
Now we introduce the most important functions, which correspond to the 
$T$- and $\Qs$-functions. 
\begin{align}
  \Ts_{\mu}^{(b_{1},b_{2},\dots,b_{m}),(f_{1},f_{2},\dots,f_{n})}(x)
  &:=\frac{{\mathcal T}_{\mu}^{(b_{1},b_{2},\dots,b_{m}),(f_{1},f_{2},\dots,f_{n})}(x)}
{{\mathcal T}_{\emptyset}^{(b_{1},b_{2},\dots,b_{m}),(f_{1},f_{2},\dots,f_{n})}(0)}, 
\label{9-thvari-1} \\[18pt]
  {\overline \Ts}_{\mu}^{(b_{1},b_{2},\dots,b_{m}),(f_{1},f_{2},\dots,f_{n})}(x)
  &:=\frac{{\mathcal {\overline T}}_{\mu}^{(b_{1},b_{2},\dots,b_{m}),(f_{1},f_{2},\dots,f_{n})}(x)}
{{\mathcal {\overline T}}_{\emptyset}^{(b_{1},b_{2},\dots,b_{m}),(f_{1},f_{2},\dots,f_{n})}(0)}.
\label{9-thvari-2} 
\end{align}
On the right hand side of \eqref{9-thvari-1} and \eqref{9-thvari-2}, a 
permutation on the components in the tuples $B_{m}=(b_{1},b_{2},\dots,b_{m})$ 
or $F_{n}=(f_{1},f_{2},\dots,f_{n})$ in 
the numerator and the denominator induces signs, but they cancel each other. 
Thus the formulae are well defined even if we treat  
the indexes in the left hand side 
just a pair of the sets $\{b_{1},b_{2},\dots,b_{m}\}, \{f_{1},f_{2},\dots,f_{n}\}$.
In short, they are invariant under the action of $S(B_{m})\times S(F_{n})$. 
We suppose that the $\Ts$-functions $\Ts_{\mu}^{(1,2,\dots,M),(M+1,M+2,\dots,M+N)}(x)$ and 
$\overline{\Ts}_{\mu}^{(1,2,\dots,M),(M+1,M+2,\dots,M+N)}(x)$ correspond 
to the eigenvalue formulae of the transfer matrices whose auxiliary space 
are evaluation representations of $U_{q}(\widehat{gl}(M|N))$. As representations 
of $U_{q}(gl(M|N))$, they are labelled by the Young diagram $\mu$. 
The Young diagram $\mu$ is related to the Kac-Dynkin label of a representation 
of $gl(M|N)$ (and thus $U_{q}(gl(M|N))$) 
as in \eqref{kacdynkin}. 
We expect that these two different expressions of the 
$\Ts$-functions are due to the existence of two kind of evaluation maps from 
$U_{q}(\widehat{gl}(M|N))$ to $U_{q}(gl(M|N))$  
(see eqs.\ (4.14)-(4.18) in \cite{BHK02}, and also \cite{Kojima08}). 
One can calculate the denominator explicitly (cf.\ \cite{BF94}).
\begin{align}
{\mathcal T}_{\emptyset}^{(b_{1},b_{2},\dots,b_{m}),(f_{1},f_{2},\dots,f_{n})}(0)
&={\mathcal {\overline T}}_{\emptyset}^{(b_{1},b_{2},\dots,b_{m}),(f_{1},f_{2},\dots,f_{n})}(0) 
\nonumber \\[5pt]
&=(-1)^{\frac{(m-n)(m+n-1)}{2}}
\Ds \bigl(
\begin{smallmatrix}
b_{1}, & b_{2}, & \dots, & b_{m} \\
f_{1}, & f_{2},& \dots, & f_{n}
\end{smallmatrix}
\bigr),
\label{bf-id}
\\[7pt]
\Ds\bigl(
\begin{smallmatrix}
b_{1}, & b_{2}, & \dots, & b_{m} \\
f_{1}, & f_{2},& \dots, & f_{n}
\end{smallmatrix}
\bigr)
&:=\frac{\prod_{1\le i<j \le m}(z_{b_{i}}-z_{b_{j}})
\prod_{1\le i<j \le n}(z_{f_{j}}-z_{f_{i}})}
{\prod_{i=1}^{m} \prod_{j=1}^{n} (z_{b_{i}}-z_{f_{j}})}.
\label{deno}
\end{align}
This is a generalization of the Cauchy identity. 
This represents a correlation function 
 of the $bc$-system of  
the conformal field theory on ${\mathbb P}^{1}$, and 
 is related to the ``boson-fermion correspondence". 
\eqref{deno} satisfies the following relations.
\begin{align}
&
\Ds\bigl(
\begin{smallmatrix}
b_{1}, & b_{2}, & \dots, & b_{m} \\
f_{1}, & f_{2},& \dots, & f_{n}
\end{smallmatrix}
\bigr)
\Ds\bigl(
\begin{smallmatrix}
b_{1}, & b_{2}, & \dots, & b_{m}, & \alpha, & \beta \\
f_{1}, & f_{2},& \dots, & f_{n}
\end{smallmatrix}
\bigr)
=
(z_{\alpha}-z_{\beta})
\Ds\bigl(
\begin{smallmatrix}
b_{1}, & b_{2}, & \dots, & b_{m}, & \alpha  \\
f_{1}, & f_{2},& \dots, & f_{n}
\end{smallmatrix}
\bigr)
\Ds\bigl(
\begin{smallmatrix}
b_{1}, & b_{2}, & \dots, & b_{m}, & \beta \\
f_{1}, & f_{2},& \dots, & f_{n}
\end{smallmatrix}
\bigr),
\label{q-sys-bf} 
\\
&
\Ds\bigl(
\begin{smallmatrix}
b_{1}, & b_{2}, & \dots, & b_{m} \\
f_{1}, & f_{2},& \dots, & f_{n}
\end{smallmatrix}
\bigr)
\Ds\bigl(
\begin{smallmatrix}
b_{1}, & b_{2}, & \dots, & b_{m} \\
f_{1}, & f_{2},& \dots,& f_{n}, & \alpha, & \beta 
\end{smallmatrix}
\bigr)
=
(z_{\beta}-z_{\alpha})
\Ds\bigl(
\begin{smallmatrix}
b_{1}, & b_{2}, & \dots, & b_{m}  \\
f_{1}, & f_{2},& \dots,& f_{n}, & \alpha
\end{smallmatrix}
\bigr)
\Ds\bigl(
\begin{smallmatrix}
b_{1}, & b_{2}, & \dots, & b_{m} \\
f_{1}, & f_{2},& \dots,& f_{n}, & \beta 
\end{smallmatrix}
\bigr),
\\
&
\Ds\bigl(
\begin{smallmatrix}
b_{1}, & b_{2}, & \dots, & b_{m}, & \alpha \\
f_{1}, & f_{2},& \dots,& f_{n}
\end{smallmatrix}
\bigr)
\Ds\bigl(
\begin{smallmatrix}
b_{1}, & b_{2}, & \dots, & b_{m}  \\
f_{1}, & f_{2},& \dots,& f_{n}, & \beta
\end{smallmatrix}
\bigr)
=
(z_{\alpha}-z_{\beta})
\Ds\bigl(
\begin{smallmatrix}
b_{1}, & b_{2}, & \dots, & b_{m}  \\
f_{1}, & f_{2},& \dots, & f_{n}
\end{smallmatrix}
\bigr)
\Ds\bigl(
\begin{smallmatrix}
b_{1}, & b_{2}, & \dots, & b_{m}, & \alpha \\
f_{1}, & f_{2},& \dots,& f_{n}, & \beta 
\end{smallmatrix}
\bigr)
.
\label{q-sys-bf3} 
\end{align}
The functional relations \eqref{QQ-rel1}-\eqref{QQ-rel4}
 may be viewed as the Yang-Baxterization
\footnote{In the context of the conformal field theory, 
the author found the identity \eqref{bf-id} in \cite{Yamada06} first. 
Then he heard from  Yasuhiko Yamada that 
this identity naturally follows from a method, for example, in \cite{Witten88}. 
One may say that the Baxter $\Qs$-functions or $\Qs$-operators are the Yang-Baxterization 
of correlation functions of the conformal field theory. 
There are higher genus analogues of correlation functions of the conformal field theory. 
They were used to calculate string amplitudes. 
Whether such correlation functions also allow the Yang-Baxterization as 
$\Qs$-functions or $\Qs$-operators (or more generally, $\Ts$-functions or $\Ts$-operators) 
 will be an interesting question.}
 of \eqref{q-sys-bf}-\eqref{q-sys-bf3}.

For generic
\footnote{without assuming (\ref{Acal-mat-el})}
 ${\mathsf Z}_{k,l},{\mathsf X}_{k,l},{\mathsf Y}_{k,l},{\mathsf W}_{k,l}$ 
in (\ref{Acal-mat}), (\ref{9-thvari-1}) is 
 a generalization of the {\em ninth variation of Schur function in terms of the first Weyl formula} \cite{Mac92,NNSY00}. In fact, $\Ts_{\mu}^{B_{M},\emptyset}(x)$ corresponds to eq. (1.7) in \cite{NNSY00}.
It is interesting to reformulate discussions in this paper 
based on the Gauss decomposition of a sub-matrix of ${\mathcal A}(x)$ 
along the line in \cite{NNSY00}. 

We will use the following lemma (cf.\ \cite{DM92,T97,T98}). 
\begin{lemma}\label{vanish-dai}
If the Young diagram $\mu $ contains a rectangle with a hight of $(m+1)$ and 
a width of $(n+1)$, 
the functions \eqref{9-thvari-1} and \eqref{9-thvari-2} vanish. 
\end{lemma}
This can be proved by the fact that 
the $(m,n)$-index \eqref{mn-index} becomes larger than or equal to $m+2$ in this case. 
We impose the normalization \eqref{tri1} (resp.\ \eqref{tri2}) 
when we use \eqref{9-thvari-1} (resp.\ \eqref{9-thvari-2}) 
 from now on. 
Now we mention one of the main results in this paper. 
\begin{theorem} \label{solution-Q}
\begin{align}
  \Qs_{B_{m}\sqcup F_{n}}(x)
  =\Qs_{B_{m}\times F_{n}}(x)
  =
\Ts_{\emptyset}^{B_{m}, F_{n}}
(xq^{\frac{m-n}{2}})
\label{q-func1}
\end{align}
solves
 the functional relations for the Baxter $\Qs$-functions \eqref{QQ-rel1}-\eqref{QQ-rel2} 
under the relation \eqref{QQ-bf1} and the normalization \eqref{tri1}, 
and 
\begin{align}
  {\overline \Qs}_{B_{m}\sqcup F_{n}}(x)
  ={\overline \Qs}_{B_{m}\times F_{n}}(x)=
 {\overline \Ts}_{\emptyset}^{B_{m}, F_{n}}
(xq^{-\frac{m-n}{2}}) 
\label{q-func2}
\end{align}
solves the functional relations for the Baxter $\Qs$-functions \eqref{QQ-rel3}-\eqref{QQ-rel4} 
under the relation \eqref{QQ-bf2} and the normalization \eqref{tri2}. 
\end{theorem}
A proof of this theorem is given in Appendix \ref{proof-Q}. 
Let us write the above formulae explicitly. 
For $m \ge n$, we have 
\begin{multline}
\overline{\Qs}_{\{b_{1},b_{2},\dots,b_{m},f_{1},f_{2},\cdots,f_{n}\}}(x) =\frac{
(-1)^{\frac{1}{2}(m-n)(m+n-1)}
\prod_{k=1}^{m}\prod_{l=1}^{n}(z_{b_{k}}-z_{f_{l}})}
{\prod_{1 \le k <l\le m}(z_{b_{k}}-z_{b_{l}})
\prod_{1 \le k <l\le n}(z_{f_{l}}-z_{f_{k}})}  \\[5pt]
 \times \det
\left(
\left(\frac{
\overline{\Qs}_{\{b_{k},f_{l}\}}(xq^{m-n})}{z_{b_{k}}-z_{f_{l}}}
\right)_{1 \le k \le m,
1 \le l \le n}
,
\left(
z_{b_{k}}^{l-1}\overline{\Qs}_{ b_{k}}(xq^{m-n -2l+1})
\right)_{1 \le k \le m,1 \le l \le m- n}
\right)
\end{multline}
since $\xi_{m,n}(\emptyset)=m-n +1$. 
And also for $m \le  n$, we have 
\begin{multline}
 \overline{\Qs}_{\{b_{1},b_{2},\dots,b_{m},f_{1},f_{2},\cdots,f_{n}\}}(x)
 =\frac{
(-1)^{\frac{1}{2}(m-n) (m+n -1)}
\prod_{k=1}^{m}\prod_{l=1}^{n}(z_{b_{k}}-z_{f_{l}})}
{\prod_{1 \le k <l\le m}(z_{b_{k}}-z_{b_{l}})
\prod_{1 \le k <l\le n}(z_{f_{l}}-z_{f_{k}})}  \\[5pt]
 \times \det
\left(
\begin{array}{c}
\left(\frac{
\overline{\Qs}_{\{b_{k},f_{l}\}}(xq^{m-n })}{z_{b_{k}}-z_{f_{l}}}
\right)_{1 \le k \le m,
1 \le l \le n}
\\[10pt]
\left(
(-z_{f_{l}})^{k-1}\overline{\Qs}_{f_{l}}(xq^{m-n +2k-1})
\right)_{1 \le k \le n-m,1 \le l \le n}
\end{array}
\right) 
\end{multline}
since $\xi_{m,n}(\emptyset )=1$. 
We remark that for $(m,n)=(M,N)$ case, the denominator in the right hand side of 
\eqref{9-thvari-1} or 
\eqref{9-thvari-2} can be replaced by 
${\mathcal T}_{\emptyset}^{B_{M},F_{N}}(x)$ or $\overline{\mathcal T}_{\emptyset}^{B_{M},F_{N}}(x)$
if both \eqref{tri1} and \eqref{tri2} are imposed at the same time. 
\eqref{q-func1} and \eqref{q-func2} for $(m,n)=(M,N)$ are supposed to 
be model dependent scalar functions (for example, \eqref{quantum-wron-ps}, 
\eqref{quantum-wron-ps2}), which correspond to the quantum Wronskian conditions. 
One can obtain other relations 
\footnote{See, eq.\ (3.1) in \cite{BT08}, where 
both \eqref{tri1} and \eqref{tri2} are imposed.} 
by applying \eqref{conj} to the determinants 
\eqref{q-func1} and \eqref{q-func2} for $(m,n)=(M,N)$.  

We have normalized the Baxter $\Qs$-functions as in \eqref{Q-norm}. 
This normalization is convenient when we realize the Baxter $\Qs$-operator 
as super-trace over a $q$-oscillator representation of 
$U_{q}(\widehat{gl}(M|N))$ \cite{workinprogress} since the final expression of the 
 $\Qs$-operator does not depend on choice of the vacuum of the Fock space
 where the super-trace is taken. 
In this normalization, one can also easy to see that the $\Ts$-function reduces to the 
(super)character formula of $gl(M|N)$ (or its subalgebras) in the limit $x\to 0$. 
Of course, one may adapt different normalizations of the Baxter $\Qs$-function. 
Let us consider the following transformation (cf.\ eq.\ (65) in \cite{Zabrodin07}): 
\begin{align}
& {\mathfrak Q}_{I}(x)={\mathfrak a}_{I}x^{{\mathsf S}_{I}}\Qs_{I}(x).
\end{align}
It will be good to put 
\begin{align}
{\mathsf S}_{I}=\sum_{i \in I}{\mathsf S}_{i}, \quad 
q^{p_{i}{\mathsf S}_{i}}=z_{i}^{\frac{1}{2}}
\label{Q-op-norma}
\end{align}
to simplify the formulae. For ${\mathfrak a}_{I}=1$, 
\eqref{QQ-rel1} and \eqref{QQ-rel2} become 
\begin{align}
& \frac{z_{i}-z_{j}}{(z_{i}z_{j})^{\frac{1}{2}}}{\mathfrak Q}_{I}(x){\mathfrak Q}_{I\sqcup \{i,j\}}(x)
={\mathfrak Q}_{I\sqcup \{i\}}(xq^{p_{i}})
{\mathfrak Q}_{I\sqcup \{j\}}(xq^{-p_{i}})-
{\mathfrak Q}_{I\sqcup \{i\}}(xq^{-p_{i}})
{\mathfrak Q}_{I\sqcup \{j\}}(xq^{p_{i}})
\nonumber \\
& \hspace{180pt} 
\text{for} \qquad p_{i}=p_{j},  \label{QQ-bt1} \\[6pt]
& 
\frac{z_{i}-z_{j}}{(z_{i}z_{j})^{\frac{1}{2}}}
{\mathfrak Q}_{I \sqcup \{i\}}(x){\mathfrak Q}_{I \sqcup \{j\}}(x)=
{\mathfrak Q}_{I}(xq^{-p_{i}})
{\mathfrak Q}_{I \sqcup \{i,j\}}(xq^{p_{i}})-
{\mathfrak Q}_{I}(xq^{p_{i}})
{\mathfrak Q}_{I \sqcup \{i,j\}}(xq^{-p_{i}}) 
\nonumber \\
& \hspace{180pt} \text{for} \qquad p_{i}=-p_{j}. \label{QQ-bt2}
\end{align}
 If one requires ${\mathsf S}_{I_{M+N} }=0$ for \eqref{Q-op-norma}, one obtains 
${\mathsf S}_{\overline{I}}=-{\mathsf S}_{I}$ and 
$\prod_{i=1}^{M+N}z_{i}^{\frac{p_{i}}{2}}=1$ since 
${\mathsf S}_{I}+{\mathsf S}_{\overline{I}}={\mathsf S}_{I_{M+N}}$, 
where $\overline{I}={\mathfrak I} \setminus I$.
We used \eqref{QQ-bt1}-\eqref{QQ-bt2} with 
 ${\mathsf S}_{I_{M+N}}=0$ and 
${\mathfrak a}_{I}=1$ for $U_{q}(\widehat{sl}(2|1))$ case \cite{BT08}.  
In the case of the Baxter $\Qs$-operators, 
the parameters ${\mathsf S}_{i}$ are certain linear
 combinations of the Cartan generators 
 of $U_{q}(\widehat{gl}(M|N))$ and external field parameters; and 
in the case of the Baxter $\Qs$-functions (eigenvalues), 
they are conserved quantum numbers (edge occupation numbers). 
In addition to \eqref{Q-op-norma}, one may require
\footnote{Here we used a notation:
$I \times (i,j)=(i_{1},i_{2},\dots,i_{a},i,j)$ etc. 
for $I=(i_{1},i_{2},\dots,i_{a})$.}
\begin{align}
{\mathfrak a}_{I \times (i,j)}=
\left(\frac{z_{i}-z_{j}}{(z_{i}z_{j})^{\frac{1}{2}}}\right)^{p_{i}p_{j}}
\frac{{\mathfrak a}_{I \times (i)}{\mathfrak a}_{I \times (j)}}{{\mathfrak a}_{I}}.
\end{align}
Thus one obtains:
\begin{align}
{\mathfrak a}_{(i_{1},i_{2},\dots,i_{a})}=
\frac{{\mathfrak a}_{(i_{1})}{\mathfrak a}_{(i_{2})}
\cdots {\mathfrak a}_{(i_{a})}}{{\mathfrak a}_{\emptyset}^{a -1}}
\prod_{1 \le j <  k \le a }
\left(\frac{z_{i_{j}}-z_{i_{k}}}{(z_{i_{j}}z_{i_{k}})^{\frac{1}{2}}}\right)^{p_{i_{j}}p_{j_{k}}}.
\end{align} 
In this case, the parameters $\{z_{j}\}$ in \eqref{QQ-rel1}-\eqref{QQ-rel2} 
disappear. In fact, \eqref{QQ-rel1} and \eqref{QQ-rel2} 
become coefficient free form
\footnote{The second functional relation in the 
coefficient free form \eqref{coe-free2}
 is discussed in \cite{KSZ07}.}
: 
\begin{align}
& {\mathfrak Q}_{I}(x){\mathfrak Q}_{I\times (i,j)}(x)
={\mathfrak Q}_{I\times (i)}(xq^{p_{i}})
{\mathfrak Q}_{I\times (j)}(xq^{-p_{i}})-
{\mathfrak Q}_{I\times (i)}(xq^{-p_{i}})
{\mathfrak Q}_{I\times (j)}(xq^{p_{i}})
\nonumber \\[2pt]
& \hspace{180pt} 
\text{for} \qquad p_{i}=p_{j},  \\[6pt]
& 
{\mathfrak Q}_{I \times (i)}(x){\mathfrak Q}_{I \times (j)}(x)=
{\mathfrak Q}_{I}(xq^{-p_{i}})
{\mathfrak Q}_{I \times (i,j)}(xq^{p_{i}})-
{\mathfrak Q}_{I}(xq^{p_{i}})
{\mathfrak Q}_{I \times (i,j)}(xq^{-p_{i}}) 
\nonumber \\[2pt]
& \hspace{180pt} \text{for} \qquad p_{i}=-p_{j}. 
\label{coe-free2}
\end{align}
Moreover, one can eliminate the parameters $\{z_{j }\}$ in the $\Ts$-functions 
by the following transformation (with ${\mathfrak a}_{\emptyset }=1$):
\begin{align}
{\mathfrak T}^{B_{m},F_{n}}_{\mu}(x)=
{\mathfrak a}_{B_{m}\times F_{n}}(q^{-\frac{m-n}{2}+\mu_{1}^{\prime}-\mu_{1}}
x)^{ {\mathsf S}_{B_{m}\sqcup F_{n}}}
\Ts^{B_{m},F_{n}}_{\mu}(x).
\end{align}
The discussion on $\overline{\Qs}_{I}(x)$ and $\overline{\Ts}^{B_{m},F_{n}}_{\mu}(x)$ 
is parallel to the one on $\Qs_{I}(x)$ and $\Ts^{B_{m},F_{n}}_{\mu}(x)$. 

Let us consider the formulae \eqref{9-thvari-1} and \eqref{9-thvari-2} 
for the rectangular Young diagram $\mu=(s^a)$, and introduce the following symbols:    
\begin{align}
& \Ts^{(a),B_{m}, F_{n}}_{s}(x):=
\begin{cases}
\Ts_{(s^{a})}^{B_{m},F_{n}}(x) & \text{for} \quad a,s \in {\mathbb Z}_{\ge 1}, \\[5pt]
\Ts_{\emptyset}^{B_{m},F_{n}}(xq^{-s}) & \text{for} \quad a=0 
\quad \text{and} \quad  s \in {\mathbb Z}, \\[5pt]
\Ts_{\emptyset}^{B_{m},F_{n}}(xq^{a}) & \text{for} \quad s=0 
\quad \text{and} \quad a \in {\mathbb Z}_{\ge 0}, \\[5pt] 
0 & \text{otherwise},
\end{cases}
\label{tfunrec1}
\end{align}
and 
\begin{align}
& \overline{\Ts}^{(a),B_{m} , F_{n}}_{s}(x):=
\begin{cases}
\overline{\Ts}_{(s^{a})}^{B_{m},F_{n}}(x) & \text{for} \quad a,s \in {\mathbb Z}_{\ge 1}, \\[5pt]
\overline{\Ts}_{\emptyset}^{B_{m},F_{n}}(xq^{s}) & \text{for} \quad a=0 
\quad \text{and} \quad  s \in {\mathbb Z}, \\[5pt]
\overline{\Ts}_{\emptyset}^{B_{m},F_{n}}(xq^{-a}) & \text{for} \quad s=0 
\quad \text{and} \quad a \in {\mathbb Z}_{\ge 0}, \\[5pt] 
0 & \text{otherwise}.
\end{cases}
\label{tfunrec2}
\end{align}
Due to Lemma \ref{vanish-dai}, 
$\Ts^{(a),B_{m} , F_{n}}_{s}(x)=\overline{\Ts}^{(a),B_{m},F_{n}}_{s}(x)=0 $ 
if $a\in {\mathbb Z}_{\ge m+1}$ and $s \in {\mathbb Z}_{\ge n+1}$. 
Taking note on Theorem \ref{solution-Q}, one finds:
\begin{align}
& \Ts^{(a),B_{m},F_{n}}_{0}(x)=\Qs_{B_{m} \sqcup F_{n}}(xq^{-\frac{m-n}{2}+a}) 
\quad \text{for} \quad a \in {\mathbb Z}_{\ge 0},
\label{t-sys-bc1} 
\\[4pt]
& \Ts^{(0),B_{m},F_{n}}_{s}(x)=\Qs_{B_{m} \sqcup F_{n}}(xq^{-\frac{m-n}{2}-s}) 
\quad \text{for} \quad s \in {\mathbb Z},
 \label{t-sys-bc2}
 \\[4pt]
& 
\overline{\Ts}^{(a),B_{m},F_{n}}_{0}(x)=\overline{\Qs}_{B_{m} \sqcup F_{n}}(xq^{\frac{m-n}{2}-a}) 
\quad \text{for} \quad a \in {\mathbb Z}_{\ge 0},
\label{t-sys-bc3} 
\\[4pt]
&
 \overline{\Ts}^{(0),B_{m},F_{n}}_{s}(x)=\overline{\Qs}_{B_{m} \sqcup F_{n}}(xq^{\frac{m-n}{2}+s}) 
\quad \text{for} \quad s \in {\mathbb Z}.
 \label{t-sys-bc4}
\end{align}

The following formulae follow from the Laplace expansion of the determinants, 
\eqref{t-sys-bc1}-\eqref{t-sys-bc4} and Lemma \ref{shift-lem}. 
\\
For $a-s \le m-n$, we have 
\begin{multline}
\Ts_{s}^{(a),B_{m},F_{n}}(x)= \sum_{I \subset B_{m},{\rm Card}(I)=a} 
\frac{(\prod_{\gamma \in I}z_{\gamma})^{s-a+m-n}
\prod_{\alpha \in I}\prod_{\beta \in F_{n}}(z_{\alpha}-z_{\beta})}
{\prod_{\alpha \in I}\prod_{\beta \in B_{m} \setminus I}(z_{\alpha}-z_{\beta})}
 \\
 \times 
\Qs_{B_{m}\sqcup F_{n} \setminus I}(xq^{-s-\frac{m-n}{2}})
\Qs_{I}(xq^{s+\frac{m-n}{2}}), 
 \label{lapQQb1}
\end{multline}
\begin{multline}
{\overline \Ts}_{s}^{(a),B_{m}, F_{n}}(x)=\sum_{I \subset B_{m},{\rm Card}(I)=a} 
\frac{(\prod_{\gamma \in I}z_{\gamma})^{s-a+m-n}
\prod_{\alpha \in I}\prod_{\beta \in F_{n}}(z_{\alpha}-z_{\beta})}
{\prod_{\alpha \in I}\prod_{\beta \in B_{m} \setminus I}(z_{\alpha}-z_{\beta})}
 \\
 \times
\overline{\Qs}_{B_{m}\sqcup F_{n} \setminus I}(xq^{s+\frac{m-n}{2}})
\overline{\Qs}_{I}(xq^{-s-\frac{m-n}{2}}),
\label{lapQQb2}
\end{multline}
where the summation is taken over all the subset $I$ of $B_{m}$ such that 
${\rm Card}(I)=a$.  
We remark that the right hand side of \eqref{lapQQb1} and \eqref{lapQQb2} 
is well defined as a function of $x$ even if $s$ is not non-negative integer. 
Based on this fact, one can derive $\Ts$-functions for conjugate representations 
by considering negative $s$ 
(as examples for $U_{q}(\widehat{sl}(2|1))$ in \cite{BT08}). 
In fact, applying \eqref{conj} to \eqref{lapQQb1}, we obtain the following relation: 
\begin{align}
{\mathfrak C}[\Ts_{s}^{(a),B_{m},F_{n}}(x)]
=(-1)^{a(n-m+1)}\frac{\prod_{b \in B_{m}}z_{b}^{a}}{\prod_{b \in F_{n}}z_{b}^{a}}
[\text{the right hand side of } \eqref{lapQQb1}]_{s \to -s-(m-n)}. 
\end{align}
\\
For $a-s \ge m-n$, we have 
\begin{multline}
\Ts_{s}^{(a),B_{m},F_{n}}(x)=\sum_{J\subset F_{n}, {\rm Card}(J)=s} 
\frac{(\prod_{\gamma \in J}(-z_{\gamma}))^{a-s+n-m}
\prod_{\alpha \in B_{m}}\prod_{\beta \in J}(z_{\alpha}-z_{\beta})}
{\prod_{\alpha \in F_{n} \setminus J}\prod_{\beta \in J}(z_{\alpha}-z_{\beta})}
 \\
 \times
\Qs_{B_{m}\sqcup F_{n} \setminus J}(xq^{a-\frac{m-n}{2}})
\Qs_{J}(xq^{-a+\frac{m-n}{2}}),
\label{lapQQb3}
\end{multline}
\begin{multline}
{\overline \Ts}_{s}^{(a),B_{m},F_{n}}(x)=\sum_{J\subset F_{n}, {\rm Card}(J)=s} 
\frac{(\prod_{\gamma \in J}(-z_{\gamma}))^{a-s+n-m}
\prod_{\alpha \in B_{m}}\prod_{\beta \in J}(z_{\alpha}-z_{\beta})}
{\prod_{\alpha \in F_{n} \setminus J}\prod_{\beta \in J}(z_{\alpha}-z_{\beta})}
 \\
 \times
\overline{\Qs}_{B_{m}\sqcup F_{n} \setminus J}(xq^{-a+\frac{m-n}{2}})
\overline{\Qs}_{J}(xq^{a-\frac{m-n}{2}}),
\label{lapQQb4}
\end{multline}
where the summation is taken over all the subset $J$ of $F_{n}$ such that 
${\rm Card}(J)=s$. 
We remark that the right hand side of \eqref{lapQQb3} and \eqref{lapQQb4} 
is well defined
\footnote{We expect that each term in the right hand side of 
 \eqref{lapQQb1}-\eqref{lapQQb4} 
corresponds to a $\Ts$-function for an infinite dimensional representation of 
$U_{q}(\widehat{gl}(M|N))$ (or $U_{q}(\widehat{sl}(M|N))$). 
See examples for some special $(M,N)$ in \cite{BLZ97,BHK02,BT08}.}
 as a function of $x$ even if $a$ is not non-negative integer. 
Note that $\overline{\Qs}_{B_{M}\sqcup F_{N} \setminus I}(x)=\Qs_{I}(x)$, 
$\overline{\Qs}_{B_{M}\sqcup F_{N} \setminus J}(x)=\Qs_{J}(x)$ for $(m,n)=(M,N)$. 
Thus the relation  $\Ts_{s}^{(a),B_{M},F_{N}}(x)={\overline \Ts}_{s}^{(a),B_{M},F_{N}}(x)$ 
holds if both \eqref{tri1} and \eqref{tri2} are imposed.
In particular for $a=m$ or $s=n$ case, the above formulae factorize with respect 
to the Baxter $\Qs$-functions: 
\begin{align}
\Ts_{s}^{(m),B_{m},F_{n}}(x)&=
(\prod_{\gamma \in B_{m}}z_{\gamma})^{s-n}
\prod_{\alpha \in B_{m}}\prod_{\beta \in F_{n}}(z_{\alpha}-z_{\beta})
\Qs_{F_{n}}(xq^{-s-\frac{m-n}{2}})
\Qs_{B_{m}}(xq^{s+\frac{m-n}{2}}) , \label{typfac1}
 \\[6pt]
\overline{\Ts}_{s}^{(m),B_{m},F_{n}}(x)&=
(\prod_{\gamma \in B_{m}}z_{\gamma})^{s-n}
\prod_{\alpha \in B_{m}}\prod_{\beta \in F_{n}}(z_{\alpha}-z_{\beta})
\overline{\Qs}_{F_{n}}(xq^{s+\frac{m-n}{2}})
\overline{\Qs}_{B_{m}}(xq^{-s-\frac{m-n}{2}}) 
\label{typfac2}
 \\
& 
\hspace{190pt}   \text{for} \qquad s \ge n, \nonumber 
\end{align}
and 
\begin{align} 
\Ts_{n}^{(a),B_{m},F_{n}}(x)&=
(\prod_{\gamma \in F_{n}}(-z_{\gamma}))^{a-m}
\prod_{\alpha \in B_{m}}\prod_{\beta \in F_{n}}(z_{\alpha}-z_{\beta})
\Qs_{B_{m}}(xq^{a-\frac{m-n}{2}})
\Qs_{F_{n}}(xq^{-a+\frac{m-n}{2}}),  
\label{typfac3} \\
\overline{\Ts}_{n}^{(a),B_{m},F_{n}}(x)&=
(\prod_{\gamma \in F_{n}}(-z_{\gamma}))^{a-m}
\prod_{\alpha \in B_{m}}\prod_{\beta \in F_{n}}(z_{\alpha}-z_{\beta})
\overline{\Qs}_{B_{m}}(xq^{-a+\frac{m-n}{2}})
\overline{\Qs}_{F_{n}}(xq^{a-\frac{m-n}{2}}) \label{typfac4} \\
& 
\hspace{190pt}  \text{for} \quad a \ge m. \nonumber 
\end{align}
In \cite{KSZ07}, 
this type of factorization formulae for the $\Ts$-functions 
 were treated as ``boundary conditions'' of the Hirota equation. 
These formulae \eqref{typfac1}-\eqref{typfac4} 
are related to the $\Ts$-functions for typical 
representations, which will be commented in section 4. 
Now we mention our main theorem. 
\begin{theorem}\label{solution-t-sys-th}
For $a,s \in {\mathbb Z}_{\ge 1}$, 
the determinant formula \eqref{tfunrec1} solves the $T$-system 
\begin{align}
\begin{split}
&  
\Ts^{(a),B_{m},F_{n}}_{s}(xq^{-1})\Ts^{(a),B_{m},F_{n}}_{s}(xq)=
\Ts^{(a),B_{m},F_{n}}_{s-1}(x)\Ts^{(a),B_{m},F_{n}}_{s+1}(x)  \\
& \hspace{180pt} +
\Ts^{(a-1),B_{m},F_{n}}_{s}(x)\Ts^{(a+1),B_{m},F_{n}}_{s}(x)  \\[2pt] 
& \hspace{75pt} \text{for} \quad 1 \le a \le m-1 \quad \text{or} \quad 1\le s \le n-1
 \quad \text{or} \quad (a,s)=(m,n),
\end{split}
\label{t-system1} 
 \\[6pt]
&  
\Ts^{(m),B_{m},F_{n}}_{s}(xq^{-1})\Ts^{(m),B_{m},F_{n}}_{s}(xq)=
\Ts^{(m),B_{m},F_{n}}_{s-1}(x)\Ts^{(m),B_{m},F_{n}}_{s+1}(x)
\quad \text{for} \; s \in {\mathbb Z}_{\ge n+1}, \label{reduc1} \\[7pt]
&  
\Ts^{(a),B_{m},F_{n}}_{n}(xq^{-1})\Ts^{(a),B_{m},F_{n}}_{n}(xq)=
\Ts^{(a-1),B_{m},F_{n}}_{n}(x)\Ts^{(a+1),B_{m},F_{n}}_{n}(x)
\; \text{for} \; a \in {\mathbb Z}_{\ge m+1}, \label{reduc2} 
\\[7pt]
& \Ts_{b+n}^{(m),B_{m},F_{n}}(x)=
\left(\frac{\prod_{\gamma \in B_{m}}z_{\gamma}}{\prod_{\gamma \in F_{n}}(-z_{\gamma})}
\right)^{b}
\Ts_{n}^{(b+m),B_{m},F_{n}}(x) \quad \text{for} \quad b \in {\mathbb Z}_{\ge 0},
 \label{dual-mn}
\end{align} 
with the boundary conditions \eqref{t-sys-bc1} and \eqref{t-sys-bc2}
 (and also \eqref{typfac1} and  \eqref{typfac3}) 
under the relation \eqref{QQ-bf1};  
and the determinant formula \eqref{tfunrec2}
 solves the $T$-system 
\begin{align}
\begin{split}
&
\overline{\Ts}^{(a),B_{m},F_{n}}_{s}(xq^{-1})\overline{\Ts}^{(a),B_{m},F_{n}}_{s}(xq)=
\overline{\Ts}^{(a),B_{m},F_{n}}_{s-1}(x)\overline{\Ts}^{(a),B_{m},F_{n}}_{s+1}(x)  \\
& \hspace{180pt} +
\overline{\Ts}^{(a-1),B_{m},F_{n}}_{s}(x)\overline{\Ts}^{(a+1),B_{m},F_{n}}_{s}(x) 
  \\[2pt]
& \hspace{75pt} \text{for} \quad 1 \le a \le m-1 \quad \text{or} \quad 1\le s \le n-1
 \quad \text{or} \quad (a,s)=(m,n),
\end{split}
 \label{t-system2}
\\[6pt]
&  
\overline{\Ts}^{(m),B_{m},F_{n}}_{s}(xq^{-1})\overline{\Ts}^{(m),B_{m},F_{n}}_{s}(xq)=
\overline{\Ts}^{(m),B_{m},F_{n}}_{s-1}(x)\overline{\Ts}^{(m),B_{m},F_{n}}_{s+1}(x)
\quad \text{for} \; s \in {\mathbb Z}_{\ge n+1}, \label{reduc3} \\[7pt]
&
\overline{\Ts}^{(a),B_{m},F_{n}}_{n}(xq^{-1})\overline{\Ts}^{(a),B_{m},F_{n}}_{n}(xq)=
\overline{\Ts}^{(a-1),B_{m},F_{n}}_{n}(x)\overline{\Ts}^{(a+1),B_{m},F_{n}}_{n}(x)
\; \text{for} \; a \in {\mathbb Z}_{\ge m+1}, \label{reduc4} 
\\[7pt]
& \overline{\Ts}_{b+n}^{(m),B_{m},F_{n}}(x)=
\left(\frac{\prod_{\gamma \in B_{m}}z_{\gamma}}{\prod_{\gamma \in F_{n}}(-z_{\gamma})}
\right)^{b}
\overline{\Ts}_{n}^{(b+m),B_{m},F_{n}}(x)
 \quad \text{for} \quad b \in {\mathbb Z}_{\ge 0} 
 \label{dual-mn2}
\end{align}
with the boundary conditions \eqref{t-sys-bc3} and \eqref{t-sys-bc4}
 (and also \eqref{typfac2} and  \eqref{typfac4})  
under the relation \eqref{QQ-bf2}.
\end{theorem}
A proof of this theorem is given in Appendix \ref{proof-t-sys}. 

Our new determinant formulae also satisfy the 
B\"{a}cklund transformations \cite{KSZ07,Zabrodin07}. 
The following theorem is also one of our main results. 
\begin{theorem}\label{solution-back}
For $a,s \in {\mathbb Z}_{\ge 0}$, 
the determinant formula \eqref{tfunrec1} 
 satisfies the B\"acklund transformations 
\begin{align}
\begin{split}
& \Ts^{(a+1),B_{m},F_{n}}_{s}(x)\Ts^{(a),B_{m-1},F_{n}}_{s}(xq^{\frac{3}{2}})-
\Ts^{(a),B_{m},F_{n}}_{s}(xq)\Ts^{(a+1),B_{m-1},F_{n}}_{s}(xq^{\frac{1}{2}})
\\
& \hspace{180pt} =
z_{b_{m}}\Ts^{(a+1),B_{m},F_{n}}_{s-1}(xq)\Ts^{(a),B_{m-1},F_{n}}_{s+1}(xq^{\frac{1}{2}}), 
\end{split}
\label{bac1} 
\\[8pt] 
\begin{split}
&  \Ts^{(a),B_{m},F_{n}}_{s+1}(xq)\Ts^{(a),B_{m-1},F_{n}}_{s}(xq^{\frac{1}{2}})-
\Ts^{(a),B_{m},F_{n}}_{s}(x)\Ts^{(a),B_{m-1},F_{n}}_{s+1}(xq^{\frac{3}{2}})
 \\ 
& \hspace{180pt} =
z_{b_{m}}\Ts^{(a+1),B_{m},F_{n}}_{s}(xq)\Ts^{(a-1),B_{m-1},F_{n}}_{s+1}(xq^{\frac{1}{2}}), 
\label{bac2}
\end{split}
\\[8pt] 
\begin{split}
&  \Ts^{(a+1),B_{m},F_{n-1}}_{s}(x)\Ts^{(a),B_{m},F_{n}}_{s}(xq^{\frac{3}{2}})-
\Ts^{(a),B_{m},F_{n-1}}_{s}(xq)\Ts^{(a+1),B_{m},F_{n}}_{s}(xq^{\frac{1}{2}})
 \\ 
& \hspace{180pt} =
z_{f_{n}}\Ts^{(a+1),B_{m},F_{n-1}}_{s-1}(xq)\Ts^{(a),B_{m},F_{n}}_{s+1}(xq^{\frac{1}{2}}), 
\label{bac3}
\end{split}
\\[8pt] 
\begin{split}
& \Ts^{(a),B_{m},F_{n-1}}_{s+1}(xq)\Ts^{(a),B_{m},F_{n}}_{s}(xq^{\frac{1}{2}})-
\Ts^{(a),B_{m},F_{n-1}}_{s}(x)\Ts^{(a),B_{m},F_{n}}_{s+1}(xq^{\frac{3}{2}})
 \\ 
& \hspace{180pt} =
z_{f_{n}}\Ts^{(a+1),B_{m},F_{n-1}}_{s}(xq)\Ts^{(a-1),B_{m},F_{n}}_{s+1}(xq^{\frac{1}{2}}) 
\label{bac4}
\end{split}
\end{align}
 under the relation \eqref{QQ-bf1}, and 
the determinant formula \eqref{tfunrec2}
 satisfies the B\"acklund transformations 
\begin{align}
\begin{split}
&  \overline{\Ts}^{(a+1),B_{m},F_{n}}_{s}(x)\overline{\Ts}^{(a),B_{m-1},F_{n}}_{s}(xq^{-\frac{3}{2}})-
\overline{\Ts}^{(a),B_{m},F_{n}}_{s}(xq^{-1})\overline{\Ts}^{(a+1),B_{m-1},F_{n}}_{s}(xq^{-\frac{1}{2}})
 \\
& \hspace{160pt} =
z_{b_{m}}\overline{\Ts}^{(a+1),B_{m},F_{n}}_{s-1}(xq^{-1})\overline{\Ts}^{(a),B_{m-1},F_{n}}_{s+1}(xq^{-\frac{1}{2}}), 
\end{split}
\label{bac5}
\\[8pt] 
\begin{split}
&  \overline{\Ts}^{(a),B_{m},F_{n}}_{s+1}(xq^{-1})\overline{\Ts}^{(a),B_{m-1},F_{n}}_{s}(xq^{-\frac{1}{2}})-
\overline{\Ts}^{(a),B_{m},F_{n}}_{s}(x)\overline{\Ts}^{(a),B_{m-1},F_{n}}_{s+1}(xq^{-\frac{3}{2}})
 \\ 
& \hspace{160pt} =
z_{b_{m}}\overline{\Ts}^{(a+1),B_{m},F_{n}}_{s}(xq^{-1})\overline{\Ts}^{(a-1),B_{m-1},F_{n}}_{s+1}(xq^{-\frac{1}{2}}), 
\label{bac6}
\end{split}
\\[8pt] 
\begin{split}
&  \overline{\Ts}^{(a+1),B_{m},F_{n-1}}_{s}(x)\overline{\Ts}^{(a),B_{m},F_{n}}_{s}(xq^{-\frac{3}{2}})-
\overline{\Ts}^{(a),B_{m},F_{n-1}}_{s}(xq^{-1})\overline{\Ts}^{(a+1),B_{m},F_{n}}_{s}(xq^{-\frac{1}{2}})
 \\ 
& \hspace{180pt} =
z_{f_{n}}\overline{\Ts}^{(a+1),B_{m},F_{n-1}}_{s-1}(xq^{-1})\overline{\Ts}^{(a),B_{m},F_{n}}_{s+1}(xq^{-\frac{1}{2}}), 
\label{bac7}
\end{split}
\\[8pt] 
\begin{split}
&  \overline{\Ts}^{(a),B_{m},F_{n-1}}_{s+1}(xq^{-1})\overline{\Ts}^{(a),B_{m},F_{n}}_{s}(xq^{-\frac{1}{2}})-
\overline{\Ts}^{(a),B_{m},F_{n-1}}_{s}(x)\overline{\Ts}^{(a),B_{m},F_{n}}_{s+1}(xq^{-\frac{3}{2}})
 \\ 
& \hspace{180pt} =
z_{f_{n}}\overline{\Ts}^{(a+1),B_{m},F_{n-1}}_{s}(xq^{-1})\overline{\Ts}^{(a-1),B_{m},F_{n}}_{s+1}
(xq^{-\frac{1}{2}})
\end{split}
\label{bac8}
\end{align}
 under the relation \eqref{QQ-bf2}.
\end{theorem}
A proof of this theorem is given in Appendix \ref{proof-back}. 
Due to Lemma \ref{vanish-dai}, some terms of the above functional relations 
\eqref{bac1}-\eqref{bac8} vanish for large $s,a $. 
As remarked in \cite{KSZ07}, this type of functional relations contain 
$\Ts\Qs$ relations as special cases. 
Due to \eqref{t-sys-bc2} and \eqref{t-sys-bc4}, the functional relations 
\eqref{bac1}, \eqref{bac3}, \eqref{bac5} and \eqref{bac7} 
reduce to $\Ts\Qs$ relations for $a=0$ and $s \in {\mathbb Z}_{\ge 1}$. 
Due to \eqref{t-sys-bc1} and \eqref{t-sys-bc3}, the functional relations 
 \eqref{bac2}, \eqref{bac4}, 
\eqref{bac6} and \eqref{bac8} also reduce to $\Ts\Qs$ relations 
for $s=0$ and $a \in {\mathbb Z}_{\ge 1}$. 
For example, \eqref{bac1} reduces to
\begin{multline}
\Ts^{(1),B_{m},F_{n}}_{s}(x)
\Qs_{B_{m-1} \sqcup F_{n}}(xq^{-\frac{m-n}{2}-s+2})-
\Qs_{B_{m} \sqcup F_{n}}(xq^{-\frac{m-n}{2}-s+1})\Ts^{(1),B_{m-1},F_{n}}_{s}(xq^{\frac{1}{2}})
\\[2pt]
=
z_{b_{m}}\Ts^{(1),B_{m},F_{n}}_{s-1}(xq)
\Qs_{B_{m-1} \sqcup F_{n}}(xq^{-s-\frac{m-n}{2}}) 
\qquad  \text{for} \quad s \in {\mathbb Z}_{\ge 1}. \label{tqeq}
\end{multline}
Considering the action of 
$S({\mathfrak B}) \times S({\mathfrak F})$, 
one sees that the above equation \eqref{tqeq} essentially corresponds to 
eq.\ (4.6) in \cite{KSZ07}.  
These are difference equations on the Baxter $\Qs$-functions 
$\{\Qs_{I}(x)\}$ or $\{\overline{\Qs}_{I}(x)\}$ for {\em different} ``levels" ${\rm Card}(I)$. 
In section 5, we will mention different type of $\Ts\Qs$ relations 
(we will call them ``Baxter equations''):  
difference equations on the Baxter $\Qs$-functions for the {\em same} level. 

Let us mention relation among the formulae in the previous section and the 
ones in this section. 
Our claim is: 
\begin{conjecture} 
We conjecture that the following relations 
 on \eqref{9-thvari-1}, \eqref{9-thvari-2}, \eqref{TF-rel03} and \eqref{TF-rel04} 
for the (non-skew) Young diagram $\mu$ hold: 
\begin{align}
 \Ts_{\mu}^{B_{m},F_{n}}(x)
 =
{\mathsf  F}_{\mu}^{B_{m}\times F_{n}}(x) 
\label{TF-rel1}
\end{align}
under the normalization \eqref{tri1} and the functional relations
 \eqref{QQ-rel1}-\eqref{QQ-rel2}, and 
\begin{align}
 {\overline \Ts}_{\mu}^{B_{m},F_{n}}(x)
 =
\overline{\mathsf  F}_{\mu}^{B_{m}\times F_{n}}(x)
\label{TF-rel2}
\end{align}
under the normalization \eqref{tri2} and the functional relations
 \eqref{QQ-rel3}-\eqref{QQ-rel4}. 
\end{conjecture}
For an empty Young diagram $\mu = \emptyset $, 
\eqref{TF-rel1} and \eqref{TF-rel2} reduce to \eqref{q-func1} and 
\eqref{q-func2}, respectively. 
We have proved the above relations for $\mu=(1)$ based on 
direct computations similar
\footnote{But, in our case, we need numerous case study 
(similar to the ones in Appendix \ref{proof-th}) depending on 
the values of $m,n$.}
 to the ones for $U_{q}(\widehat{gl}(M))$ case 
in section 7 of \cite{Suzuki01}. 
As for the rectangular Young diagram $\mu=(s^{a})$
 ($a,s \in {\mathbb Z}_{\ge 1}$), 
the above relations follow from the fact that  
 $\Ts_{s}^{(a),B_{m},F_{n}}(x)$ and $\mathsf F_{s}^{(a),B_{m}\times F_{n}}(x)$ 
(resp.\ $\overline{\Ts}_{s}^{(a),B_{m},F_{n}}(x)$ and
 $\overline{\mathsf F}_{s}^{(a),B_{m} \times F_{n}}(x)$) satisfy 
the same the functional relations with the same boundary conditions. 
As for $U_{q}(\widehat{gl}(M))$ case, the relation \eqref{TF-rel1} 
 follows from Theorem 3.2 in \cite{NNSY00}. 
The relations \eqref{TF-rel1}-\eqref{TF-rel2} 
for general case are conjectures based on numerous 
check by Mathematica. And the proof is under investigation.  

We have other expressions of the Wronskian-like 
formulae \eqref{9-thvari-1}, and \eqref{9-thvari-2}. 
Let us introduce the following functions: 
\begin{align}
& \hspace{-15pt} {\mathsf t}_{\mu}^{(b_{1},b_{2},\dots,b_{m}), (f_{1},f_{2},\dots,f_{n})}(x)
:= 
 \prod_{i=1}^{\xi_{m,n}-1}z_{b_{i}}^{\mu_{i}+m-n-i}
\Qs_{b_{i}}(xq^{2\mu_{i}-2i+1+\frac{m-n}{2}+\mu_{1}^{\prime}-\mu_{1}})
\nonumber \\
&\times \prod_{i=1}^{l_{m,n}-1}(-z_{f_{i}})^{\mu_{i}^{\prime}+n-m-i}
\Qs_{f_{i}}(xq^{-2\mu_{i}^{\prime}+2i-1+\frac{m-n}{2}+\mu_{1}^{\prime}-\mu_{1}})
\nonumber \\ 
&\times \prod_{i=\xi_{m,n}}^{m}
\frac{(-z_{f_{l_{m,n}+i-\xi_{m,n}}})^{r_{m,n}}}
{z_{b_{i}}^{r_{m,n}}(z_{b_{i}}-z_{f_{l_{m,n}+i-\xi_{m,n}}})}
\Qs_{\{b_{i},f_{l_{m,n}+i-\xi_{m,n}}\}}(xq^{-2(r_{m,n}+m-n)+\frac{m-n}{2}+\mu_{1}^{\prime}-\mu_{1}})
 \nonumber \\ 
& \times \prod_{i=l_{m,n}+m+1-\xi_{m,n}}^{n}
(-z_{f_{i}})^{n-i}
\Qs_{f_{i}}(xq^{-2(m-i)-1+\frac{m-n}{2}+\mu_{1}^{\prime}-\mu_{1}}) 
\label{infiniteT1} 
\end{align}
and
\begin{align}
& \hspace{-15pt} 
\overline{\mathsf t}_{\mu}^{(b_{1},b_{2},\dots,b_{m}), (f_{1},f_{2},\dots,f_{n})}(x)
:=
 \prod_{i=1}^{\xi_{m,n}-1}z_{b_{i}}^{\mu_{i}+m-n-i}
\overline{\Qs}_{b_{i}}(xq^{-2\mu_{i}+2i-1-\frac{m-n}{2}-\mu_{1}^{\prime}+\mu_{1}})
 \nonumber \\
& \times  \prod_{i=1}^{l_{m,n}-1}(-z_{f_{i}})^{\mu_{i}^{\prime}+n-m-i}
\overline{\Qs}_{f_{i}}(xq^{2\mu_{i}^{\prime}-2i+1-\frac{m-n}{2}-\mu_{1}^{\prime}+\mu_{1}})
 \nonumber \\
& 
\times \prod_{i=\xi_{m,n}}^{m}
\frac{(-z_{f_{l_{m,n}+i-\xi_{m,n}}})^{r_{m,n}}}
{z_{b_{i}}^{r_{m,n}}(z_{b_{i}}-z_{f_{l_{m,n}+i-\xi_{m,n}}})}
\overline{\Qs}_{\{ b_{i},f_{l_{m,n}+i-\xi_{m,n}}\}}
(xq^{2(r_{m,n}+m-n)-\frac{m-n}{2}-\mu_{1}^{\prime}+\mu_{1}})
 \nonumber \\
& 
\times \prod_{i=l_{m,n}+m+1-\xi_{m,n}}^{n}
(-z_{f_{i}})^{n-i}
\overline{\Qs}_{f_{i}}(xq^{2(m-i)+1-\frac{m-n}{2}-\mu_{1}^{\prime}+\mu_{1}}), 
\label{infiniteT2}
\end{align}
where $l_{m,n}=\mu_{\xi_{m,n}}+1$, $r_{m,n}=n-m+\xi_{m,n}(\mu)-l_{m,n}$. 
Then we find 
\begin{proposition}\label{infi-pro} 
The determinant formula \eqref{9-thvari-1} can be expressed as 
\begin{align}
\Ts_{\mu}^{(b_{1},b_{2},\dots,b_{m}),(f_{1},f_{2},\dots,f_{n})}(x)
 &=
 \frac{\ds 
\sum_{w \in S(B_{m}) \times S(F_{n})}
{\rm sgn}(w)
w
 \left[
{\mathsf t}_{\mu}^{(b_{1},b_{2},\dots,b_{m}), (f_{1},f_{2},\dots,f_{n})}(x)
 \right]
 }
{(m-\xi_{m,n}(\mu)+1)!
\Ds\bigl(
\begin{smallmatrix}
b_{1}, & b_{2}, & \dots, & b_{m} \\
f_{1}, & f_{2},& \dots,& f_{n}
\end{smallmatrix}
\bigr)
} 
\label{weyl1}
\\[8pt]
 &=
\frac{\ds \sum_{w \in S(B_{m}) \times S(F_{n})/H}
{\rm sgn}(w)w
 \left[
{\mathsf t}_{\mu}^{(b_{1},b_{2},\dots,b_{m}), (f_{1},f_{2},\dots,f_{n})}(x)
 \right]}
{\Ds\bigl(
\begin{smallmatrix}
b_{1}, & b_{2}, & \dots, & b_{m} \\
f_{1}, & f_{2},& \dots,& f_{n}
\end{smallmatrix}
\bigr)}
\end{align}
under the functional relation \eqref{QQ-bf1}; 
and the determinant formula \eqref{9-thvari-2} can be expressed as 
\begin{align}
 {\overline \Ts}_{\mu}^{(b_{1},b_{2},\dots,b_{m}),(f_{1},f_{2},\dots,f_{n})}(x)
 &=
 \frac{\ds 
\sum_{w \in S(B_{m}) \times S(F_{n})}
{\rm sgn}(w)
w
 \left[
\overline{\mathsf t}_{\mu}^{(b_{1},b_{2},\dots,b_{m}),( f_{1},f_{2},\dots,f_{n})}(x)
 \right]
 }
{(m-\xi_{m,n}(\mu)+1)!
\Ds\bigl(
\begin{smallmatrix}
b_{1}, & b_{2}, & \dots, & b_{m} \\
f_{1}, & f_{2},& \dots,& f_{n}
\end{smallmatrix}
\bigr)
} 
\label{weyl2}
\\[8pt]
 &=
\frac{ \ds \sum_{w \in S(B_{m}) \times S(F_{n})/H}
{\rm sgn}(w)w
 \left[
\overline{\mathsf t}_{\mu}^{(b_{1},b_{2},\dots,b_{m}), (f_{1},f_{2},\dots,f_{n})}(x)
 \right]}
{\Ds\bigl(
\begin{smallmatrix}
b_{1}, & b_{2}, & \dots, & b_{m} \\
f_{1}, & f_{2},& \dots,& f_{n}
\end{smallmatrix}
\bigr)}
\end{align}
under the functional relation \eqref{QQ-bf2}, 
where  ${\rm sgn}(w)$ is the signature of $w$;  
 $H$ is a subgroup of $S(B_{m}) \times S(F_{n})$ whose 
elements $w=\sigma_{b}\times \sigma_{f}$ are such that
 $\sigma_{b}$ is a permutation of 
$(b_{\xi_{m,n}},b_{\xi_{m,n}+1},\dots, b_{m })$ and $\sigma_{f}$ is the same  
permutation of $(f_{l_{m,n}},f_{l_{m,n}+1},\dots,f_{l_{m,n}+m-\xi_{m,n}})$. 
\end{proposition}
Note that ${\mathsf t}_{\mu}^{B_{m}, F_{n}}(x)$ is invariant under the action of $H$. 
For $(m,n)=(M,N)$, \eqref{weyl1} and \eqref{weyl2} are 
invariant under the action of the Weyl group $S(B_{M})\times S(F_{N})$ of $gl(M|N)$.
Proofs of the above formulae are similar to $x=0$ case 
(see Theorem 3.4 in \cite{MV03}), where the following relation will be used:
\begin{align}
\frac{z_{b}^r\Qs_{\{b,f\}}(x)}{z_{b}-z_{f}}
=\frac{z_{f}^r\Qs_{\{b,f\}}(xq^{-2r})}{z_{b}-z_{f}}
+\sum_{s=0}^{r-1}z_{b}^{s}z_{f}^{r-s-1}
\Qs_{b}(xq^{-2r+2s+1})\Qs_{f}(xq^{-2r+2s+1}),
\end{align}
where $b \in {\mathfrak B}, f \in {\mathfrak F}$. 
This relation follows from \eqref{QQ-bf1}. 
There is a similar relation among 
$\overline{\Qs}_{\{b,f\}}(x),\overline{\Qs}_{b}(x),\overline{\Qs}_{f}(x)$.
The finite dimensional modules can be written 
in terms of a direct sum of infinite dimensional highest weight modules 
due to the Bernstein-Gel'fand-Gel'fand (BGG) resolution. 
In this respect, 
the function (\ref{infiniteT1}) and (\ref{infiniteT2}) divided by (\ref{deno}) 
for $(m,n)=(M,N)$ may be interpreted
\footnote{A merit of this type of formula is that it contains only one term, 
and thus is easy to evaluate. 
The traditional form of the eigenvalue formula by the Bethe ansatz 
(such as the one in section 2) contains  
an infinite number of terms if the auxiliary space is an infinite dimensional space, 
and thus is not always easy to evaluate. 
Some discussions on $\Ts$-functions for infinite dimensional representations 
(in relation to the AdS/CFT correspondence) can be 
seen, for example, in Appendix B of \cite{BKSZ05} and section 6.4 of \cite{Beisert07}.}
 as $\Ts$-functions for infinite dimensional 
highest weight representations of $U_{q}(gl(M|N))$ (compare the above formula with 
eq.\ (5.21) in \cite{BHK02} for $U_{q}(gl(3))$, and also formulae in \cite{BLZ97,BDKM06,BT08,DM08}). 
Of course, this needs further research in view of the fact that the 
BGG-type resolution of representations of superalgebras is more involved than 
that of bosonic algebras 
(cf.\ \cite{CKL08}). 

Let us introduce linear operators, which act on the variables $\{z_{1},z_{2},\dots, z_{M+N} \}$ as
\begin{align}
{\mathbf B}_{a}(x) \cdot 
z_{1}^{k_{1}}z_{2}^{k_{2}} \cdots z_{M+N}^{k_{M+N}}&=
z_{1}^{k_{1}}z_{2}^{k_{2}} \cdots z_{M+N}^{k_{M+N}}
 \Qs_{a}(xq^{(2k_{a}+1)p_{a}}), 
 \\[6pt] 
\begin{split}
\overline{\mathbf B}_{a}(x) \cdot 
z_{1}^{k_{1}}z_{2}^{k_{2}} \cdots z_{M+N}^{k_{M+N}}&=
z_{1}^{k_{1}}z_{2}^{k_{2}} \cdots z_{M+N}^{k_{M+N}}
 \overline{\Qs}_{a}(xq^{-(2k_{a}+1)p_{a}}) \\[4pt]
& \qquad \text{for} \quad a \in \{1,2,\dots M+N \}, 
\end{split}
\end{align}
where one should write any term like 
$\frac{1}{z_{b}-z_{f}}$ as $\sum_{k=0}^{\infty} z_{b}^{-k-1}z_{f}^{k}$ before 
the actions of these operators. For example, for $b \in {\mathfrak B}$, $f \in {\mathfrak F}$ and 
$a \in {\mathfrak I}$ ($a \ne b,f$), we have 
\begin{align}
{\mathbf B}_{a}(x){\mathbf B}_{b}(x){\mathbf B}_{f}(x) \cdot \frac{1}{z_{b}-z_{f}}
&= \sum_{k=0}^{\infty}
{\mathbf B}_{a}(x){\mathbf B}_{b}(x){\mathbf B}_{f}(x) \cdot 
z_{b}^{-k-1}z_{f}^{k} 
\nonumber \\
&= \sum_{k=0}^{\infty} z_{b}^{-k-1}z_{f}^{k}
 \Qs_{a}(xq^{p_{a}})\Qs_{b}(xq^{-2k-1})\Qs_{f}(xq^{-2k-1})
\nonumber \\
&= \frac{1}{z_{b}-z_{f}} \Qs_{a}(xq^{p_{a}})\Qs_{\{b,f\}}(x),
\end{align}
where we used the relation (\ref{expan-aij}).
Then we find 
\begin{align}
& \prod_{a \in B_{m} \sqcup F_{n}}
{\mathbf B}_{a}(xq^{-\frac{3(m-n)}{2}+\mu_{1}^{\prime}-\mu_{1}})
\cdot 
{\mathsf t}_{\mu}^{(b_{1},b_{2},\dots,b_{m}), (f_{1},f_{2},\dots,f_{n})}(0)
=
{\mathsf t}_{\mu}^{(b_{1},b_{2},\dots,b_{m}), (f_{1},f_{2},\dots,f_{n})}(x),
\\
& \prod_{a \in B_{m} \sqcup F_{n}}
\overline{\mathbf B}_{a}(xq^{\frac{3(m-n)}{2}-\mu_{1}^{\prime}+\mu_{1}})
\cdot 
\overline{\mathsf t}_{\mu}^{(b_{1},b_{2},\dots,b_{m}), (f_{1},f_{2},\dots,f_{n})}(0)
=
\overline{\mathsf t}_{\mu}^{(b_{1},b_{2},\dots,b_{m}), (f_{1},f_{2},\dots,f_{n})}(x).
\end{align}
Thus our formulae in this paper are the Yang-Baxterization of the 
(super)characters of representations of the subalgebras 
$gl(m|n)$ of $gl(M|N)$ by the operators
 $\{{\mathbf B}_{a}(x) \}$ and $\{\overline{\mathbf B}_{a}(x) \}$. 
The $q$-characters, which correspond to the $\Ts$-functions, 
 appear in the kernel of the Frenkel-Reshetikhin screening operators \cite{FR99}. 
The transfer matrices for $Y(gl(M|N))$ were also written in terms of a certain 
``group derivative'' \cite{KV07} on the (super)characters of representations of $gl(M|N)$. 
It will be interesting to investigate
\footnote{One thing we should do may be to apply the ``group derivative'' in \cite{KV07} 
to the formulae for $x=0$ in this paper.}
 connection among 
our operators and the operators in \cite{FR99,KV07}. 
\section{$\Ts$-functions for typical representations}\label{T-fun-sec}
The type 1 quantum superalgebra $U_{q}(gl(M|N))$ (or $U_{q}(sl(M|N))$) admits a one-parameter
family of finite-dimensional irreducible representations, 
which correspond to typical representations (see Appendix \ref{representation}).  
Thus the evaluation representations of $U_{q}(\widehat{gl}(M|N))$ (or $U_{q}(\widehat{sl}(M|N))$) 
based on the above mentioned representations depend not only on the spectral parameter 
but also on another continuous parameter. 
There are $R$-matrices for this family of representations \cite{DGLZ95}. 
In \cite{T98-2}, we proposed $\Ts$-functions 
for a wide class of such representations from the analytic Bethe ansatz 
(see also, \cite{Maassarani95,RM96,KSZ07}). 
Here we briefly comment on how to obtain the $\Ts$-functions for such representations 
in the auxiliary space from our Wronskian-like formulae. 

The determinant formula \eqref{9-thvari-2} factorizes
 if $\xi_{m,n}=m+1$ ($\mu_{m+1} \le n \le \mu_{m}$): 
\begin{align}
\overline{\Ts}^{B_{m},F_{n}}_{\mu}(x)
=\prod_{\alpha \in B_{m}}\prod_{\beta \in F_{n}}(z_{\alpha}-z_{\beta})
\overline{\Ts}^{B_{m},\emptyset}_{\tau}(xq^{-\frac{n}{2}+\mu_{n+1}^{\prime}-\mu_{1}^{\prime}})
\overline{\Ts}^{\emptyset,F_{n}}_{\eta}(xq^{\frac{m}{2}-\mu_{m+1}+\mu_{1}}), 
\label{typical-BF}
\end{align}
where $\tau :=(\tau_{1},\tau_{2},\dots,\tau_{m})=
(\mu_{1}-n,\mu_{2}-n,\dots, \mu_{m}-n)$ and 
$\eta :=(\eta_{1},\eta_{2},\dots,\eta_{\eta_{1}^{\prime}})=
(\mu_{m+1},\mu_{m+2},\dots, \mu_{\mu_{1}^{\prime }})$. 
Let us consider a generalization of the above relation. 
Due to the relations \eqref{del-shift1} and \eqref{del-shift2}, 
 the following relations hold for $c_{1},c_{2} \in {\mathbb Z}_{\ge 0}$: 
\begin{align}
& \overline{\Ts}^{B_{m},\emptyset}_{\tau_{c_{1}}}(x)
=\left(\prod_{\alpha \in B_{m}}z_{\alpha}\right)^{c_{1}}
\overline{\Ts}^{B_{m},\emptyset}_{\tau}
(xq^{-c_{1}+(1-\delta_{c_{1},0})(\tau_{1}^{\prime}-m)}), \label{typical-Bm} \\[7pt]
& \overline{\Ts}^{\emptyset,F_{n}}_{\eta_{c_{2}}}(x)
=\left(\prod_{\beta \in F_{n}}(-z_{\beta })\right)^{c_{2}}
\overline{\Ts}^{\emptyset,F_{n}}_{\eta}(xq^{c_{2}-(1-\delta_{c_{2},0})(\eta_{1}-n)}),  \label{typical-Fn}
\end{align}
where $\tau_{c_{1}}:=(\tau_{1}+c_{1},\tau_{2}+c_{1},\dots,\tau_{m}+c_{1})$ and 
$\eta_{c_{2}}:=(\underbrace{n,n,\dots,n }_{c_{2}},\eta_{1},\eta_{2},\dots,\eta_{\eta_{1}^{\prime}})$. 
Let us assume $m,n \in {\mathbb Z}_{\ge 1}$. 
Combining (\ref{typical-BF}), (\ref{typical-Bm}) and (\ref{typical-Fn}), we obtain 
\begin{multline}
\overline{\Ts}^{B_{m},F_{n}}_{\mu_{c_{1},c_{2}}}(x)
=
\left(\prod_{\alpha \in B_{m}}z_{\alpha}\right)^{c_{1}}
\left(\prod_{\beta \in F_{n}}(-z_{\beta })\right)^{c_{2}}
\prod_{\alpha \in B_{m}}\prod_{\beta \in F_{n}}(z_{\alpha}-z_{\beta})
 \\[2pt]
 \times 
\overline{\Ts}^{B_{m},\emptyset}_{\tau}(xq^{-\frac{n}{2}+\mu_{n+1}^{\prime}
-\mu_{1}^{\prime}-c_{1}-c_{2}})
\overline{\Ts}^{\emptyset,F_{n}}_{\eta}(xq^{\frac{m}{2}-\mu_{m+1}+\mu_{1}+c_{1}+c_{2}}), 
\label{typical-BF2}
\end{multline}
where $\mu_{c_{1},c_{2}}:=(\mu_{1}+c_{1},\mu_{2}+c_{1},\dots,\mu_{m}+c_{1},
\underbrace{n,n,\dots,n}_{c_{2}},\mu_{m+1},\mu_{m+2},\dots, \mu_{\mu_{1}^{\prime}})$. 
Similarly we obtain
\begin{multline}
\Ts^{B_{m},F_{n}}_{\mu_{c_{1},c_{2}}}(x)
=
\left(\prod_{\alpha \in B_{m}}z_{\alpha}\right)^{c_{1}}
\left(\prod_{\beta \in F_{n}}(-z_{\beta })\right)^{c_{2}}
\prod_{\alpha \in B_{m}}\prod_{\beta \in F_{n}}(z_{\alpha}-z_{\beta})
 \\[2pt]
 \times 
\Ts^{B_{m},\emptyset}_{\tau}(xq^{\frac{n}{2}-\mu_{n+1}^{\prime}
+\mu_{1}^{\prime}+c_{1}+c_{2}})
\Ts^{\emptyset,F_{n}}_{\eta}(xq^{-\frac{m}{2}+\mu_{m+1}-\mu_{1}-c_{1}-c_{2}}).
\label{typical-BF22}
\end{multline}
In particular for $\mu=(n^{m})$ ($\tau =\eta =\emptyset$) case, these relations reduce to 
\begin{multline}
\Ts^{B_{m},F_{n}}_{((n+c_{1})^{m},n^{c_{2}})}(x)
=
\left(\prod_{\alpha \in B_{m}}z_{\alpha}\right)^{c_{1}}
\left(\prod_{\beta \in F_{n}}(-z_{\beta })\right)^{c_{2}}
\prod_{\alpha \in B_{m}}\prod_{\beta \in F_{n}}(z_{\alpha}-z_{\beta})
 \\[2pt]
 \times 
\Qs_{B_{m}}(xq^{\frac{m+n}{2}+c_{1}+c_{2}})
\Qs_{F_{n}}(xq^{-\frac{m+n}{2}-c_{1}-c_{2}}), 
\label{typical-rek1} 
\end{multline}
\begin{multline}
\overline{\Ts}^{B_{m},F_{n}}_{((n+c_{1})^{m},n^{c_{2}})}(x)
=
\left(\prod_{\alpha \in B_{m}}z_{\alpha}\right)^{c_{1}}
\left(\prod_{\beta \in F_{n}}(-z_{\beta })\right)^{c_{2}}
\prod_{\alpha \in B_{m}}\prod_{\beta \in F_{n}}(z_{\alpha}-z_{\beta})
 \\[2pt]
 \times 
\overline{\Qs}_{B_{m}}(xq^{-\frac{m+n}{2}-c_{1}-c_{2}})
\overline{\Qs}_{F_{n}}(xq^{\frac{m+n}{2}+c_{1}+c_{2}}).
\label{typical-rek2} 
\end{multline}
There relations (\ref{typical-rek1}) and (\ref{typical-rek2}) 
are generalization of the relations (\ref{typfac1})-(\ref{typfac4}). 
The right hand side of the above relations (\ref{typical-BF2}), 
(\ref{typical-BF22}), (\ref{typical-rek1}) and (\ref{typical-rek2}) make sense as a function of 
$x$, $c_{1}$ and $c_{2}$ even when $c_{1}$ or $c_{2}$ are not integers. 
And for $(m,n)=(M,N)$, these should be interpreted as a $\Ts$-functions for the 
typical representations (see Appendix \ref{representation}). 

Next we shall mention relation between (\ref{typical-BF2}) and our previous results \cite{T98-2}.
Combining (\ref{typical-BF}) for $\mu=\check{\mu}=(\mu_{1},\mu_{2},\dots,\mu_{m})$ 
and (\ref{typical-BF2}), we obtain
\begin{multline}
\overline{\Ts}^{B_{m},F_{n}}_{\mu_{c_{1},c_{2}}}(x)
=
\left(\prod_{\alpha \in B_{m}}z_{\alpha}\right)^{c_{1}}
\left(\prod_{\beta \in F_{n}}(-z_{\beta })\right)^{c_{2}} 
 \\
 \times 
\frac{
\overline{\Ts}^{B_{m},F_{n}}_{\check{\mu}}
(xq^{m-\mu_{1}^{\prime}-c_{1}-c_{2}})
\overline{\Ts}^{\emptyset,F_{n}}_{\eta}(xq^{\frac{m}{2}+\mu_{1}-\mu_{m+1}+c_{1}+c_{2}})
}
{\overline{\Qs}_{F_{n}}(xq^{\frac{3m}{2}-\frac{n}{2}+\mu_{1}-\mu_{1}^{\prime}-c_{1}-c_{2}})}
\label{typical-BF3},
\end{multline}
where we used the relation 
$\overline{\Ts}_{\emptyset}^{\emptyset, F_{n}}(x)
=\overline{\Qs}_{F_{n}}(xq^{-\frac{n}{2}})$. 
One can rewrite the above relation via the relations 
\eqref{TF-rel04} and \eqref{TF-rel2} as 
\begin{multline}
\overline{{\mathcal F}}^{B_{m}\times F_{n}}_{\mu_{c_{1},c_{2}}}(x)
=
\left(\prod_{\alpha \in B_{m}}z_{\alpha}\right)^{c_{1}}
\left(\prod_{\beta \in F_{n}}(-z_{\beta })\right)^{c_{2}}
\frac{
\overline{\Qs}_{B_{m}\sqcup F_{n}}(xq^{\frac{m-n}{2}+\mu_{1}-\mu_{1}^{\prime}-c_{1}-c_{2}})
}
{\overline{\Qs}_{B_{m}\sqcup F_{n}}(xq^{\frac{m-n}{2}+\mu_{1}-\mu_{1}^{\prime}+c_{1}-c_{2}})
}
 \\[3pt]
 \times 
\frac{
\overline{\Qs}_{F_{n}}(xq^{\frac{3m}{2}-\frac{n}{2}+\mu_{1}-\mu_{1}^{\prime}+c_{1}+c_{2}})
}
{
\overline{\Qs}_{F_{n}}(xq^{\frac{3m}{2}-\frac{n}{2}+\mu_{1}-\mu_{1}^{\prime}-c_{1}-c_{2}})
}
\overline{{\mathcal F}}^{B_{m}\times F_{n}}_{\check{\mu}}
(xq^{m-\mu_{1}^{\prime}-c_{1}-c_{2}})
\overline{{\mathcal F}}^{F_{n}}_{\eta}(xq^{\frac{m}{2}+\mu_{1}-\mu_{m+1}+c_{1}+c_{2}}).
\label{typical-BF4}
\end{multline}
We find that (\ref{typical-BF4}) for $(m,n)=(M,N)$ and $c_{2}=0$ 
essentially corresponds to 
the $\Ts$-function for the typical representation written in eq. (3.23) in \cite{T98-2}.
\section{Baxter equations}
The Baxter equations are difference equations whose solutions give 
the Baxter $\Qs$-functions. The Baxter equations in question 
are the equations for the $\Qs$-functions $\{\Qs_{I}(x)\},\{\overline{\Qs}_{I}(x)\}$ 
of the same ``level'' ${\rm Card}(I)=1$.
Discussions on the Baxter equations 
for superalgebra related models can also be seen in \cite{BDKM06,B07-1,B07-2}. 

Once Wronskian-type determinant formulae are established, 
the Baxter equations can be derived easily. 
In essence, the following Baxter equations are based on a basic 
theorem on the linear algebra: if two different lows or columns of a matrix coincide, 
the corresponding determinant vanishes. 
Thus one can derive the Baxter equations 
directly from the Wronskian-like formulae
 (\ref{9-thvari-1})-(\ref{9-thvari-2}) 
related
\footnote{if $(m,n)=(M,N)$}
 to typical representations as follows:
\begin{align}
&
\sum_{a=0}^{m}(-z_{k})^{-a}
\Ts_{((n+1)^{a},n^{m-a})}^{B_{m},F_{n}}
(xq^{-(a-m)\delta_{n,0}-\delta_{a,0}})\Qs_{k}(xq^{-2a+\frac{3m}{2}+\frac{n}{2}})=0 
\quad \text{for} \quad k \in B_{m}, \label{baxtereq1} \\[6pt]
&
\sum_{a=0}^{n}z_{k}^{-a}
\Ts_{(n^{m},a)}^{B_{m},F_{n}}(xq^{(a-n)\delta_{m,0}+\delta_{a,0}})
\Qs_{k}(xq^{2a-\frac{m}{2}-\frac{3n}{2}})=0 
\quad \text{for} \quad k \in F_{n} \label{baxtereq2},
\end{align}
 and 
\begin{align}
& 
\sum_{a=0}^{m}(-z_{k})^{-a}
\overline{\Ts}_{((n+1)^{a},n^{m-a})}^{B_{m},F_{n}}
(xq^{(a-m)\delta_{n,0}+\delta_{a,0}})\overline{\Qs}_{k}(xq^{2a-\frac{3m}{2}-\frac{n}{2}})=0 
\quad \text{for} \quad k \in B_{m}, \label{baxtereq3} \\[6pt]
&
\sum_{a=0}^{n}z_{k}^{-a}
\overline{\Ts}_{(n^{m},a)}^{B_{m},F_{n}}(xq^{-(a-n)\delta_{m,0}-\delta_{a,0}})
\overline{\Qs}_{k}(xq^{-2a+\frac{m}{2}+\frac{3n}{2}})=0 
\quad \text{for} \quad k \in F_{n} \label{baxtereq4},
\end{align}
where $((n+1)^{a},n^{m-a})=((n+1)^{m})$ at $a=m$, and $(n^{m})$ at $a=0$; 
$(n^{m},a)=(a)$ at $m=0$; 
$(n^{m})=\emptyset$ at $n=0$ or $m=0$.
These are finite order linear difference equations on 
the Baxter $\Qs$-functions. 
Note that 
\eqref{baxtereq1} and \eqref{baxtereq2} (resp.\ \eqref{baxtereq3} and \eqref{baxtereq4})
 share the 
functions $\Ts_{(n^{m})}^{B_{m},F_{n}}(x)$ and 
$\Ts_{((n+1)^{m})}^{B_{m},F_{n}}(x)=
\left\{\frac{\prod_{\gamma \in B_{m}}z_{\gamma }}{\prod_{\gamma \in F_{n}}(-z_{\gamma })}
\right\}
\Ts_{(n^{m+1})}^{B_{m},F_{n}}(x)$
(resp.\ $\overline{\Ts}_{(n^{m})}^{B_{m},F_{n}}(x)$ and 
$\overline{\Ts}_{((n+1)^{m})}^{B_{m},F_{n}}(x)=
\left\{\frac{\prod_{\gamma \in B_{m}}z_{\gamma }}{\prod_{\gamma \in F_{n}}(-z_{\gamma })}
\right\}
\overline{\Ts}_{(n^{m+1})}^{B_{m},F_{n}}(x)$). 
The functions $\Ts_{(n^{m})}^{B_{m},F_{n}}(x), \Ts_{(n^{m+1})}^{B_{m}, F_{n}}(x)$ 
and $\Ts_{((n+1)^{m})}^{B_{m},F_{n}}(x)$ can also be written 
in terms of $\Ts^{(a),B_{m},F_{n}}_{1}(x)$ 
 or $\Ts^{(1),B_{m},F_{n}}_{s}(x)$ through (\ref{superJT1}), (\ref{superJT2}), 
(\ref{TF-rel1}). 
The equations \eqref{baxtereq1}-\eqref{baxtereq4} for $(m,n)=(M,N)$ are 
the Baxter equations for $U_{q}(\widehat{gl}(M|N))$. 
These equations for $m<M, n<N$ are in the intermediate steps of the B\"{a}cklund flows. 
In addition, $N=0$ and $(m,n)=(M,0)$ 
case of (\ref{baxtereq1}) and (\ref{baxtereq3}) correspond to the Baxter equations 
for $U_{q}(\widehat{gl}(M))$ (cf.\ \cite{KLWZ97,Kojima08}). 

One can also factor out overall factors in  \eqref{baxtereq1}-\eqref{baxtereq4}
 based on the formula \eqref{typical-BF} (and the same type of formula for 
$\Ts^{B_{m}, F_{n}}_{\mu }(x)$) as follows:  
\begin{align}
&
\sum_{a=0}^{m}(-z_{k})^{-a}
\Ts_{(1^{a})}^{B_{m},\emptyset}
(xq^{-a-\delta_{a,0}})\Qs_{k}(xq^{-2a+\frac{m}{2}})=0 
\quad \text{for} \quad k \in B_{m}, \label{baxtereq1r} \\[6pt]
&
\sum_{a=0}^{n}z_{k}^{-a}
\Ts_{(a)}^{\emptyset,F_{n}}(xq^{a+\delta_{a,0}})
\Qs_{k}(xq^{2a-\frac{n}{2}})=0 
\quad \text{for} \quad k \in F_{n} \label{baxtereq2r},
\end{align}
 and 
\begin{align}
&
\sum_{a=0}^{m}(-z_{k})^{-a}
\overline{\Ts}_{(1^{a})}^{B_{m},\emptyset}
(xq^{a+\delta_{a,0}})\overline{\Qs}_{k}(xq^{2a-\frac{m}{2}})=0 
\quad \text{for} \quad k \in B_{m}, \label{baxtereq1r} \\[6pt]
&
\sum_{a=0}^{n}z_{k}^{-a}
\overline{\Ts}_{(a)}^{\emptyset,F_{n}}(xq^{-a-\delta_{a,0}})
\overline{\Qs}_{k}(xq^{-2a+\frac{n}{2}})=0 
\quad \text{for} \quad k \in F_{n} \label{baxtereq2r}.
\end{align}
These are also kinds of Baxter equations, but in the intermediate steps of the 
B\"{a}cklund flows except for the case $N=0$ and $(m,n)=(M,0)$, or $M=0$ and $(m,n)=(0,N)$. 

There is another type of Baxter equations, which follows from 
 the kernel of the non-commutative generating series (\ref{gene1})-(\ref{gene4}). 
In contrast to the above equations, this type of Baxter equations 
has an infinite number of terms.  
This is because that the non-commutative generating series of 
the $\Ts$-functions  
are infinite order difference operators. 
This situation is similar to other quantum affine (super)algebras cases such as 
 $U_{q}(B_{M}^{(1)})$, $U_{q}(D_{M}^{(1)})$ \cite{KOS95,KOSY}, 
 $U_{q}(D_{M+1}^{(2)})$, $U_{q}(D_{4}^{(3)})$ \cite{T02}, 
  $U_{q}(\widehat{osp}(M|2N))$ \cite{T99}
\footnote{$U_{q}(\widehat{osp}(1|2N))$ case is an exception, where 
the $\Ts$-function is similar to $U_{q}(A_{2N}^{(2)})$ case.},  
where the non-commutative generating series of the $\Ts$-functions are infinite order 
difference operators and 
Wronskian-like formulae for the $\Ts$-functions have not been established. 
In this paper, we have overcome this difficulty for $U_{q}(\widehat{gl}(M|N))$ case. 
Thus our example on $U_{q}(\widehat{gl}(M|N))$ will be a good reference to 
find \cite{workinprogress} 
finite order Baxter equations and Wronskian-like formulae for $\Ts$-functions for 
more general quantum affine (super)algebras. 

We have also derived a determinant form of finite order Baxter equations based on 
conserved quantities in Appendix \ref{Conserved}. 
\section{Discussions}
In this paper, we assume that the deformation parameter $q$ is generic. 
If $q$ is a root of unity, the Baxter 
$\Qs$-function becomes a periodic function of $x$. 
Thus, through our new Wronskian-like determinant formulae of the $\Ts$-functions, 
one sees that 
the quantum group truncation occurs in a very different way than for the bosonic algebra case. 
Namely, the $\Ts$-function $\Ts^{(a)B_{m}, F_{n}}_{s}(x)$ may vanish for some $s$ {\em or} $a$ 
for $m,n >0$ case if the twist parameters are also proportional to appropriate roots of unity. 
Note that truncation occurs only on the variable $s$ for $n=0$ case. 

We have proposed various formulae for the Baxter $\Qs$-functions and the $\Ts$-functions. 
It is important to realize these as operators. 
Of course, the same formulae hold true for both operators and their eigenvalues 
since the Baxter $\Qs$-operators and the $T$-operators belong to the same commuting 
family of operators. 
So far, realizations of the Baxter $\Qs$-operators based on the 
$q$-oscillator representations of the quantum affine (super)algebra, 
which are relevant to our formulation in this paper,   
are available for $U_{q}(\widehat{sl}(2))$ \cite{BLZ97}, 
$U_{q}(\widehat{sl}(3))$ \cite{BHK02}, 
$U_{q}(\widehat{sl}(M))$ \cite{Kojima08}, 
$U_{q}(\widehat{sl}(2|1))$ \cite{BT08}, and the general case 
$U_{q}(\widehat{gl}(M|N))$ (or $U_{q}(\widehat{sl}(M|N))$) 
is under investigation \cite{workinprogress}. 
Although the identities among $T$ and $\Qs$-functions in this paper are combinatorial ones and 
hold true {\em independent} of the representation theory, 
it is also important to establish connection to the universal $R$-matrix and to 
prove our new formulae from the point of 
view of the representation theory. 

The supersymmetric Bazhanov-Reshetikhin formulae 
(eqs. (\ref{superJT1})-(\ref{superJT2}) for $K=M+N$) \cite{T97,T98,T98-2} played an important 
role to derive \cite{Tsuboi06} (see also \cite{Tsuboi04}) 
a system of nonlinear integral equations (NLIE) for the 
free energy of integrable spin chains at finite temperature, which is equivalent to 
 the thermodynamic Bethe ansatz (TBA) equation
\footnote{TBA equations for $gl(M|N)$ related rational models were 
written in \cite{Saleur00}.}, from the 
$T$-system (\ref{ori-t-system1})-(\ref{ori-t-sys-bc1}) \cite{T97,T98,T98-2}. 
Then a natural question is whether the Wronskian-like determinant formulae
 are also useful to derive the NLIE. 
 There is another type of NLIE \cite{DD92}, 
which is equivalent to the TBA equation. 
To generalize this type of NLIE to models 
 whose underlying algebras have arbitrary rank is not always 
an easy problem as one needs considerable trial and errors 
to find auxiliary functions with good analytical properties which are needed to derive the NLIE.  
Although the $Y$-system itself is {\em not} truncated to a finite set of functional relations for 
{\em generic} $q$, an introduction of appropriate auxiliary functions modifies \cite{Suzuki99}
 the $Y$-system to a finite set of functional relations, which is suited for the NLIE. 
So far this type of NLIE is known for at most rank 4 case \cite{SDK08}. 
We expect that some of our formulae in this paper are also useful to 
construct auxiliary functions for this type of NLIE for arbitrary rank. 
One can easily transform the $T$-system into the $Y$-system based on 
a transformation on the $\Ts$-functions (to the ``$Y$-functions''), as in \cite{JKS98}. 
This type of transformation 
(applied for \eqref{bac1}-\eqref{bac8}) also induces a system of B\"{a}cklund transformations for 
the $Y$-system. Thereby B\"{a}cklund transformations for 
the TBA equations are obtained. Our formulae for the $\Ts$-functions 
$\Ts_{s}^{(a),B_{m},F_{n}}(x)$ and  $\overline{\Ts}_{s}^{(a),B_{m},F_{n}}(x)$  
(or $F_{s}^{(a),I}(x)$ and  $\overline{F}_{s}^{(a),I}(x)$  ) 
will be convenient 
 to check analyticity of the $Y$-functions for these B\"{a}cklund transformations. 

Our new formulae in this paper have potential applicability to
 the analysis of the AdS/CFT $T$-system (or the $Y$-system) \cite{GKV09,BFT09},
 after some modifications. 
 Note that many formulae in this paper do not depend on precise function 
form of the Baxter $\Qs$-functions. 
The modifications should be done so that the Baxter $\Qs$-functions 
are compatible with the Bethe ansatz equation in \cite{BS05} (for large $L$). 
We also expect that this is also the case with the Hubbard model. 

To generalize our formulae to 
mixed representations is also an important problem. 
So far, the supersymmetric Bazhanov-Reshetikhin type formula 
for a simplest case was written long time ago (cf.\ eq.\ (4.1) in \cite{T97}),
 but more general case 
 is under investigation \cite{workinprogress}. 
We expect that the (super)character formulae for the 
mixed representations of $gl(M|N)$ also allow 
the Yang-Baxterization by the Baxter $\Qs$-functions (or $\Qs$-operators). 

The representation theory of the quantum affine superalgebras 
 is not always well understood even in finite dimensional representations case. 
In this concern, to generalize our formulae in this paper to other quantum affine superalgebras 
 such as $U_{q}(\widehat{osp}(M|2N))$, 
$U_{q}(D^{(1)}(2,1;\alpha))$ etc. is interesting not only in the study of the integrable system 
 but also in the representation theory in mathematics. 
Tableaux sum expressions for $\Ts$-functions, some supersymmetric Bazhanov-Reshetikhin-like  
formulae for fusion models and 
 the $T$-system are available 
 for $U_{q}(\widehat{osp}(M|2N))$ \cite{T99} case, and this will be a starting point   
to establish \cite{workinprogress} Wronskian-like formulae for $\Ts$-functions or   
 $q$-(super)character formulae for the general quantum affine superalgebras. 
\section*{Acknowledgments}  
The author would like to thank Vladimir Bazhanov for collaboration in Ref.\ \cite{BT08} 
and a comment. 
He also thanks Yasuhiko Yamada for bringing Ref.\ \cite{Witten88} to his attention, 
and Vladimir Kazakov and Anton Zabrodin for discussions on mixed representations. 
He is supported by 
Grant-in-Aid for Young Scientists, B \#19740244 from 
The Ministry of Education, Culture, Sports, Science and Technology in Japan. 
The main results of this paper for $(m,n)=(M,N)$  
have been previously announced on a
number of conferences%
\footnote{
``Finite-size Technology in Low Dimensional Quantum System (IV)'',  
APCTP (Pohang, Korea), June 2008,
[http://wimn.ewha.ac.kr/focus2008/]; 
a meeting of the Physical Society of Japan, September 23, 2008.
}
%

\app{Representations of the superalgebra}\label{representation}
In this section, we briefly mention representations of the superalgebra  
$gl(M|N)$ and their characters. There are a lot of literatures on this subject 
(see for example,  \cite{Kac78,BB81,Pragacz91,MV03}). 

There are several choices of simple root systems 
$\{\alpha_{1},\alpha_{2},\dots,\alpha_{M+N-1} \}$ depending on 
 the choices of the Borel subalgebras.
The simplest system of simple roots is the so called 
 distinguished one. It leads as follows 
\begin{align}
   &\alpha_i = \epsilon_{i}-\epsilon_{i+1}, 
    \qquad i\in \{1,2,\dots, M-1\}, \nonumber \\
   &\alpha_{M} = \epsilon_{M}-\delta_{1}  \\ 
   & \alpha_{j+M} = \delta_{j}-\delta_{j+1} ,
    \quad j \in \{ 1,2,\dots,N-1\}, \nonumber   
\end{align}
where  
 $\epsilon_{1},\dots,\epsilon_{M},\delta_{1},\dots,\delta_{N}$ 
are the basis of the dual space of the Cartan subalgebra with the bilinear 
form $(\ |\ )$ such that 
\begin{align}
 (\epsilon_{i}|\epsilon_{j})=\delta_{i\, j}, \quad 
 (\epsilon_{i}|\delta_{j})=(\delta_{i}|\epsilon_{j})=0 , \quad 
 (\delta_{i}|\delta_{j})=-\delta_{i\, j} .
\end{align}
 $\{\alpha_i \}_{i \ne M}$ are even roots and $\alpha_{M}$ 
is an odd root with $(\alpha_{M} | \alpha_{M})=0$.

Any weight can be expressed in the following form:
\begin{align}
  \Lambda=\sum_{i=1}^{M}\Lambda_{i} \epsilon_{i}
         +\sum_{j=1}^{N}\overline{\Lambda}_{j} \delta_{j},
\quad \Lambda_{i},\bar{\Lambda_{j}}\in {\mathbb C}.
\end{align}
There is a class of irreducible tensor representations of $gl(M|N)$ whose 
highest weight $\Lambda$ is characterized by the 
Young diagram $\mu=(\mu_{1},\mu_{2},\dots)$,  
($\mu_{1} \ge \mu_{2} \ge \dots  \ge 0$, $\mu_{M+1} \le N$) in the $(M,N)$-hook (see 
Figure \ref{MN-hook}): 
\begin{align}
\begin{split}
& \Lambda_{i}=\mu_{i} \qquad \text{for} \quad  
  1 \le i \le M,  \\[4pt]
& \overline{\Lambda}_{j}=\eta_{j} \qquad \text{for} \quad   
 1 \le j \le N,
\end{split}
\end{align}
where $\eta_{j}={\rm max}\{\mu_{j}^{\prime}-M,0\}$. 
There is a set of parameters $[b_{1},b_{2},\dots,b_{M+N-1}]$, 
called the {\em Kac-Dynkin label} of $\Lambda$, defined by 
\begin{align}
 b_{j}=
\begin{cases}
\frac{2(\Lambda|\alpha_{j})}{(\alpha_{j}|\alpha_{j})} & 
\text{for} \quad   (\alpha_{j}|\alpha_{j}) \ne 0, \\[6pt]
 (\Lambda|\alpha_{j}) & 
\text{for}  \quad (\alpha_{j}|\alpha_{j})=0.
\end{cases} 
\end{align}
An irreducible representation with the highest weight $\Lambda $ is 
finite dimensional if $b_{j} \in {\mathbb Z}_{\ge 0}$ for all $j \ne M$, 
and $b_{M} \in {\mathbb C}$.
The Young diagram is related to the Kac-Dynkin label 
 as follows: 
\begin{align}
& b_{j}=\mu_{j}-\mu_{j+1} \quad \text{for} \quad  
  1 \le j \le M-1, \nonumber\\[4pt]
& b_{M}=\mu_{M}+\eta_{1},  \label{kacdynkin}    \\[4pt]
& b_{j+M}=\eta_{j}-\eta_{j+1} \quad \text{for} \quad  
 1 \le j \le N-1. \nonumber
\end{align}
Let $\Lambda$ be a real dominant weight. 
 An irreducible representation of $gl(M|N)$ with the highest weight 
 $\Lambda$ is  called {\em atypical}  if 
\begin{align}
(\Lambda+\rho,\epsilon_{i}-\delta_{j})=0 \quad 
\text{for some} \quad 
(i,j): \quad 
1\le i \le M, \quad 1\le j\le N,
\label{aty}
\end{align}
 and 
 {\em typical} if there is no such $(i,j)$. 
 Here $\rho $ is the graded half sum of the positive roots: 
\begin{align}
 \rho=\frac{1}{2}\sum_{i=1}^{M}(M-N-2i+1)\epsilon_{i}+
      \frac{1}{2}\sum_{j=1}^{N}(M+N-2j+1)\delta_{j}. 
\end{align}
Thus the aforementioned irreducible tensor 
representation is typical if $\mu_{M} \ge N$, and  
atypical if $\mu_{M} < N$.
 
There is a large class of finite dimensional typical representations, 
which is not tensor-like. 
For example, for the above mentioned typical irreducible tensor 
representations with the 
highest weight $\Lambda$, there is a one parameter family of irreducible typical 
\footnote{If the atypicality condition \eqref{aty} holds, this representation 
will be reducible (but in general indecomposable).} 
representations with the highest weight  
\begin{align}
\Lambda (c)=\Lambda +c \omega_{1} ,\label{cweight}
\end{align}
where $\omega_{1} =\epsilon_{1}+\epsilon_{2}+\cdots +\epsilon_{M}$. 
Note that the $M$-th Kac-Dynkin label $b_{M}=\mu_{M}+\eta_{1}+c$ is not  
 an integer if the parameter $c$ is  not an integer. 
One may generalize the above representation to the one with the highest weight 
$\Lambda (c_{1},c_{2})=\Lambda +c_{1} \omega_{1}+c_{2} \omega_{2}$, where  
$\omega_{2} =\delta_{1}+\delta_{2}+\cdots +\delta_{N}$. 
The Kac-Dynkin label of $\Lambda (c_{1},c_{2})$ depends only on $c_{1}+c_{2}$, and 
coincides with that of $\Lambda (c_{1}+c_{2})$. In this sense, 
this representation essentially has only one parameter. 
These parameters $c_{1}$ and $c_{2}$ correspond to the ones in section \ref{T-fun-sec}. 

There is a $gl(M|N)$ analog of the first Weyl formula, called 
the {\em Sergeev-Pragacz formula} \cite{Pragacz91}. 
It is an alternative representation of the supersymmetric Jacobi-Trudi formula, 
which corresponds to a $gl(M|N)$ analog of the second Weyl formula.  
Let us take a non-skew Young diagram $\mu$. Then the Sergeev-Pragacz formula has the 
following form
\footnote{We note that the sign of the  variables $y_{1},\dots,y_{N}$ 
in \cite{Pragacz91} is opposite to our definition.}
\begin{align}
\hspace{-3pt}
S_{\mu}(\{ x_{j} \}_{j=1}^{M}|\{ y_{k} \}_{k=1}^{N})
=\frac{\ds \sum_{\sigma \in S_{M}\times S_{N}}
{\rm sgn}(\sigma) \ \sigma    
\left[\ds
\prod_{i=1}^{M-1}x_{i}^{M-i}\prod_{j=1}^{N-1}y_{j}^{N-j}
\prod_{(i,j)\in \mu}(x_{i}-y_{j}) 
\right]
}{\ds \prod_{i<j}(x_{i}-x_{j})\prod_{i<j}(y_{i}-y_{j})},
\label{Sergeev-Pragacz}
\end{align}
where the third product in the numerator is taken over all boxes 
${(i,j)\in \mu}$ of the Young diagram $\mu $. 
We assumed that $x_{i}=0$ if $i \ge M+1$ and $y_{j}=0$ if $j \ge N+1 $. 
The notation $S_{M}$ (resp.\ $S_{N}$) denotes the symmetric group of order $M$ (resp.\ $N$). 
The symbol $\sigma[\ldots]$ stands for the action of the 
permutation $\sigma$, namely the Weyl group of $gl(M|N)$, 
 on the variables $x_1,\ldots,x_M,y_1,\ldots,y_N$
 in the square brackets. 
This formula \eqref{Sergeev-Pragacz} gives the (super)character of the above mentioned 
tensor representation labeled by the Young diagram $\mu$   
 if the variables are identified with the formal exponentials: 
$x_{j}={\mathrm e}^{\epsilon_{j}}$ for $j \in \{1,2,\dots M \}$ and 
$y_{j}={\mathrm e}^{\delta_{j}}$ for $j \in \{1,2,\dots N \}$. 
The Wronskian-like formula 
\eqref{9-thvari-1} (or \eqref{9-thvari-2}) at $(m,n)=(M,N)$, 
$(b_{1},\dots,b_{M},f_{1},\dots,f_{N})=(1,2,\dots,M+N)$ and $x=0$ 
coincides with the Sergeev-Pragacz formula \eqref{Sergeev-Pragacz}, where 
the variables $x_{1},\dots,x_{M}$ and $y_{1},\dots,y_{N}$ are related to 
the variables $z_{1},\dots, z_{M+N}$ in the main text as 
$x_{j}=z_{j}$ for $j \in \{1,2,\dots M \}$ and 
$y_{j}=z_{j+M}$ for $j \in \{1,2,\dots N \}$. 
This coincidence was proved in section 5 in \cite{MV03}. 

The representation theory of the quantum superalgebra $U_{q}(gl(M|N))$ 
for generic $q$ is similar 
(see for example, \cite{Zhang93}) to 
that of $gl(M|N)$. The atypicality conditions for both cases are the same. 
And the (super)characters are also the same. 

To make discussions complete, we should mention the representation theory of 
$U_{q}(\widehat{gl}(M|N))$ (or its Borel subalgebra). This is realized in combination of 
the evaluation map from (a Borel subalgebra of) $U_{q}(\widehat{gl}(M|N))$ to $U_{q}(gl(M|N))$ and 
 representations of $U_{q}(gl(M|N))$. 
We expect that the functions $\Ts^{B_{M}\sqcup F_{N}}_{\mu}(x)$ and 
$\overline{\Ts}^{B_{M}\sqcup F_{N}}_{\mu}(x)$ come from two different 
evaluation maps (see eqs.\ (4.14)-(4.18) in \cite{BHK02}, and also \cite{Kojima08}). 
$q$-oscillator representations are also needed to describe the Baxter $\Qs$-operators 
(cf.\ \cite{BLZ97,BHK02,Kojima08,BT08}). 
We would like to postpone these issues to a separate publication \cite{workinprogress}, 
where the functions in this paper will be realized as operators. 

\app{On the normalization of the Baxter $\Qs$-functions}\label{vaumm}
Let us renormalize the Baxter $\Qs$-function for any $I \subset {\mathfrak I}$ as: 
\begin{align}
\Qs_{I}(x)=\frac{Q_{I}(x)}{\phi_{I}(x)},
\end{align}
so that $Q_{I}(x)$ is a polynomial of 
finite degree (or an entire function of $x$):
\begin{align}
Q_{I}(x)=
\begin{cases}
\ds 
\prod_{j=1}^{n_{I}}\left(1-\frac{x}{x_{j}^{I}}\right) & 
\text{for} \quad I \ne \emptyset,\ {\mathfrak I}, \\[18pt]
1 & \text{for} \quad I=\emptyset,\ {\mathfrak I} .
\end{cases}
 \label{Q-fun-app}
\end{align}
Here the function $\phi_{I}(x)$ (vacuum part), 
whose precise form will be fixed in what follows,  
does not contain the roots $x_{j}^{I}$ of the Bethe ansatz equation. 
In our normalization, it satisfies $\phi_{I}(0)=1$.  
Then the functional relations \eqref{QQ-rel1}-\eqref{QQ-rel2} are modified to
\footnote{Compare \eqref{QQ-relre} with eq. (7.18) in \cite{KSZ07}.}
:  
\begin{multline}
(z_{i}-z_{j})Q_{I}(x)Q_{I\sqcup \{i,j\}}(x)  
=z_{i} \Phi_{I}^{(i,j)}(x)
Q_{I\sqcup \{i\}}(xq^{p_{i}})
Q_{I\sqcup \{j\}}(xq^{-p_{i}})
\\
-
z_{j} \Phi_{I}^{(j,i)}(x)
Q_{I\sqcup \{i\}}(xq^{-p_{i}})
Q_{I\sqcup \{j\}}(xq^{p_{i}})  
 \quad \text{for} \quad p_{i}=p_{j}, \label{QQ-rel1re} 
\end{multline}
\begin{multline}
 (z_{i}-z_{j})
Q_{I\sqcup \{i\}}(x)Q_{I \sqcup \{j \}}(x)=
z_{i} \Phi_{I}^{(i,j)}(x)^{-1}
Q_{I}(xq^{-p_{i}})
Q_{I\sqcup \{i,j\}}(xq^{p_{i}})
\\
-
z_{j} \Phi_{I}^{(j,i)}(x)^{-1}
Q_{I}(xq^{p_{i}})
Q_{I\sqcup \{i,j\}}(xq^{-p_{i}}) 
 \quad \text{for} \quad p_{i}=-p_{j}, \label{QQ-relre}
\end{multline}
where we set 
\begin{align}
\Phi_{I}^{(i,j)}(x)=\frac{\phi_{I}(xq^{-\frac{p_{i}-p_{j}}{2}})
\phi_{I \sqcup \{i,j \}}(xq^{\frac{p_{i}-p_{j}}{2}})}
{\phi_{I\sqcup \{i \}}(xq^{\frac{p_{i}+p_{j}}{2}})
\phi_{I\sqcup \{j \}}(xq^{-\frac{p_{i}+p_{j}}{2}})}.
\end{align}
Taking note on the zeros of the Baxter $Q$-functions \eqref{Q-fun-app}, one obtains  
Bethe ansatz equations from the functional relations \eqref{QQ-rel1re}-\eqref{QQ-relre}: 
\begin{multline}
 -\frac{\phi_{I_{a-1}}(x_{j}^{I_{a}}q^{p_{i_{a}}})
\phi_{I_{a}}(x_{j}^{I_{a}}q^{-2p_{i_{a+1}}})
\phi_{I_{a+1}}(x_{j}^{I_{a}}q^{p_{i_{a+1}}})}
{\phi_{I_{a-1}}(x_{j}^{I_{a}}q^{-p_{i_{a}}})
\phi_{I_{a}}(x_{j}^{I_{a}}q^{2p_{i_{a}}})
\phi_{I_{a+1}}(x_{j}^{I_{a}}q^{-p_{i_{a+1}}})} 
\\[3pt]
 =\frac{p_{i_{a+1}}z_{i_{a+1}}}{p_{i_{a}}z_{i_{a}}}
\frac{Q_{I_{a-1}}(x_{j}^{I_{a}}q^{p_{i_{a}}})
Q_{I_{a}}(x_{j}^{I_{a}}q^{-2p_{i_{a+1}}})
Q_{I_{a+1}}(x_{j}^{I_{a}}q^{p_{i_{a+1}}})}
{Q_{I_{a-1}}(x_{j}^{I_{a}}q^{-p_{i_{a}}})
Q_{I_{a}}(x_{j}^{I_{a}}q^{2p_{i_{a}}})
Q_{I_{a+1}}(x_{j}^{I_{a}}q^{-p_{i_{a+1}}})} 
\label{BAE2}
\\[4pt]
\text{for} \quad j\in \{1,2,\dots, n_{I_{a}}\} \quad \text{and} \quad a \in \{1,2,\dots, M+N-1 \} . 
\end{multline}
Left hand side of the above equation \eqref{BAE2} should have a standard form 
of the Bethe ansatz equation 
 (
cf. \cite{Sc83,Kul85,OW86,T98-2,BR08,KSZ07}) 
\begin{multline}
 -\frac{\Phi_{a}(x_{j}^{I_{a}}q^{\sum_{i\in I_{a}}p_{i}})}
 {\Phi_{a+1}(x_{j}^{I_{a}}q^{\sum_{i\in I_{a}}p_{i}})}
 =\frac{p_{i_{a+1}}z_{i_{a+1}}}{p_{i_{a}}z_{i_{a}}}
\frac{Q_{I_{a-1}}(x_{j}^{I_{a}}q^{p_{i_{a}}})
Q_{I_{a}}(x_{j}^{I_{a}}q^{-2p_{i_{a+1}}})
Q_{I_{a+1}}(x_{j}^{I_{a}}q^{p_{i_{a+1}}})}
{Q_{I_{a-1}}(x_{j}^{I_{a}}q^{-p_{i_{a}}})
Q_{I_{a}}(x_{j}^{I_{a}}q^{2p_{i_{a}}})
Q_{I_{a+1}}(x_{j}^{I_{a}}q^{-p_{i_{a+1}}})} 
\label{BAE3}
\\[4pt]
\text{for} \quad j\in \{1,2,\dots, n_{I_{a}}\} \quad \text{and} \quad a \in \{1,2,\dots, M+N-1 \} ,
\end{multline}
where the function $\Phi_{a}(x)$ has, 
for example for the Perk-Schultz model \cite{Sc83,Perk:1981}, 
the following form
\footnote{For more general quantum space with the 
highest weight $\sum_{j=1}^{M}\lambda_{j}^{(k)}\epsilon_{j} 
+\sum_{j=1}^{N} \lambda_{j+M}^{(k)} \delta_{j}$ 
for each site ($k=1,2,\dots, L$) of the quantum space, 
we expect that the function has the following form: 
$\Phi_{a}(x)=\prod_{k=1}^{L}
\left(1-\frac{xq^{-2p_{i_{a}} \lambda_{i_{a}}^{(k)}}}{w_{k}}\right)$. 
There will be an another form of the vacuum part, 
where the left hand side of \eqref{BAE3} has the form 
$-\frac{\overline{\Phi}_{a+1}(x_{j}^{I_{a}}q^{-\sum_{i\in I_{a}}p_{i}})}
 {\overline{\Phi}_{a}(x_{j}^{I_{a}}q^{-\sum_{i\in I_{a}}p_{i}})} $. 
Here we set 
$\overline{\Phi}_{a}(x)=\prod_{k=1}^{L}\left(1-\frac{xq^{2p_{i_{a}} \lambda_{i_{a}}^{(k)}}}{w_{k}}\right)$. 
The left hand side of the Bethe ansatz equation for 
$\overline{\Phi}_{a}(x)$ is speculated by applying 
a method in \cite{KOS95} to a (non-skew) Young diagram, where Drinfeld polynomials 
can be obtained; 
while the one for  $\Phi_{a}(x)$ is 
speculated by applying 
a method in \cite{KOS95} to 180 degrees rotation of
 this Young diagram. See a remark in section 7 in \cite{KSZ07}. 
We expect these come from two different evaluation representations of 
$U_{q}(\widehat{gl}(M|N))$, commented in Appendix \ref{representation},  
for the quantum space of the models. 
A precise description of these issue will need detailed Bethe ansatz 
calculations based on the representation theory of the quantum 
affine superalgebra, and 
is beyond the scope of this paper (cf. \cite{BR08}).}
\begin{align}
\Phi_{a}(x)=\prod_{k=1}^{L}\left(1-\frac{xq^{-2p_{i_{1}}\delta_{a,1}}}{w_{k}}\right).
\end{align}
Here $L$ is the number of the lattice site 
and $ w_{k}$ is an inhomogeneity on the spectral parameter. 
Thus $\phi_{I_{a}}(x)$ must satisfy the following functional equations. 
\begin{multline}
 \frac{\Phi_{a}(xq^{\sum_{i\in I_{a}}p_{i}})}
 {\Phi_{a+1}(xq^{\sum_{i\in I_{a}}p_{i}})}
 =
 \frac{\phi_{I_{a-1}}(xq^{p_{i_{a}}})
\phi_{I_{a}}(xq^{-2p_{i_{a+1}}})
\phi_{I_{a+1}}(xq^{p_{i_{a+1}}})}
{\phi_{I_{a-1}}(xq^{-p_{i_{a}}})
\phi_{I_{a}}(xq^{2p_{i_{a}}})
\phi_{I_{a+1}}(xq^{-p_{i_{a+1}}})} 
\\[4pt]
\text{for} \quad  a \in \{1,2,\dots, M+N-1 \} .
\label{funceq}
\end{multline}
One may solve \eqref{funceq} under the condition
\footnote{In general, one will need the inverse matrix of a $q$-deformed Cartan matrix of
$gl(M|N)$ to solve \eqref{funceq}. However, the Cartan matrix of $gl(M|M)$ is degenerate. 
Then we imposed a condition $M \ne N$. 
As for the case $M=N$, one will have to relax the condition 
$\phi_{I_{0}}(x)=1$ or $\phi_{I_{M+N}}(x)=1$.} 
$\phi_{I_{0}}(x)=\phi_{I_{M+N}}(x)=1$ and $\phi_{I_{a}}(0)=1$ for $0 \le a \le M+N$, 
$M \ne N$. This corresponds to impose both \eqref{tri1} and \eqref{tri2}. 
For example for the Perk-Schultz model, we  have 
the following solution of \eqref{funceq}.
\begin{align}
\phi_{I_{a}}(x)=\prod_{k=1}^{L}\frac{\left(\frac{xq^{\sum_{j \in I_{a}}p_{j}}}
{w_{k}};q^{2(M-N)}\right)_{\infty}}
{\left(\frac{xq^{2(M-N)-\sum_{j \in I_{a}}p_{j}}}
{w_{k}};q^{2(M-N)}\right)_{\infty}} 
\quad \text{for} \quad a \in \{1,2,\dots, N+N-1 \},
\label{sol-special}
\end{align}
where we assumed $|q^{M-N}|<1$ (the case $|q^{M-N}|>1$ will be similar) and introduced a 
symbol 
$\left(a;q \right)_{\infty}=\prod_{k=0}^{\infty}(1-aq^{k})$. It is interesting 
to see that cancellation occurs in \eqref{sol-special} for any $a$ 
if $M-N =\pm 1$. For example, for $(M,N)=(2,1)$, 
$p_{1}=p_{2}=1,p_{3}=-1$ case,
 we have 
\begin{align}
&\phi_{\{1 \}}(x)=\phi_{\{2 \}}(x)=1, 
& & \phi_{\{3 \}}(x)=f(xq^{-1})f(xq), \nonumber  \\
&\phi_{\{1,2\}}(x)=\frac{1}{f(x)},
& & \phi_{\{1,3 \}}(x)=\phi_{\{2,3\}}(x)=f(x), 
\label{norsl21}
\end{align}
where $f(x):=\prod_{k=1}^{L}(1-x/w_{k})$. 
This agrees with the result in \cite{BT08} (see eqs. (2.87) in \cite{BT08}). 
Now the Baxter $\Qs$-functions $\Qs_{I}(x)$ are meromorphic functions of $x$, while 
 $Q_{I}(x)$ are polynomials (or entire functions) of $x$. 
  The Baxter $\Qs$-functions $\Qs_{I}(x)$ with both the conditions 
\eqref{tri1} and \eqref{tri2} for $M \ne N$ will be in the normalization of the 
universal $R$-matrix.  
\\\\
\app{Proof of the functional relations}
\label{proof-th}
We will use the following lemma in the proof. 
\begin{lemma}\label{shift-lem}
The following relations hold for the determinant \eqref{minor1}, where  
$m+\alpha=n+\beta$.
If there is $k$ ($1 \le k \le \beta $) such that  $s_{k}=0$, then we have 
\begin{align}
& \Delta^{(b_{1},b_{2},\dots,b_{m} ), (r_{1},r_{2},\dots,r_{\alpha})}
_{(f_{1},f_{2},\dots,f_{n} ), (s_{1},s_{2},\dots,s_{\beta})}
(xq^{2})
=
(-1)^{\alpha}
\frac{z_{f_{1}}z_{f_{2}}\cdots z_{f_{n}}}{z_{b_{1}}z_{b_{2}}\cdots z_{b_{m}}}
\Delta^{(b_{1},b_{2},\dots,b_{m} ), (r_{1}-1,r_{2}-1,\dots,r_{\alpha}-1)}
_{(f_{1},f_{2},\dots,f_{n} ), (s_{1}+1,s_{2}+1,\dots,s_{\beta}+1)}
(x)
\label{defe1}
\end{align}
under the relation \eqref{QQ-bf1}.
If there is $k$ ($1 \le k \le \alpha $) such that $r_{k}=0$, then we have 
\begin{align}
& \Delta^{(b_{1},b_{2},\dots, b_{m} ), (r_{1},r_{2},\dots, r_{\alpha})}
_{(f_{1},f_{2},\dots,f_{n} ), (s_{1},s_{2},\dots,s_{\beta})}
(xq^{-2})
=
(-1)^{\alpha}
\frac{z_{b_{1}}z_{b_{2}}\cdots z_{b_{m}}}{z_{f_{1}}z_{f_{2}}\cdots z_{f_{n}}}
\Delta^{(b_{1},b_{2},\dots,b_{m} ), (r_{1}+1,r_{2}+1,\dots,r_{\alpha}+1)}
_{(f_{1},f_{2},\dots,f_{n} ), (s_{1}-1,s_{2}-1,\dots,s_{\beta}-1)}
(x)
\label{defe2}
\end{align}
under the relation \eqref{QQ-bf1}. One can also derive similar relations for 
\eqref{minor2} by $q \to q^{-1}$. 
\end{lemma}
In particular for $m=\beta =0$ or $\alpha=n=0$, 
 the following relations hold for any $c \in {\mathbb C}$: 
\begin{align}
& \Delta^{\emptyset, (r_{1}+c,r_{2}+c,\dots,r_{n}+c)}
_{(f_{1},f_{2},\dots,f_{n} ), \emptyset}
(x)
=
((-1)^{n}z_{f_{1}}z_{f_{2}}\cdots z_{f_{n}})^{c}
\Delta^{\emptyset , (r_{1},r_{2},\dots,r_{n})}
_{(f_{1},f_{2},\dots,f_{n} ), \emptyset}
(xq^{-2c}), \label{del-shift1}
\\[6pt]
& \Delta^{(b_{1},b_{2},\dots, b_{m} ), \emptyset }
_{\emptyset , (s_{1}+c,s_{2}+c,\dots,s_{m}+c)}
(x)
=
(z_{b_{1}}z_{b_{2}}\cdots z_{b_{m}})^{c}
\Delta^{(b_{1},b_{2},\dots,b_{m} ),\emptyset }
_{\emptyset , (s_{1},s_{2},\dots,s_{m})}
(xq^{2c}). \label{del-shift2}
\end{align}
For any $m \times n$ matrix, we will write a minor determinant whose 
$j_{1},j_{2},\dots,j_{\alpha} $-th rows and $k_{1},k_{2},\dots,k_{\beta}$-th columns removed from 
it as 
$D\begin{bmatrix}
j_{1}, & j_{2}, & \dots, j_{\alpha} \\
k_{1}, & k_{2}, & \dots, k_{\beta}
\end{bmatrix}$,
where $m-\alpha =n -\beta$, $j_{1} <j_{2}\cdots < j_{\alpha} $ and 
$k_{1} < k_{2} < \dots < k_{\beta}$. 
We will use the following identities for determinants. 
\begin{align}
&
D\begin{bmatrix}
j_{1}, &  j_{2}\\
{} 
\end{bmatrix}
D\begin{bmatrix}
j_{3}, &  j_{4}\\
{} 
\end{bmatrix}
-
D\begin{bmatrix}
j_{1}, &  j_{3}\\
{} 
\end{bmatrix}
D\begin{bmatrix}
j_{2}, &  j_{4}\\
{} 
\end{bmatrix}
+
D\begin{bmatrix}
j_{1}, &  j_{4}\\
{} 
\end{bmatrix}
D\begin{bmatrix}
j_{2}, &  j_{3}\\
{} 
\end{bmatrix}
=0,  \label{plucker}
\\
&
D\begin{bmatrix}
{} \\
k_{1} ,&  k_{2}
\end{bmatrix}
D\begin{bmatrix}
{} \\
k_{3}, &  k_{4}
\end{bmatrix}
-
D\begin{bmatrix}
{} \\ 
k_{1}, &  k_{3} 
\end{bmatrix}
D\begin{bmatrix}
{} \\
k_{2}, &  k_{4}
\end{bmatrix}
+
D\begin{bmatrix}
{} \\
k_{1}, &  k_{4}
\end{bmatrix}
D\begin{bmatrix}
{} \\ 
k_{2} ,&  k_{3}
\end{bmatrix}
=0,  \label{plucker2}
\\
&
D\begin{bmatrix}
j_{1} \\
{} 
\end{bmatrix}
D\begin{bmatrix}
j_{2}, &  j_{3} \\
k_{1} 
\end{bmatrix}
-
D\begin{bmatrix}
j_{2} \\
{} 
\end{bmatrix}
D\begin{bmatrix}
j_{1},&  j_{3} \\
k_{1} 
\end{bmatrix}
+
D\begin{bmatrix}
j_{3} \\
{} 
\end{bmatrix}
D\begin{bmatrix}
j_{1}, &  j_{2} \\
k_{1} 
\end{bmatrix}
=0, 
\label{plucker3}
\\[8pt]
&
D\begin{bmatrix}
{} \\
k_{1} 
\end{bmatrix}
D\begin{bmatrix}
j_{1} \\
k_{2} ,&  k_{3}  
\end{bmatrix}
-
D\begin{bmatrix}
{} \\
k_{2}  
\end{bmatrix}
D\begin{bmatrix}
j_{1}  \\ 
k_{1} ,&  k_{3} 
\end{bmatrix}
+
D\begin{bmatrix}
{} \\
k_{3} 
\end{bmatrix}
D\begin{bmatrix}
j_{1} \\
k_{1}, &  k_{2} 
\end{bmatrix}
=0, 
\label{plucker4}
\\[8pt]
& 
D\begin{bmatrix}
\quad \\
\quad 
\end{bmatrix}
D\begin{bmatrix}
j_{1}, &  j_{2} \\
k_{1}, &  k_{2}
\end{bmatrix}
-
D\begin{bmatrix}
j_{1}  \\
k_{1} 
\end{bmatrix}
D\begin{bmatrix}
 j_{2} \\
 k_{2}
\end{bmatrix}
+
D\begin{bmatrix}
j_{1}  \\
k_{2} 
\end{bmatrix}
D\begin{bmatrix}
 j_{2} \\
 k_{1}
\end{bmatrix}
=0. 
\label{jacobi}
\end{align}
\eqref{plucker}-\eqref{plucker4} are specialization of the so-called Pl\"ucker identity, 
and 
\eqref{jacobi} is the Jacobi identity. 
We remark that  some Pl\"ucker identities on supersymmetric polynomials  
were also discussed in \cite{GPS05}. 

\sapp{Proof of Theorem \ref{solution-Q}}\label{proof-Q}
We will prove Theorem \ref{solution-Q} for \eqref{q-func1}. 
The proof for \eqref{q-func2} is similar to the one for \eqref{q-func1}. 
Let us introduce a notation:
\begin{align}
I_{m,n }=
\begin{cases}
(1,2,\dots, m -n ) & \text{for} \quad m -n >0, \\[3pt]
\emptyset & \text{for}  \quad  m -n \le 0 .
\end{cases}
\end{align}
That \eqref{q-func1} satisfies \eqref{QQ-rel1}-\eqref{QQ-rel2} for $I=B_{m} \sqcup F_{n}$
is equivalent to:
\begin{multline}
 -\Delta^{B_{m},I_{n,m}}_{F_{n},I_{m,n}}(xq)
\Delta^{B_{m+2} ,I_{n,m +2}}_{F_{n},I_{m +2,n}}
(xq^{-1})
=z_{b_{m+1}} 
\Delta^{B_{m+1},I_{n,m +1}}_{F_{n},I_{m +1,n}}(xq)
\Delta^{B_{m}\times (b_{m+2}) ,I_{n,m +1}}_{F_{n},I_{m +1,n}}
(xq^{-1})
   \\
-
z_{b_{m+2}} 
\Delta^{B_{m+1},I_{n,m +1}}_{F_{n},I_{m +1,n}}
(xq^{-1})
\Delta^{B_{m}\times (b_{m+2}) ,I_{n,m +1}}_{F_{n},I_{m +1,n}}
(xq)  \label{qq-app1}
\end{multline}
for $i=b_{m+1},j=b_{m+2}$; 
\begin{multline}
 \Delta^{B_{m},I_{n,m}}_{F_{n},I_{m,n}}(xq^{-1})
\Delta^{B_{m} ,I_{n +2,m}}_{F_{n+2},I_{m ,n +2}}
(xq)
=z_{f_{n+1}} 
\Delta^{B_{m},I_{n+1,m }}_{F_{n+1},I_{m ,n+1}}
(xq^{-1})
\Delta^{B_{m} ,I_{n+1,m }}_{F_{n}\times (f_{n+2}),I_{m ,n+1}}
(xq)
  \\
-
z_{f_{n+2}} 
\Delta^{B_{m},I_{n+1,m }}_{F_{n+1},I_{m ,n+1}}
(xq)
\Delta^{B_{m} ,I_{n+1,m }}_{F_{n}\times (f_{n+2}),I_{m ,n+1}}
(xq^{-1}) 
  \label{qq-app2}
\end{multline}
for $i=f_{n+1},j=f_{n+2}$;
\begin{multline}
 \Delta^{B_{m+1},I_{n,m +1}}_{F_{n},I_{m +1,n}}
(xq^{-1})
\Delta^{B_{m} ,I_{n +1,m}}_{F_{n+1},I_{m ,n +1}}
(xq)
=z_{b_{m+1}} 
\Delta^{B_{m},I_{n,m }}_{F_{n},I_{m ,n}}
(xq^{-1})
\Delta^{B_{m+1} ,I_{n +1,m +1}}_{F_{n+1},I_{m +1,n +1}}
(xq)
   \\
 -
z_{f_{n+1}} 
\Delta^{B_{m},I_{n,m}}_{F_{n},I_{m,n}}
(xq)
\Delta^{B_{m+1} ,I_{n +1,m +1}}_{F_{n+1},I_{m +1,n +1}}
(xq^{-1}) 
 \label{qq-app3}
\end{multline}
for $i=b_{m+1},j=f_{n+1}$. 

We have to consider the following nine cases since the $(m,n)$-index 
\eqref{mn-index} for $\mu=\emptyset $ depends on 
$m,n$: 
for \eqref{qq-app1}: (i.1) $m \ge n $, 
(i.2) $m = n -1$, 
(i.3) $m \le n -2$; 
for \eqref{qq-app2}: (ii.1) $m \le n $,  
(ii.2) $m = n +1$, 
(ii.3) $m \ge n +2$; 
for \eqref{qq-app3}: (iii.1) $m \ge n +1 $, 
(iii.2) $m = n $, 
(iii.3) $m \le n -1$. 
We find that 
the proof of the case (ii.1) is similar to the case (i.1);
the case (ii.2) is similar to the case (i.2);
the case (ii.3) is similar to the case (i.3). So we will 
treat (i.1), (i.2), (i.3),(iii.1), (iii.2), (iii.3) in what follows.\\\\
{\it (i.1) \eqref{qq-app1} for the case} $m \ge n $

In this case \eqref{qq-app1} reduces to 
\begin{multline}
 -\Delta^{B_{m},\emptyset }_{F_{n},(1,2,\dots, m -n )}(xq)
\Delta^{B_{m+2},\emptyset }_{F_{n},(1,2,\dots, m -n+2 )}
(xq^{-1})
\\
\qquad =z_{b_{m+1}} 
\Delta^{B_{m+1},\emptyset }_{F_{n},(1,2,\dots, m -n +1)}(xq)
\Delta^{B_{m}\times (b_{m+2}) ,\emptyset }_{F_{n},(1,2,\dots, m -n +1)}
(xq^{-1})
   \\
 -
z_{b_{m+2}} 
\Delta^{B_{m+1},\emptyset }_{F_{n},(1,2,\dots, m -n +1)}
(xq^{-1})
\Delta^{B_{m}\times (b_{m+2}) ,\emptyset }_{F_{n},(1,2,\dots, m -n +1)}
(xq).
\end{multline}
Due to \eqref{defe1}, this is identical to the following relation:
\begin{multline}
 -\Delta^{B_{m},\emptyset }_{F_{n},(1,2,\dots, m -n )}(x)
\Delta^{B_{m+2} ,\emptyset }_{F_{n},(0,1,\dots, m -n+1 )}
(x)
 =
\Delta^{B_{m+1},\emptyset }_{F_{n},(1,2,\dots, m -n +1)}(x)
\Delta^{B_{m}\times (b_{m+2}) ,\emptyset }_{F_{n},(0,1,\dots, m -n )}
(x)
   \\
 -
\Delta^{B_{m+1},\emptyset }_{F_{n},(0,1,\dots, m -n )}
(x)
\Delta^{B_{m}\times (b_{m+2}) ,\emptyset }_{F_{n},(1,2,\dots, m -n +1)}
(x).
\end{multline}
This is nothing but the Jacobi identity \eqref{jacobi}. \\\\
{\it (i.2)  \eqref{qq-app1} for the case} $m = n -1$  

In this case \eqref{qq-app1} reduces to 
\begin{multline}
-\Delta^{B_{m},(1)}_{F_{m +1},\emptyset }(xq)
\Delta^{B_{m+2} ,\emptyset }_{F_{m +1},(1)}
(xq^{-1})
%
 =z_{b_{m+1}} 
\Delta^{B_{m+1},\emptyset }_{F_{m +1},\emptyset }(xq)
\Delta^{B_{m}\times (b_{m+2}) , \emptyset }_{F_{m +1},\emptyset }
(xq^{-1})
  \\
 -
z_{b_{m+2}} 
\Delta^{B_{m+1},\emptyset }_{F_{m +1},\emptyset }
(xq^{-1})
\Delta^{B_{m}\times (b_{m+2}) ,\emptyset }_{F_{m +1},\emptyset }
(xq) .
\label{toprovei-2}
\end{multline} 
Let us expand the determinants
\footnote{
In Appendix \ref{proof-th}, we will use the following notation for matrices in the 
determinant.
We use $k$ for the row index and $l$ for the 
column index. Thus  
$(a_{k,l})$ is a matrix whose $k$-th row and $l$-th column element is 
$a_{k,l}$;
 $(a_{k})$ is a column vector; $(a_{l})$ is a row vector.
}
 in the left hand side of the above relation:
\begin{align}
& -\Delta^{B_{m},(1)}_{F_{m +1},\emptyset }(xq)
\Delta^{B_{m+2} ,\emptyset }_{F_{m +1},(1)}
(xq^{-1})
=
-\sum_{\alpha =1}^{m +1}(-1)^{m +1 +\alpha}
\Qs_{f_{\alpha}}(x) 
\Delta^{B_{m},\emptyset}_{F_{m +1}\backslash f_{\alpha},\emptyset }(xq)
\nonumber \\ 
& \hspace{180pt} \times \sum_{\beta =1}^{m +2}(-1)^{\beta +m +2}
\Qs_{b_{\beta}}(x)
\Delta^{B_{m+2}\backslash b_{\beta} ,\emptyset }_{F_{m +1},\emptyset}(xq^{-1})
\nonumber \\[3pt] &
=\sum_{\alpha =1}^{m +1}\sum_{\beta =1}^{m +2}
(-1)^{\alpha+\beta}
\frac{z_{b_{\beta}}\Qs_{\{b_{\beta},f_{\alpha}\}}(xq)-
z_{f_{\alpha}}\Qs_{\{b_{\beta},f_{\alpha}\}}(xq^{-1})}{z_{b_{\beta}}-z_{f_{\alpha}}}
\Delta^{B_{m},\emptyset}_{F_{m +1}\backslash f_{\alpha},\emptyset }(xq)
\Delta^{B_{m+2}\backslash b_{\beta} ,\emptyset }_{F_{m +1},\emptyset}(xq^{-1}) 
\nonumber \\[3pt]
& =-
\sum_{\beta =1}^{m +2}(-1)^{\beta+m}z_{b_{\beta}}
\Delta^{B_{m+2}\backslash b_{\beta} ,\emptyset }_{F_{m +1},\emptyset}(xq^{-1}) 
\begin{vmatrix}
\left(\frac{\Qs_{\{b_{k},f_{l}\}}(xq)}{z_{b_{k}}-z_{f_{l}}}
 \right)_{\genfrac{}{}{0pt}{2}{1 \le k \le m,}{1 \le l \le m+1}}
 \\[10pt]
\left(\frac{\Qs_{\{b_{\beta },f_{l}\}}(xq)}{z_{b_{\beta}}-z_{f_{l}}}\right)_{1 \le l \le m+1}
\end{vmatrix} 
\nonumber \\[2pt] 
&+
\sum_{\alpha =1}^{m+1}
(-1)^{m+1+\alpha}z_{f_{\alpha}}
\Delta^{B_{m},\emptyset}_{F_{m +1}\backslash f_{\alpha},\emptyset }(xq)
\begin{vmatrix}
\left(\frac{\Qs_{\{b_{k},f_{l}\}}(xq^{-1})}{z_{b_{k}}-z_{f_{l}}}
 \right)_{\genfrac{}{}{0pt}{2}{1 \le k \le m+2,}{1 \le l \le m+1}}
 & 
\left(\frac{\Qs_{\{b_{k},f_{\alpha}\}}(xq^{-1})}{z_{b_{k}}-z_{f_{\alpha}}}\right)_{1 \le k \le m+2}
\end{vmatrix}
,
\label{showi-2}
\end{align}
where we introduced a notation $B_{m+2} \backslash b_{\beta}=
(b_{1},b_{2},\dots,b_{\beta-1},b_{\beta+1},\dots,b_{m+2})$, and used the functional 
relation \eqref{QQ-bf1}. 
The determinant $\begin{vmatrix} \cdots \end{vmatrix}$ 
in the first summand in the right hand side of \eqref{showi-2} vanishes for 
$\beta \ne m+1,m+2$ as it has two identical rows. 
The determinant in the second summand vanishes as it has two identical columns. 
Then the right hand side of \eqref{showi-2} reduces to the right hand side of \eqref{toprovei-2}.
\\\\
{\it (i.3)  \eqref{qq-app1} for the case} $m \le n -2$ 

In this case \eqref{qq-app1} reduces to 
\begin{multline}
-\Delta^{B_{m},(1,2,\dots, n - m )}_{F_{n},\emptyset }(xq)
\Delta^{B_{m+2},(1,2,\dots, n - m -2)}_{F_{n},\emptyset }
(xq^{-1})
  \\
 =z_{b_{m+1}} 
\Delta^{B_{m+1},(1,2,\dots, n - m-1 )}_{F_{n},\emptyset }(xq)
\Delta^{B_{m}\times (b_{m+2}) ,(1,2,\dots, n - m -1)}_{F_{n},\emptyset }
(xq^{-1})
  \\
 -
z_{b_{m+2}} 
\Delta^{B_{m+1},(1,2,\dots, n - m -1)}_{F_{n},\emptyset }
(xq^{-1})
\Delta^{B_{m}\times (b_{m+2}) ,(1,2,\dots, n - m -1)}_{F_{n},\emptyset }
(xq).
\end{multline}
Due to \eqref{defe2}, this relation can be modified to:
\begin{multline}
\Delta^{B_{m},(0,1,\dots, n - m-1 )}_{F_{n},\emptyset }(x)
\Delta^{B_{m+2},(1,2,\dots, n - m -2)}_{F_{n},\emptyset }
(x)
  \\
 =
\Delta^{B_{m+1},(0,1,\dots, n - m-2 )}_{F_{n},\emptyset }(x)
\Delta^{B_{m}\times (b_{m+2}) ,(1,2,\dots, n - m -1)}_{F_{n},\emptyset }
(x)
  \\
 - 
\Delta^{B_{m+1},(1,2,\dots, n - m -1)}_{F_{n},\emptyset }
(x)
\Delta^{B_{m}\times (b_{m+2}) ,(0,1,\dots, n - m -2)}_{F_{n},\emptyset }
(x).
\end{multline}
This is nothing but the Pl\"ucker relation \eqref{plucker}. \\\\
{\it (iii.1) \eqref{qq-app3} for the case} $m \ge n +1 $ 

In this case, \eqref{qq-app3} has the form:
\begin{multline}
 \Delta^{B_{m+1},\emptyset }_{F_{n},(1,2,\dots,m-n+1)}
(xq^{-1})
\Delta^{B_{m} ,\emptyset }_{F_{n+1},(1,2,\dots,m-n-1)}
(xq)
\\
=z_{b_{m+1}} 
\Delta^{B_{m},\emptyset }_{F_{n},(1,2,\dots,m-n)}
(xq^{-1})
\Delta^{B_{m+1} ,\emptyset }_{F_{n+1},(1,2,\dots,m-n)}
(xq)
   \\
 -
z_{f_{n+1}} 
\Delta^{B_{m},\emptyset }_{F_{n},(1,2,\dots,m-n)}
(xq)
\Delta^{B_{m+1} ,\emptyset }_{F_{n+1},(1,2,\dots,m-n)}
(xq^{-1}) .
\end{multline}
Due to \eqref{defe1}, this relation reduces to:
\begin{multline}
 \Delta^{B_{m+1},\emptyset }_{F_{n},(0,1,\dots,m-n)}
(x)
\Delta^{B_{m} ,\emptyset }_{F_{n+1},(1,2,\dots,m-n-1)}
(x)
=
\Delta^{B_{m},\emptyset }_{F_{n},(0,1,\dots,m-n-1)}
(x)
\Delta^{B_{m+1} ,\emptyset }_{F_{n+1},(1,2,\dots,m-n)}
(x)
   \\
 -
\Delta^{B_{m},\emptyset }_{F_{n},(1,2,\dots,m-n)}
(x)
\Delta^{B_{m+1} ,\emptyset }_{F_{n+1},(0,1,\dots,m-n-1)}
(x) .
\end{multline}
This is nothing but the Pl\"ucker identity \eqref{plucker4}.
\\\\
{\it (iii.2) \eqref{qq-app3} for the case} $m = n $ 

In this case, \eqref{qq-app3} has the form: 
\begin{multline}
 \Delta^{B_{m+1},\emptyset}_{F_{m},(1)}
(xq^{-1})
\Delta^{B_{m} ,(1)}_{F_{m+1},\emptyset}
(xq)
=z_{b_{m+1}} 
\Delta^{B_{m},\emptyset}_{F_{m},\emptyset}
(xq^{-1})
\Delta^{B_{m+1} ,\emptyset}_{F_{m+1},\emptyset}
(xq)
   \\
 -
z_{f_{m+1}} 
\Delta^{B_{m},\emptyset}_{F_{m},\emptyset}
(xq)
\Delta^{B_{m+1} ,\emptyset}_{F_{m+1},\emptyset}
(xq^{-1}) .
\label{toproveiii2}
\end{multline}
Let us expand the left hand side of \eqref{toproveiii2}.
\begin{align}
& \Delta^{B_{m+1},\emptyset}_{F_{m},(1)}
(xq^{-1})
\Delta^{B_{m} ,(1)}_{F_{m+1},\emptyset}
(xq)
=\sum_{\alpha=1}^{m+1}(-1)^{\alpha +m+1}
 \Qs_{b_{\alpha }}(x)
 \Delta^{B_{m+1}\backslash b_{\alpha },\emptyset}_{F_{m},\emptyset }
(xq^{-1})
\nonumber \\ 
& \hspace{150pt} \times 
\sum_{\beta=1}^{m+1}(-1)^{m+1+\beta }
\Qs_{f_{\beta}}(x)
\Delta^{B_{m} ,\emptyset}_{F_{m+1}\backslash f_{\beta},\emptyset}
(xq)
\nonumber \\[3pt] 
&=\sum_{\alpha=1}^{m+1}\sum_{\beta=1}^{m+1}(-1)^{\alpha +\beta }
\frac{z_{b_{\alpha}}\Qs_{\{b_{\alpha},f_{\beta}\}}(xq)- 
z_{f_{\beta}}\Qs_{\{b_{\alpha},f_{\beta}\}}(xq^{-1})}{z_{b_{\alpha}}-z_{f_{\beta}}}
\Delta^{B_{m+1}\backslash b_{\alpha },\emptyset}_{F_{m},\emptyset }
(xq^{-1})
\Delta^{B_{m} ,\emptyset}_{F_{m+1}\backslash f_{\beta},\emptyset}
(xq) 
\nonumber \\[3pt]
&= 
\sum_{\alpha=1}^{m+1}(-1)^{\alpha +m+1}
z_{b_{\alpha}}
\Delta^{B_{m+1}\backslash b_{\alpha },\emptyset}_{F_{m},\emptyset }(xq^{-1}) 
\begin{vmatrix}
\left(\frac{\Qs_{\{b_{k},f_{l}\}}(xq)}{z_{b_{k}}-z_{f_{l}}}
 \right)_{\genfrac{}{}{0pt}{2}{1 \le k \le m,}{1 \le l \le m+1}}
 \\[10pt]
\left(\frac{\Qs_{\{b_{\alpha },f_{l}\}}(xq)}{z_{b_{\alpha}}-z_{f_{l}}}\right)_{1 \le l \le m+1}
\end{vmatrix}
\nonumber \\[2pt]
& -\sum_{\beta=1}^{m+1}(-1)^{\beta +m+1} 
z_{f_{\beta}}
\Delta^{B_{m} ,\emptyset}_{F_{m+1}\backslash f_{\beta},\emptyset}
(xq) 
\begin{vmatrix}
\left(\frac{\Qs_{\{b_{k},f_{l}\}}(xq^{-1})}{z_{b_{k}}-z_{f_{l}}}
 \right)_{\genfrac{}{}{0pt}{2}{1 \le k \le m+1,}{1 \le l \le m}}
 &
\left(\frac{\Qs_{\{b_{k },f_{\beta}\}}(xq^{-1})}{z_{b_{k}}-z_{f_{\beta}}}\right)_{1 \le k \le m+1}
\end{vmatrix}
, \label{shimeseiii2}
\end{align}
where we used the functional relation \eqref{QQ-bf1}. 
The determinant  $\begin{vmatrix}\cdots  \end{vmatrix}$ 
in the first summand in the right hand side of \eqref{shimeseiii2} 
vanishes for $\alpha \ne m+1$. 
The determinant in the second summand  
vanishes for $\beta \ne m+1$. Thus the right hand side of \eqref{shimeseiii2} 
coincides with the right hand side of \eqref{toproveiii2}. \\\\
{\it (iii.3) \eqref{qq-app3} for the case} $m \le n -1$ 

In this case  \eqref{qq-app3} has the form: 
\begin{multline}
 \Delta^{B_{m+1},(1,2,\dots,n-m-1)}_{F_{n},\emptyset}
(xq^{-1})
\Delta^{B_{m} ,(1,2,\dots,n-m+1)}_{F_{n+1},\emptyset}
(xq)
\\
=z_{b_{m+1}} 
\Delta^{B_{m},(1,2,\dots,n-m)}_{F_{n},\emptyset}
(xq^{-1})
\Delta^{B_{m+1} ,(1,2,\dots,n-m)}_{F_{n+1},\emptyset}
(xq)
   \\
 -
z_{f_{n+1}} 
\Delta^{B_{m},(1,2,\dots,n-m)}_{F_{n},\emptyset}
(xq)
\Delta^{B_{m+1} ,(1,2,\dots,n-m)}_{F_{n+1},\emptyset}
(xq^{-1}) .
\end{multline}
Due to \eqref{defe2}, this reduces to: 
\begin{multline}
\hspace{-5pt}
 - \Delta^{B_{m+1},(1,2,\dots,n-m-1)}_{F_{n},\emptyset}
(x)
\Delta^{B_{m} ,(0,1,\dots,n-m)}_{F_{n+1},\emptyset}
(x)
=
\Delta^{B_{m},(1,2,\dots,n-m)}_{F_{n},\emptyset}
(x)
\Delta^{B_{m+1} ,(0,1,\dots,n-m-1)}_{F_{n+1},\emptyset}
(x)
   \\
 -
\Delta^{B_{m},(0,1,\dots,n-m-1)}_{F_{n},\emptyset}
(x)
\Delta^{B_{m+1} ,(1,2,\dots,n-m)}_{F_{n+1},\emptyset}
(x) .
\end{multline}
This is nothing but the Pl\"ucker identity \eqref{plucker3}.

\sapp{Proof of Theorem \ref{solution-t-sys-th}}\label{proof-t-sys}
We will prove Theorem \ref{solution-t-sys-th} for \eqref{tfunrec1}. 
The proof for \eqref{tfunrec2} is similar to the one for \eqref{tfunrec1}.
For $a,s \in {\mathbb Z}_{\ge 1}$, we will prove that \eqref{tfunrec1} 
satisfies 
\begin{multline}
\Ts^{(a),B_{m},F_{n}}_{s}(xq^{-1})\Ts^{(a),B_{m},F_{n}}_{s}(xq)
=
\Ts^{(a),B_{m},F_{n}}_{s-1}(x) \Ts^{(a),B_{m},F_{n}}_{s+1}(x)
\\ +
\Ts^{(a-1),B_{m},F_{n}}_{s}(x) \Ts^{(a+1),B_{m},F_{n}}_{s}(x).
\end{multline}
This is identical to the following functional relation. 
\begin{align}
{\mathcal T}^{B_{m}, F_{n}}_{(s^a)}(xq^{-1}){\mathcal T}^{B_{m}, F_{n}}_{(s^a)}(xq)
=
{\mathcal T}^{B_{m}, F_{n}}_{((s-1)^a)}(x){\mathcal T}^{B_{m}, F_{n}}_{((s+1)^a)}(x)
 +
{\mathcal T}^{B_{m}, F_{n}}_{(s^{a-1})}(x){\mathcal T}^{B_{m}, F_{n}}_{(s^{a+1})}(x). 
\label{toprove}
\end{align}
We have to consider the following seven cases since the $(m,n)$-index \eqref{mn-index} depends on 
$m,n,a,s$: 
(i) $a < m-n$, 
(ii) $a = m-n$, 
(iii)  $a-s < m-n<a$, 
(iv) $a-s = m-n $, 
(v) $-s < m-n<a-s$, 
(vi) $m-n=-s$, 
(vii) $m-n<-s$. 
The proof for the case (v) is similar to the case (iii);  
the case (vi) is similar to the case (ii);  
the case (vii) is similar to the case (i).
So we will treat four cases (i), (ii), (iii), (iv) from now on. 
 
The ``duality relations'' \eqref{dual-mn} and \eqref{dual-mn2} follow 
from the formulae \eqref{typfac1}-\eqref{typfac4}. 
One also have to take into account reduction of the functional relation
 \eqref{toprove} based on 
Lemma \ref{vanish-dai} for 
\eqref{reduc1}, \eqref{reduc2}, \eqref{reduc3} and \eqref{reduc4}, 
and Theorem \ref{solution-Q} for the boundary conditions 
\eqref{t-sys-bc1}, \eqref{t-sys-bc2}, \eqref{t-sys-bc3} and \eqref{t-sys-bc4}. 
\\\\
{\it (i) The case} $a < m-n$ 

Due to \eqref{defe1}, 
\eqref{toprove} is identical to the following relation.
\begin{multline}
\Delta^{B_{m},\emptyset}_{F_{n},(1,2,\dots,m-n-a,m-n-a+s+1,\dots,m-n+s)}(x)
\Delta^{B_{m},\emptyset}_{F_{n},(0,1,\dots,m-n-a-1,m-n-a+s,\dots,m-n+s-1)}(x)
 \\
=
\Delta^{B_{m},\emptyset}_{F_{n},(1,2,\dots,m-n-a,m-n-a+s,\dots,m-n+s-1)}(x)
\Delta^{B_{m},\emptyset}_{F_{n},(0,1,\dots,m-n-a-1,m-n-a+s+1,\dots,m-n+s)}(x)
 \\
+
\Delta^{B_{m},\emptyset}_{F_{n},(1,2,\dots,m-n-a-1,m-n-a+s,\dots,m-n+s)}(x)
\Delta^{B_{m},\emptyset}_{F_{n},(0,1,\dots,m-n-a,m-n-a+s+1,\dots,m-n+s-1)}(x).
\end{multline}
This is nothing but the Pl\"ucker identity \eqref{plucker2}. \\\\
{\it (ii) The case} $a = m-n$ 

\eqref{toprove} is identical to the following relation.
\begin{multline}
\Delta^{B_{m},\emptyset}_{F_{n},(s+1,s+2,\dots,s+a)}(x)
\Delta^{B_{m},\emptyset}_{F_{n},(s+1,s+2,\dots,s+a)}(xq^{-2})
\\
=
\Delta^{B_{m},\emptyset}_{F_{n},(s,s+1,\dots,s+a-1)}(x)
\Delta^{B_{m},\emptyset}_{F_{n},(s+2,s+3,\dots,s+a+1)}(xq^{-2})
\\
+(-1)^{a+1}
\Delta^{B_{m},(1)}_{F_{n},(s,s+1,\dots,s+a)}(x)
\Delta^{B_{m},\emptyset}_{F_{n},(1,s+2,s+3,\dots,s+a)}(xq^{-2}).
\label{toproveii}
\end{multline}
Let us expand the determinants in the second term of the right hand side of
 the above relation.
\begin{align}
&\hspace{-15pt} 
\Delta^{B_{m},(1)}_{F_{n},(s,s+1,\dots,s+a)}(x)
\Delta^{B_{m},\emptyset}_{F_{n},(1,s+2,s+3,\dots,s+a)}(xq^{-2})
 =
\sum_{\beta=1}^{n}(-1)^{\beta +m+1}
\Qs_{f_{\beta}}(xq^{-1})
\nonumber \\
& \times 
\Delta^{B_{m},\emptyset}_{F_{n}\backslash f_{\beta},(s,s+1,\dots,s+a)}(x)
\sum_{\alpha=1}^{m}(-1)^{\alpha +n+1}
\Qs_{b_{\alpha}}(xq^{-1})
\Delta^{B_{m}\backslash b_{\alpha},\emptyset}_{F_{n},(s+2,s+3,\dots,s+a)}(xq^{-2})
\\[4pt]
&\hspace{-10pt} 
=\sum_{\alpha=1}^{m}\sum_{\beta=1}^{n}(-1)^{\alpha+\beta + n+m} 
\frac{z_{b_{\alpha}}\Qs_{\{b_{\alpha},f_{\beta} \}}(x)-
z_{f_{\beta}}\Qs_{\{b_{\alpha},f_{\beta} \}}(xq^{-2})}{z_{b_{\alpha}}-z_{f_{\beta}}}
\nonumber \\
& \times 
\Delta^{B_{m}\backslash b_{\alpha},\emptyset}_{F_{n},(s+2,s+3,\dots,s+a)}(xq^{-2})
\Delta^{B_{m},\emptyset}_{F_{n}\backslash f_{\beta},(s,s+1,\dots,s+a)}(x)
\\[4pt]
& \hspace{-10pt} =-\sum_{\beta=1}^{n}(-1)^{\beta +m+1} z_{f_{\beta}}
\Delta^{B_{m},\emptyset}_{F_{n}\backslash f_{\beta},(s,s+1,\dots,s+a)}(x)
\nonumber \\
& \times 
\begin{vmatrix}
\left(\frac{\Qs_{\{b_{k},f_{l}\}}(xq^{-2})}{z_{b_{k}}-z_{f_{l}}}
 \right)_{\genfrac{}{}{0pt}{2}{1 \le k \le m,}{1 \le l \le n}}
 & 
\left(\frac{\Qs_{\{b_{k},f_{\beta}\}}(xq^{-2})}{z_{b_{k}}-z_{f_{\beta}}}\right)_{1 \le k \le m}
 & 
\left(z_{b_{k}}^{s+l}\Qs_{b_{k}}(xq^{2(s+l)-1}) 
\right)_{\genfrac{}{}{0pt}{2}{1 \le k \le m,}{1 \le l \le a-1}}
\end{vmatrix} 
\nonumber \\[3pt]
& + \sum_{\alpha=1}^{m} (-1)^{\alpha + n+1} z_{b_{\alpha }}
\Delta^{B_{m}\backslash b_{\alpha},\emptyset}_{F_{n},(s+2,s+3,\dots,s+a)}(xq^{-2})
\nonumber \\
& \times 
\begin{vmatrix}
\left(\frac{\Qs_{\{b_{k},f_{l}\}}(x)}{z_{b_{k}}-z_{f_{l}}}
 \right)_{\genfrac{}{}{0pt}{2}{1 \le k \le m,}{1 \le l \le n}}
 & 
\left(z_{b_{k}}^{s+l-2}\Qs_{b_{k}}(xq^{2(s+l)-3}) 
\right)_{\genfrac{}{}{0pt}{2}{1 \le k \le m,}{1 \le l \le a+1}}\\[10pt]
\left(\frac{\Qs_{\{b_{\alpha},f_{l}\}}(x)}{z_{b_{\alpha}}-z_{f_{l}}}\right)_{1 \le l \le n}
& (0)_{1 \times (a+1)}
\end{vmatrix} 
, \label{defoii}
\end{align}
where we introduced notations 
$B_{m}\backslash b_{\alpha}=(b_{1},\dots,b_{\alpha-1},b_{\alpha+1},\dots,b_{m})$, 
$F_{n}\backslash f_{\beta}=(f_{1},\dots,f_{\beta-1},f_{\beta+1},\dots,f_{n})$, and 
used the functional relation (\ref{QQ-bf1}) . 
The determinant $\begin{vmatrix}\cdots  \end{vmatrix}$ in the first summand in the right hand side of \eqref{defoii} vanishes as it has two identical columns. 
Subtracting the $\alpha$-th row from the $(m+1)$-st row in 
the determinant $\begin{vmatrix}\cdots  \end{vmatrix}$ in the second summand in 
the right hand side of \eqref{defoii}, we get: 
\begin{align}
& 
\begin{vmatrix}
\left(\frac{\Qs_{\{b_{k},f_{l}\}}(x)}{z_{b_{k}}-z_{f_{l}}}
 \right)_{\genfrac{}{}{0pt}{2}{1 \le k \le m,}{1 \le l \le n}}
 & 
\left(z_{b_{k}}^{s+l-2}\Qs_{b_{k}}(xq^{2(s+l)-3}) 
\right)_{\genfrac{}{}{0pt}{2}{1 \le k \le m,}{1 \le l \le a+1}}\\[12pt]
\left(\frac{\Qs_{\{b_{\alpha},f_{l}\}}(x)}{z_{b_{\alpha}}-z_{f_{l}}}\right)_{1 \le l \le n}
& (0)_{1 \times (a+1)}
\end{vmatrix}
\nonumber \\[5pt]
& \quad =
-
\begin{vmatrix}
\left(\frac{\Qs_{\{b_{k},f_{l}\}}(x)}{z_{b_{k}}-z_{f_{l}}}
 \right)_{\genfrac{}{}{0pt}{2}{1 \le k \le m,}{1 \le l \le n}}
 & 
\left(z_{b_{k}}^{s+l-2}\Qs_{b_{k}}(xq^{2(s+l)-3}) 
\right)_{\genfrac{}{}{0pt}{2}{1 \le k \le m,}{1 \le l \le a+1}}\\[12pt]
(0)_{1 \times n}
&
\left(z_{b_{\alpha}}^{s+l-2}\Qs_{b_{\alpha}}(xq^{2(s+l)-3}) 
\right)_{1 \le l \le a+1}
\end{vmatrix} 
\nonumber \\[5pt]
& \quad 
=-\sum_{\gamma=1}^{a+1}(-1)^{m+1+n+\gamma}
z_{b_{\alpha}}^{s+\gamma-2}\Qs_{b_{\alpha}}(xq^{2(s+\gamma)-3})
\Delta^{B_{m},\emptyset}_{F_{n},(s,s+1,\dots, s+a) \backslash (s+\gamma-1)}(x).
\end{align}
Here we expanded the $(m+1)$-st row. 
Thus \eqref{defoii} reduces to 
\begin{align}
& - \sum_{\alpha=1}^{m} (-1)^{\alpha + n+1} z_{b_{\alpha }}
\Delta^{B_{m}\backslash b_{\alpha},\emptyset}_{F_{n},(s+2,s+3,\dots,s+a)}(xq^{-2})
\nonumber \\
& \quad \times 
\sum_{\gamma=1}^{a+1}(-1)^{m+1+n+\gamma}
z_{b_{\alpha}}^{s+\gamma-2}\Qs_{b_{\alpha}}(xq^{2(s+\gamma)-3})
\Delta^{B_{m},\emptyset}_{F_{n},(s,s+1,\dots, s+a) \backslash (s+\gamma-1)}(x)
\nonumber \\
&= 
\sum_{\gamma=1}^{a+1}(-1)^{m+1+n+\gamma+1}
\Delta^{B_{m},\emptyset}_{F_{n},(s,s+1,\dots, s+a) \backslash (s+\gamma-1)}(x)
\nonumber \\
& 
\times 
\begin{vmatrix}
\left(\frac{\Qs_{\{b_{k},f_{l}\}}(xq^{-2})}{z_{b_{k}}-z_{f_{l}}}
 \right)_{\genfrac{}{}{0pt}{2}{1 \le k \le m,}{1 \le l \le n}}
 & \!
\left(
z_{b_{k}}^{s+\gamma-1}\Qs_{b_{k}}(xq^{2(s+\gamma )-3}) 
\right)_{1 \le k \le m}
 & \! \! 
\left(z_{b_{k}}^{s+l}\Qs_{b_{k}}(xq^{2(s+l)-1}) 
\right)_{\genfrac{}{}{0pt}{2}{1 \le k \le m,}{1 \le l \le a-1}}
\end{vmatrix} .
\label{defoii-2}
\end{align}
The determinant $\begin{vmatrix}\cdots \end{vmatrix}$
 in the right hand side of the above relation vanishes 
for $\gamma \ne 1,a+1$ as it has two identical columns. 
Then the right hand side of \eqref{defoii-2} reduces to 
\begin{multline}
(-1)^{a+1}\Delta^{B_{m},\emptyset}_{F_{n},(s+1,s+2,\dots, s+a)}(x)
\Delta^{B_{m},\emptyset}_{F_{n},(s+1,s+2,\dots, s+a)}(xq^{-2})
\\ -
(-1)^{a+1}\Delta^{B_{m},\emptyset}_{F_{n},(s,s+1,\dots, s+a-1)}(x)
\Delta^{B_{m},\emptyset}_{F_{n},(s+2,s+3,\dots, s+a+1)}(xq^{-2}). 
\end{multline}
This proves \eqref{toproveii}. \\\\
{\it (iii) The case} $a-s < m-n<a$ 

Due to \eqref{defe2}, 
\eqref{toprove} is identical to the following relation.
\begin{multline}
\Delta^{B_{m},(0,1,\dots,n-m+a-1)}_{F_{n},(s-a+m-n+2,s-a+m-n+3,\dots ,s+m-n+1)}(x)
\Delta^{B_{m},(1,2,\dots,n-m+a)}_{F_{n},(s-a+m-n+1,s-a+m-n+2,\dots ,s+m-n)}(x)
\\[2pt]
=
\Delta^{B_{m},(0,1,\dots,n-m+a-1)}_{F_{n},(s-a+m-n+1,s-a+m-n+2,\dots ,s+m-n)}(x)
\Delta^{B_{m},(1,2,\dots,n-m+a)}_{F_{n},(s-a+m-n+2,s-a+m-n+3,\dots ,s+m-n+1)}(x)
\\[2pt]
-
\Delta^{B_{m},(0,1,\dots,n-m+a)}_{F_{n},(s-a+m-n+1,s-a+m-n+2,\dots ,s+m-n+1)}(x)
\Delta^{B_{m},(1,2,\dots,n-m+a-1)}_{F_{n},(s-a+m-n+2,s-a+m-n+3,\dots ,s+m-n)}(x).
\end{multline}
This is nothing but the Jacobi identity \eqref{jacobi}. \\\\
{\it (iv) The case} $a-s = m-n $ 

\eqref{toprove} is identical to the following relation.
\begin{multline}
\Delta^{B_{m},(1,2,\dots,s)}_{F_{n},(1,2,\dots ,a)}(x)
\Delta^{B_{m},(1,2,\dots,s)}_{F_{n},(1,2,\dots ,a)}(xq^{-2})\\
=(-1)^{a+s+1}
\Delta^{B_{m},(2,3,\dots,s)}_{F_{n},(1,2,\dots ,a-1)}(x)
\Delta^{B_{m},(1,2,\dots,s)}_{F_{n},(2,3,\dots ,a+1)}(xq^{-2}) \\
+(-1)^{a+s+1}
\Delta^{B_{m},(2,3,\dots,s+1)}_{F_{n},(1,2,\dots ,a)}(x)
\Delta^{B_{m},(1,2,\dots,s-1)}_{F_{n},(2,3,\dots ,a)}(xq^{-2}).
\label{toproveiv}
\end{multline}
Let us expand the determinants in the left hand side of the above relation \eqref{toproveiv}, 
and apply \eqref{QQ-bf1}.
\begin{align}
& \hspace{-10pt}
\Delta^{B_{m},(1,2,\dots,s)}_{F_{n},(1,2,\dots ,a)}(x)
\Delta^{B_{m},(1,2,\dots,s)}_{F_{n},(1,2,\dots ,a)}(xq^{-2})  
=\sum_{\alpha=1}^{n}(-1)^{\alpha +m+1}\Qs_{f_{\alpha}}(xq^{-1})
\Delta^{B_{m},(2,3,\dots,s)}_{F_{n}\backslash f_{\alpha},(1,2,\dots ,a)}(x)
\nonumber \\
& \hspace{170pt} \times 
\sum_{\beta=1}^{m}(-1)^{\beta +n+1}\Qs_{b_{\beta}}(xq^{-1})
\Delta^{B_{m}\backslash b_{\beta},(1,2,\dots,s)}_{F_{n},(2,3,\dots ,a)}(xq^{-2})
 \nonumber \\[3pt]
& =\sum_{\alpha=1}^{n} \sum_{\beta=1}^{m}(-1)^{\alpha+\beta +m+n} 
\frac{z_{b_{\beta}}\Qs_{\{b_{\beta},f_{\alpha}\}}(x)-
z_{f_{\alpha}}\Qs_{\{b_{\beta},f_{\alpha}\}}(xq^{-2})}{z_{b_{\beta}}-z_{f_{\alpha}}}
\nonumber \\
& \qquad \times 
\Delta^{B_{m},(2,3,\dots,s)}_{F_{n}\backslash f_{\alpha},(1,2,\dots ,a)}(x)
\Delta^{B_{m}\backslash b_{\beta},(1,2,\dots,s)}_{F_{n},(2,3,\dots ,a)}(xq^{-2})
 \nonumber \\[3pt]
& = 
\sum_{\beta=1}^{m}(-1)^{\beta +n+1} z_{ b_{\beta}}
\Delta^{B_{m}\backslash b_{\beta},(1,2,\dots,s)}_{F_{n},(2,3,\dots ,a)}(xq^{-2})
\nonumber \\
& \qquad \times 
\begin{vmatrix}
\left(\frac{\Qs_{\{b_{k},f_{l}\}}(x)}{z_{b_{k}}-z_{f_{l}}}
 \right)_{\genfrac{}{}{0pt}{2}{1 \le k \le m,}{1 \le l \le n}}
 & 
\left(z_{b_{k}}^{l-1}\Qs_{b_{k}}(xq^{2l-1}) 
\right)_{\genfrac{}{}{0pt}{2}{1 \le k \le m,}{1 \le l \le a}}\\[12pt]
\left(\frac{\Qs_{\{b_{\beta},f_{l}\}}(x)}{z_{b_{\beta}}-z_{f_{l}}}
 \right)_{1 \le l \le n}
&
(0)_{1 \times a} \\[12pt]
\left((-z_{f_{l}})^{k-1}\Qs_{f_{l}}(xq^{-2k+1}) 
\right)_{\genfrac{}{}{0pt}{2}{2 \le k \le s,}{1 \le l \le n}}
&
(0)_{(s-1) \times a}
\end{vmatrix}
\nonumber \\[2pt]
& \quad -\sum_{\alpha=1}^{n} (-1)^{\alpha +m+1} z_{f_{\alpha}}
\Delta^{B_{m},(2,3,\dots,s)}_{F_{n}\backslash f_{\alpha},(1,2,\dots ,a)}(x)
\nonumber \\
& \times 
\begin{vmatrix}
\left(\frac{\Qs_{\{b_{k},f_{l}\}}(xq^{-2})}{z_{b_{k}}-z_{f_{l}}}
 \right)_{\genfrac{}{}{0pt}{2}{1 \le k \le m,}{1 \le l \le n}}
 & 
\left(\frac{\Qs_{\{b_{k},f_{\alpha}\}}(xq^{-2})}{z_{b_{k}}-z_{f_{\alpha}}}
 \right)_{1 \le k \le m}
 & 
\left(z_{b_{k}}^{l-1}\Qs_{b_{k}}(xq^{2l-3}) 
\right)_{\genfrac{}{}{0pt}{2}{1 \le k \le m,}{2 \le l \le a}}\\[14pt]
\left((-z_{f_{l}})^{k-1}\Qs_{f_{l}}(xq^{-2k-1}) 
\right)_{\genfrac{}{}{0pt}{2}{1 \le k \le s,}{1 \le l \le n}}
&
(0)_{s \times 1}
&
(0)_{s \times (a-1)}
\end{vmatrix}.
\label{defoiii}
\end{align}
Subtracting the $\beta $-th row from the $(m+1)$-st row in 
the determinant $\begin{vmatrix} \cdots \end{vmatrix}$ in the first summand in the right hand side of 
\eqref{defoiii}, we obtain: 
\begin{align}
& \hspace{-15pt}
\begin{vmatrix}
\left(\frac{\Qs_{\{b_{k},f_{l}\}}(x)}{z_{b_{k}}-z_{f_{l}}}
 \right)_{\genfrac{}{}{0pt}{2}{1 \le k \le m,}{1 \le l \le n}}
 & 
\left(z_{b_{k}}^{l-1}\Qs_{b_{k}}(xq^{2l-1}) 
\right)_{\genfrac{}{}{0pt}{2}{1 \le k \le m,}{1 \le l \le a}}\\[12pt]
\left(\frac{\Qs_{\{b_{\beta},f_{l}\}}(x)}{z_{b_{\beta}}-z_{f_{l}}}
 \right)_{1 \le l \le n}
&
(0)_{1 \times a} \\[12pt]
\left((-z_{f_{l}})^{k-1}\Qs_{f_{l}}(xq^{-2k+1}) 
\right)_{\genfrac{}{}{0pt}{2}{2 \le k \le s,}{1 \le l \le n}}
&
(0)_{(s-1) \times a}
\end{vmatrix}
\nonumber \\[8pt]
&= -
\begin{vmatrix}
\left(\frac{\Qs_{\{b_{k},f_{l}\}}(x)}{z_{b_{k}}-z_{f_{l}}}
 \right)_{\genfrac{}{}{0pt}{2}{1 \le k \le m,}{1 \le l \le n}}
 & 
\left(z_{b_{k}}^{l-1}\Qs_{b_{k}}(xq^{2l-1}) 
\right)_{\genfrac{}{}{0pt}{2}{1 \le k \le m,}{1 \le l \le a}}\\[12pt]
(0)_{1 \times n}
&
\left(z_{b_{\beta }}^{l-1}\Qs_{b_{\beta }}(xq^{2l-1}) 
\right)_{1 \le l \le a}
 \\[12pt]
\left((-z_{f_{l}})^{k-1}\Qs_{f_{l}}(xq^{-2k+1}) 
\right)_{\genfrac{}{}{0pt}{2}{2 \le k \le s,}{1 \le l \le n}}
&
(0)_{(s-1) \times a}
\end{vmatrix}
\nonumber \\[4pt]
&= -
\sum_{\gamma=1}^{a}(-1)^{m+1+n+\gamma} z_{ b_{\beta}}^{\gamma-1}
\Qs_{b_{\beta }}(xq^{2\gamma-1}) 
\Delta^{B_{m},(2,3,\dots,,s)}_{F_{n},(1,2,\dots ,a)\backslash \gamma }(x).
\end{align}
Subtracting the $\alpha $-th column from the $(n+1)$-st column in 
the determinant $\begin{vmatrix} \cdots \end{vmatrix}$ in the second summand in the right hand side of 
\eqref{defoiii}, we obtain
\begin{align}
& 
\begin{vmatrix}
\left(\frac{\Qs_{\{b_{k},f_{l}\}}(xq^{-2})}{z_{b_{k}}-z_{f_{l}}}
 \right)_{\genfrac{}{}{0pt}{2}{1 \le k \le m,}{1 \le l \le n}}
 & 
\left(\frac{\Qs_{\{b_{k},f_{\alpha}\}}(xq^{-2})}{z_{b_{k}}-z_{f_{\alpha}}}
 \right)_{1 \le k \le m}
 & 
\left(z_{b_{k}}^{l-1}\Qs_{b_{k}}(xq^{2l-3}) 
\right)_{\genfrac{}{}{0pt}{2}{1 \le k \le m,}{2 \le l \le a}}\\[14pt]
\left((-z_{f_{l}})^{k-1}\Qs_{f_{l}}(xq^{-2k-1}) 
\right)_{\genfrac{}{}{0pt}{2}{1 \le k \le s,}{1 \le l \le n}}
&
(0)_{s \times 1}
&
(0)_{s \times (a-1)}
\end{vmatrix}
=
\nonumber \\[10pt]
&-
\begin{vmatrix}
\left(\frac{\Qs_{\{b_{k},f_{l}\}}(xq^{-2})}{z_{b_{k}}-z_{f_{l}}}
 \right)_{\genfrac{}{}{0pt}{2}{1 \le k \le m,}{1 \le l \le n}}
 & \! \! 
(0)_{m \times 1}
 & 
 \hspace{-27pt}
\left( z_{b_{k}}^{l-1}\Qs_{b_{k}}(xq^{2l-3}) 
\right)_{\genfrac{}{}{0pt}{2}{1 \le k \le m,}{2 \le l \le a}}\\[14pt]
\left((-z_{f_{l}})^{k-1}\Qs_{f_{l}}(xq^{-2k-1}) 
\right)_{\genfrac{}{}{0pt}{2}{1 \le k \le s,}{1 \le l \le n}}
& \! \! 
\left((-z_{f_{\alpha }})^{k-1}\Qs_{f_{\alpha }}(xq^{-2k-1}) 
\right)_{1 \le k \le s}
& \! \! 
(0)_{s \times (a-1)}
\end{vmatrix}
\nonumber \\[10pt]
&=-\sum_{\gamma=1}^{s}(-1)^{m+\gamma+n+1}
(-z_{f_{\alpha }})^{\gamma-1}\Qs_{f_{\alpha }}(xq^{-2\gamma -1})
\Delta^{B_{m},(1,2,\dots,s)\backslash \gamma}_{F_{n},(2,3,\dots ,a) }(xq^{-2}).
\end{align}
Then \eqref{defoiii} reduces to 
\begin{align}
 &-
\sum_{\beta=1}^{m}(-1)^{\beta +n+1} z_{ b_{\beta}}
\Delta^{B_{m}\backslash b_{\beta},(1,2,\dots,s)}_{F_{n},(2,3,\dots ,a)}(xq^{-2})
\nonumber \\
& \qquad \times 
\sum_{\gamma=1}^{a}(-1)^{m+1+n+\gamma} z_{ b_{\beta}}^{\gamma-1}
\Qs_{b_{\beta }}(xq^{2\gamma-1}) 
\Delta^{B_{m},(2,3,\dots,,s)}_{F_{n},(1,2,\dots ,a)\backslash \gamma }(x)
\nonumber \\
 & \quad +\sum_{\alpha=1}^{n} (-1)^{\alpha +m+1} z_{f_{\alpha}}
\Delta^{B_{m},(2,3,\dots,s)}_{F_{n}\backslash f_{\alpha},(1,2,\dots ,a)}(x)
\nonumber \\
& \qquad \times 
\sum_{\gamma=1}^{s}(-1)^{m+\gamma+n+1}
(-z_{f_{\alpha }})^{\gamma-1}\Qs_{f_{\alpha }}(xq^{-2\gamma -1})
\Delta^{B_{m},(1,2,\dots,s)\backslash \gamma}_{F_{n},(2,3,\dots ,a) }(xq^{-2})
\nonumber \\[4pt]
& =
\sum_{\gamma=1}^{a}(-1)^{m+n+\gamma} 
\Delta^{B_{m},(2,3,\dots,,s)}_{F_{n},(1,2,\dots ,a)\backslash \gamma }(x)
\nonumber \\[2pt] 
& \times
\begin{vmatrix}
\left(\frac{\Qs_{\{b_{k},f_{l}\}}(xq^{-2})}{z_{b_{k}}-z_{f_{l}}}
 \right)_{\genfrac{}{}{0pt}{2}{1 \le k \le m,}{1 \le l \le n}}
 & 
\left( z_{b_{k}}^{\gamma }\Qs_{b_{k}}(xq^{2\gamma-1}) 
 \right)_{1 \le k \le m}
 & 
\left(z_{b_{k}}^{l-1}\Qs_{b_{k}}(xq^{2l-3}) 
\right)_{\genfrac{}{}{0pt}{2}{1 \le k \le m,}{2 \le l \le a}}\\[14pt]
\left((-z_{f_{l}})^{k-1}\Qs_{f_{l}}(xq^{-2k-1}) 
\right)_{\genfrac{}{}{0pt}{2}{1 \le k \le s,}{1 \le l \le n}}
&
(0)_{s \times 1}
&
(0)_{s \times (a-1)}
\end{vmatrix}
\nonumber \\[3pt]
& \quad -
\sum_{\gamma=1}^{s}(-1)^{m+\gamma+n+1}
\Delta^{B_{m},(1,2,\dots,s)\backslash \gamma}_{F_{n},(2,3,\dots ,a) }(xq^{-2})
\nonumber \\
& \qquad \times 
\begin{vmatrix}
\left(\frac{\Qs_{\{b_{k},f_{l}\}}(x)}{z_{b_{k}}-z_{f_{l}}}
 \right)_{\genfrac{}{}{0pt}{2}{1 \le k \le m,}{1 \le l \le n}}
 & 
\left(z_{b_{k}}^{l-1}\Qs_{b_{k}}(xq^{2l-1}) 
\right)_{\genfrac{}{}{0pt}{2}{1 \le k \le m,}{1 \le l \le a}}\\[12pt]
\left( (-z_{f_{l}})^{\gamma }\Qs_{f_{l}}(xq^{-2\gamma -1}) 
 \right)_{1 \le l \le n}
&
(0)_{1 \times a} \\[10pt]
\left((-z_{f_{l}})^{k-1}\Qs_{f_{l}}(xq^{-2k+1}) 
\right)_{\genfrac{}{}{0pt}{2}{2 \le k \le s,}{1 \le l \le n}}
&
(0)_{(s-1) \times a}
\end{vmatrix}
. \label{defoiii-2}
\end{align}
The determinant $\begin{vmatrix} \cdots \end{vmatrix} $ in the first summand in the
 right hand side of \eqref{defoiii-2} vanishes for $\gamma \ne a $ as it has two identical  
 columns. 
 The determinant $\begin{vmatrix} \cdots \end{vmatrix} $ in the second summand in the
 right hand side of \eqref{defoiii-2} vanishes for $\gamma \ne s $.  
Then  the right hand side of \eqref{defoiii-2}  reduces to 
\begin{multline}
(-1)^{m+n+1} 
\Delta^{B_{m},(2,3,\dots,s)}_{F_{n},(1,2,\dots ,a-1) }(x)
\Delta^{B_{m},(1,2,\dots,s)}_{F_{n},(2,3,\dots ,a+1) }(xq^{-2})
 \\
 -(-1)^{m+n}
\Delta^{B_{m},(1,2,\dots,s-1)}_{F_{n},(2,3,\dots ,a) }(xq^{-2})
\Delta^{B_{m},(2,3,\dots,s+1)}_{F_{n},(1,2,\dots ,a) }(x)
. 
\end{multline}
This coincides with the right hand side of \eqref{toproveiv} since $a-s=m-n$. \\

\sapp{Proof of Theorem \ref{solution-back}}\label{proof-back}
We will prove Theorem \ref{solution-back} for \eqref{tfunrec1}. 
The proof for \eqref{tfunrec2} is similar to the one for \eqref{tfunrec1}.
For $a,s \in {\mathbb Z}_{\ge 0}$, we will prove that \eqref{tfunrec1}
 satisfies \eqref{bac1}. The case $s=0 $ is trivial. Then we consider 
$s \ge 1$ case. 
We have to consider the following six cases since the $(m,n)$-index \eqref{mn-index} depends on 
$m,n,a,s$: 
(i) $a+2 \le m-n$, 
(ii) $a+1 = m-n$, 
(iii)  $a-s+2 \le m-n \le a$, 
(iv) $a-s+1 = m-n $, 
(v) $-s+1 \le m-n \le a-s$,  
(vi) $m-n \le -s$. 
For $a=0$, the case (v) is void.  
\\
The proof of the other functional relations is similar to that of 
\eqref{bac1}. 
\\\\
{\it (i) The case} $a+2 \le m-n$ 

Due to \eqref{defe1}, 
\eqref{bac1} is identical to the following relation.
\begin{align}
&\Delta^{B_{m},\emptyset}_{F_{n},(0,1,\dots,m-n-a-2,m-n-a+s-1,\dots,m-n+s-1)}(x)
\Delta^{B_{m-1},\emptyset}_{F_{n},(1,2,\dots,m-n-a-1,m-n-a+s,\dots,m-n+s-1)}(x)
 \nonumber \\[2pt]
&-
\Delta^{B_{m},\emptyset}_{F_{n},(0,1,\dots,m-n-a-1,m-n-a+s,\dots,m-n+s-1)}(x)
\Delta^{B_{m-1},\emptyset}_{F_{n},(1,2,\dots,m-n-a-2,m-n-a+s-1,\dots,m-n+s-1)}(x)
\nonumber \\[2pt]
&=
\Delta^{B_{m},\emptyset}_{F_{n},(1,2,\dots,m-n-a-1,m-n-a+s-1,\dots,m-n+s-1)}(x)
\Delta^{B_{m-1},\emptyset}_{F_{n},(0,1,\dots,m-n-a-2,m-n-a+s,\dots,m-n+s-1)}(x)
.
\end{align}
For $a=0$ case, we must interpret
 $(1,2,\dots,m-n-a-1,m-n-a+s,\dots,m-n+s-1) $ as 
 $(1,2,\dots,m-n-1)$ etc. 
The above relation is nothing but the Pl\"ucker identity \eqref{plucker4}.
\\\\
{\it (ii) The case} $a+1 = m-n$

In this case, \eqref{bac1} has the form:  
\begin{multline}
\Delta^{B_{m},\emptyset}_{F_{n},(s+1,s+2,\dots,a+s+1)}(x)
\Delta^{B_{m-1},\emptyset}_{F_{n},(s+1,s+2,\dots,a+s)}(xq^{2})
\\
-(-1)^{m+n}
\Delta^{B_{m},\emptyset}_{F_{n},(1,s+2,s+3,\dots,a+s+1)}(x)
\Delta^{B_{m-1},(1)}_{F_{n},(s,s+1,\dots,a+s)}(xq^{2})
\\
= z_{b_{m}}
\Delta^{B_{m},\emptyset}_{F_{n},(s,s+1,\dots,a+s)}(xq^{2})
\Delta^{B_{m-1},\emptyset}_{F_{n},(s+2,s+3,\dots,a+s+1)}(x)
. \label{bac-ii}
\end{multline}
Let us expand the determinants in the second term of the left hand side of 
\eqref{bac-ii}:
\begin{align}
& \Delta^{B_{m},\emptyset}_{F_{n},(1,s+2,s+3,\dots,a+s+1)}(x)
\Delta^{B_{m-1},(1)}_{F_{n},(s,s+1,\dots,a+s)}(xq^{2})
\nonumber \\ 
&
=\sum_{\alpha =1}^{m}(-1)^{\alpha +n+1}\Qs_{b_{\alpha}}(xq)
\Delta^{B_{m} \backslash b_{\alpha},\emptyset}_{F_{n},(s+2,s+3,\dots,a+s+1)}(x)
\nonumber \\ 
& \qquad \times 
\sum_{\beta =1}^{n}(-1)^{m+\beta }\Qs_{f_{\beta}}(xq)
\Delta^{B_{m-1},\emptyset }_{F_{n} \backslash f_{\beta},(s,s+1,\dots,a+s)}(xq^{2})
\nonumber \\[4pt] 
&
= \sum_{\alpha =1}^{m}(-1)^{\alpha +n+1}
\sum_{\beta =1}^{n}(-1)^{m+\beta }
\frac{z_{b_{\alpha}}\Qs_{\{b_{\alpha},f_{\beta}\}}(xq^{2})-
z_{f_{\beta}} \Qs_{\{b_{\alpha},f_{\beta}\}}(x)}
{z_{b_{\alpha}}-z_{f_{\beta}}}
\nonumber \\
& \qquad \times 
\Delta^{B_{m} \backslash b_{\alpha},\emptyset}_{F_{n},(s+2,s+3,\dots,a+s+1)}(x)
\Delta^{B_{m-1},\emptyset }_{F_{n} \backslash f_{\beta},(s,s+1,\dots,a+s)}(xq^{2})
\nonumber \\[3pt]
&=
\sum_{\alpha=1}^{m} (-1)^{\alpha + n+1} z_{b_{\alpha }}
\Delta^{B_{m}\backslash b_{\alpha},\emptyset}_{F_{n},(s+2,s+3,\dots,s+a+1)}(x)
\nonumber \\
& \qquad \times 
\begin{vmatrix}
\left(\frac{\Qs_{\{b_{k},f_{l}\}}(xq^{2})}{z_{b_{k}}-z_{f_{l}}}
 \right)_{\genfrac{}{}{0pt}{2}{1 \le k \le m-1,}{1 \le l \le n}}
 & 
\left(z_{b_{k}}^{s+l-2}\Qs_{b_{k}}(xq^{2(s+l)-1}) 
\right)_{\genfrac{}{}{0pt}{2}{1 \le k \le m-1,}{1 \le l \le a+1}}\\[12pt]
\left(\frac{\Qs_{\{b_{\alpha},f_{l}\}}(xq^{2})}{z_{b_{\alpha}}-z_{f_{l}}}\right)_{1 \le l \le n}
& (0)_{1 \times (a+1)}
\end{vmatrix} 
\nonumber \\[2pt]
&\quad -
\sum_{\beta =1}^{n}(-1)^{m+\beta }
z_{f_{\beta }}
\Delta^{B_{m-1},\emptyset }_{F_{n} \backslash f_{\beta},(s,s+1,\dots,a+s)}(xq^{2})
\nonumber \\
& \times
\begin{vmatrix}
\left(\frac{\Qs_{\{b_{k},f_{l}\}}(x)}{z_{b_{k}}-z_{f_{l}}}
 \right)_{\genfrac{}{}{0pt}{2}{1 \le k \le m,}{1 \le l \le n}}
 & 
\left(\frac{\Qs_{\{b_{k},f_{\beta}\}}(x)}{z_{b_{k}}-z_{f_{\beta}}}\right)_{1 \le k \le m}
 & 
\left(z_{b_{k}}^{s+l}\Qs_{b_{k}}(xq^{2(s+l)+1}) 
\right)_{\genfrac{}{}{0pt}{2}{1 \le k \le m,}{1 \le l \le a}}
\end{vmatrix} 
, \label{bac-defoii}
\end{align}
where we introduced notations 
$B_{m}\backslash b_{\alpha}=(b_{1},\dots,b_{\alpha-1},b_{\alpha+1},\dots,b_{m})$, 
$F_{n}\backslash f_{\beta}=(f_{1},\dots,f_{\beta-1},f_{\beta+1},\dots,f_{n})$, and 
used the functional relation (\ref{QQ-bf1}) . 
The determinant $\begin{vmatrix}\cdots  \end{vmatrix}$ in the second summand in
 the right hand side of \eqref{bac-defoii} vanishes. 
Subtracting the $\alpha$-th row ($1 \le \alpha \le m-1$) from the $m$-th row in 
the determinant $\begin{vmatrix}\cdots  \end{vmatrix}$ in the first summand in 
the right hand side of \eqref{bac-defoii}, we get: 
\begin{align}
& \eqref{bac-defoii}=
\sum_{\alpha=1}^{m-1} (-1)^{\alpha + n+1} z_{b_{\alpha }}
\Delta^{B_{m}\backslash b_{\alpha},\emptyset}_{F_{n},(s+2,s+3,\dots,s+a+1)}(x)
\nonumber \\
& \qquad 
\times 
\begin{vmatrix}
\left(\frac{\Qs_{\{b_{k},f_{l}\}}(xq^{2})}{z_{b_{k}}-z_{f_{l}}}
 \right)_{\genfrac{}{}{0pt}{2}{1 \le k \le m-1,}{1 \le l \le n}}
 & 
\left(z_{b_{k}}^{s+l-2}\Qs_{b_{k}}(xq^{2(s+l)-1}) 
\right)_{\genfrac{}{}{0pt}{2}{1 \le k \le m-1,}{1 \le l \le a+1}}\\[12pt]
\left(\frac{\Qs_{\{b_{\alpha},f_{l}\}}(xq^{2})}{z_{b_{\alpha}}-z_{f_{l}}}\right)_{1 \le l \le n}
& (0)_{1 \times (a+1)}
\end{vmatrix} 
\nonumber \\[4pt]
& \quad  +
(-1)^{m + n+1} z_{b_{m }}
\Delta^{B_{m-1},\emptyset}_{F_{n},(s+2,s+3,\dots,s+a+1)}(x)
\nonumber \\
& \qquad \times 
\begin{vmatrix}
\left(\frac{\Qs_{\{b_{k},f_{l}\}}(xq^{2})}{z_{b_{k}}-z_{f_{l}}}
 \right)_{\genfrac{}{}{0pt}{2}{1 \le k \le m-1,}{1 \le l \le n}}
 & 
\left(z_{b_{k}}^{s+l-2}\Qs_{b_{k}}(xq^{2(s+l)-1}) 
\right)_{\genfrac{}{}{0pt}{2}{1 \le k \le m-1,}{1 \le l \le a+1}}\\[12pt]
\left(\frac{\Qs_{\{b_{m},f_{l}\}}(xq^{2})}{z_{b_{m}}-z_{f_{l}}}\right)_{1 \le l \le n}
& (0)_{1 \times (a+1)}
\end{vmatrix} 
\nonumber \\[5pt]
& =
-\sum_{\alpha=1}^{m-1} (-1)^{\alpha + n+1} z_{b_{\alpha }}
\Delta^{B_{m}\backslash b_{\alpha},\emptyset}_{F_{n},(s+2,s+3,\dots,s+a+1)}(x)
\nonumber \\
& \qquad 
\times 
\begin{vmatrix}
\left(\frac{\Qs_{\{b_{k},f_{l}\}}(xq^{2})}{z_{b_{k}}-z_{f_{l}}}
 \right)_{\genfrac{}{}{0pt}{2}{1 \le k \le m-1,}{1 \le l \le n}}
 & 
\left(z_{b_{k}}^{s+l-2}\Qs_{b_{k}}(xq^{2(s+l)-1}) 
\right)_{\genfrac{}{}{0pt}{2}{1 \le k \le m-1,}{1 \le l \le a+1}}
\\[12pt]
(0)_{1 \times n}
& 
\left(z_{b_{\alpha }}^{s+l-2}\Qs_{b_{\alpha }}(xq^{2(s+l)-1}) 
\right)_{1 \le l \le a+1}
\end{vmatrix} 
\nonumber  \\[4pt]
& \quad  +
(-1)^{m + n+1} z_{b_{m }}
\Delta^{B_{m-1},\emptyset}_{F_{n},(s+2,s+3,\dots,s+a+1)}(x)
\nonumber \\
&  \times 
\left\{
\Delta^{B_{m},\emptyset}_{F_{n},(s,s+1,\dots,s+a)}(xq^{2})-
\begin{vmatrix}
\left(\frac{\Qs_{\{b_{k},f_{l}\}}(xq^{2})}{z_{b_{k}}-z_{f_{l}}}
 \right)_{\genfrac{}{}{0pt}{2}{1 \le k \le m-1,}{1 \le l \le n}}
 & 
\left(z_{b_{k}}^{s+l-2}\Qs_{b_{k}}(xq^{2(s+l)-1}) 
\right)_{\genfrac{}{}{0pt}{2}{1 \le k \le m-1,}{1 \le l \le a+1}}\\[12pt]
(0)_{1 \times n}
& 
\left(z_{b_{m }}^{s+l-2}\Qs_{b_{m}}(xq^{2(s+l)-1}) 
\right)_{1 \le l \le a+1}
\end{vmatrix} 
\right\}
\nonumber \\[5pt]
& 
=
-\sum_{\alpha=1}^{m} (-1)^{\alpha + n+1} z_{b_{\alpha }}
\Delta^{B_{m}\backslash b_{\alpha},\emptyset}_{F_{n},(s+2,s+3,\dots,s+a+1)}(x)
\nonumber \\
& \qquad 
\times 
\begin{vmatrix}
\left(\frac{\Qs_{\{b_{k},f_{l}\}}(xq^{2})}{z_{b_{k}}-z_{f_{l}}}
 \right)_{\genfrac{}{}{0pt}{2}{1 \le k \le m-1,}{1 \le l \le n}}
 & 
\left(z_{b_{k}}^{s+l-2}\Qs_{b_{k}}(xq^{2(s+l)-1}) 
\right)_{\genfrac{}{}{0pt}{2}{1 \le k \le m-1,}{1 \le l \le a+1}}
\\[12pt]
(0)_{1 \times n}
& 
\left(z_{b_{\alpha }}^{s+l-2}\Qs_{b_{\alpha }}(xq^{2(s+l)-1}) 
\right)_{1 \le l \le a+1}
\end{vmatrix} 
\nonumber \\[4pt]
& \quad  +
(-1)^{m + n+1} z_{b_{m }}
\Delta^{B_{m-1},\emptyset}_{F_{n},(s+2,s+3,\dots,s+a+1)}(x)
\Delta^{B_{m},\emptyset}_{F_{n},(s,s+1,\dots,s+a)}(xq^{2})
.
\label{bac-defoii2}
\end{align}
Expanding the $m$-th row in the determinant in the right hand side of \eqref{bac-defoii2}, 
we obtain:
\begin{align}
& 
-\sum_{\alpha=1}^{m} (-1)^{\alpha + n+1} z_{b_{\alpha }}
\Delta^{B_{m}\backslash b_{\alpha},\emptyset}_{F_{n},(s+2,s+3,\dots,s+a+1)}(x)
\nonumber \\
& \qquad 
\times 
\sum_{\gamma =1}^{a+1}(-1)^{m+n+\gamma}
z_{b_{\alpha }}^{s+\gamma -2}\Qs_{b_{\alpha }}(xq^{2(s+\gamma)-1})
\Delta^{B_{m-1},\emptyset}_{F_{n},(s,s+1,\dots,s+a) \backslash (s+\gamma -1)}(xq^{2})
\nonumber \\
& \quad  +
(-1)^{m + n+1} z_{b_{m }}
\Delta^{B_{m-1},\emptyset}_{F_{n},(s+2,s+3,\dots,s+a+1)}(x)
\Delta^{B_{m},\emptyset}_{F_{n},(s,s+1,\dots,s+a)}(xq^{2})
\nonumber \\[5pt]
&= - \sum_{\gamma =1}^{a+1}(-1)^{m+n+\gamma}
\Delta^{B_{m-1},\emptyset}_{F_{n},(s,s+1,\dots,s+a) \backslash (s+\gamma -1)}(xq^{2})
\nonumber \\
& \times 
\begin{vmatrix}
\left(\frac{\Qs_{\{b_{k},f_{l}\}}(x)}{z_{b_{k}}-z_{f_{l}}}
 \right)_{\genfrac{}{}{0pt}{2}{1 \le k \le m,}{1 \le l \le n}}
& 
\left(z_{b_{k }}^{s+\gamma-1}\Qs_{b_{k }}(xq^{2(s+\gamma )-1}) 
\right)_{1 \le k \le m}
 & 
\left(z_{b_{k}}^{s+l}\Qs_{b_{k}}(xq^{2(s+l)+1}) 
\right)_{\genfrac{}{}{0pt}{2}{1 \le k \le m,}{1 \le l \le a}}
\end{vmatrix} 
\nonumber \\[3pt]
& \quad  +
(-1)^{m + n+1} z_{b_{m }}
\Delta^{B_{m-1},\emptyset}_{F_{n},(s+2,s+3,\dots,s+a+1)}(x)
\Delta^{B_{m},\emptyset}_{F_{n},(s,s+1,\dots,s+a)}(xq^{2})
\label{bac-ii-saigo} \\[5pt]
&= 
 (-1)^{m+n}
\Delta^{B_{m-1},\emptyset}_{F_{n},(s+1,\dots,s+a)}(xq^{2})
\Delta^{B_{m},\emptyset}_{F_{n},(s+1,s+2,\dots,s+a+1)}(x)
\nonumber \\
& 
\quad -
(-1)^{m + n} z_{b_{m }}
\Delta^{B_{m-1},\emptyset}_{F_{n},(s+2,s+3,\dots,s+a+1)}(x)
\Delta^{B_{m},\emptyset}_{F_{n},(s,s+1,\dots,s+a)}(xq^{2})
.
\end{align}
The determinant $\begin{vmatrix}\cdots  \end{vmatrix}$ 
in the right hand side of \eqref{bac-ii-saigo} vanished for 
$\gamma \ne 1$. 
Then we obtain \eqref{bac-ii}.
\\\\
{\it (iii) The case} $a-s+2 \le m-n \le a$ 

Due to \eqref{defe2}, 
\eqref{bac1} is identical to the following relation.
\begin{multline}
\Delta^{B_{m},(1,2,\dots, n-m+a+1)}_{F_{n},(m-n-a+s,m-n-a+s+1,\dots,m-n+s)}(x)
\Delta^{B_{m-1},(0,1,\dots, n-m+a)}_{F_{n},(m-n-a+s+1,m-n-a+s+2,\dots,m-n+s)}(x)
\\
-
\Delta^{B_{m},(1,2,\dots, n-m+a)}_{F_{n},(m-n-a+s+1,m-n-a+s+2,\dots,m-n+s)}(x)
\Delta^{B_{m-1},(0,1,\dots, n-m+a+1)}_{F_{n},(m-n-a+s,m-n-a+s+1,\dots,m-n+s)}(x)
\\[3pt]
=
\Delta^{B_{m},(0,1,\dots, n-m+a)}_{F_{n},(m-n-a+s,m-n-a+s+1,\dots,m-n+s)}(x)
\Delta^{B_{m-1},(1,2,\dots, n-m+a+1)}_{F_{n},(m-n-a+s+1,m-n-a+s+2,\dots,m-n+s)}(x)
.
\end{multline}
This is nothing but the Pl\"ucker identity \eqref{plucker3}.
\\\\
{\it (iv) The case} $a-s+1 = m-n $ 

In this case, \eqref{bac1} has the form:  
\begin{multline}
\Delta^{B_{m},(1,2,\dots,s)}_{F_{n},(1,2,\dots,a+1)}(x)
\Delta^{B_{m-1},(1,2,\dots,s)}_{F_{n},(1,2,\dots,a)}(xq^{2})
=
(-1)^{a+s}
\Delta^{B_{m},(1,2,\dots,s-1)}_{F_{n},(2,3,\dots,a+1)}(x)
\Delta^{B_{m-1},(2,3,\dots,s+1)}_{F_{n},(1,2,\dots,a)}(xq^{2})
\\
+(-1)^{a+s}z_{b_{m}}
\Delta^{B_{m},(2,3,\dots,s)}_{F_{n},(1,2,\dots,a)}(xq^{2})
\Delta^{B_{m-1},(1,2,\dots,s)}_{F_{n},(2,3,\dots,a+1)}(x)
. \label{bac-pro-iv}
\end{multline}
Let us expand the determinant 
$\Delta^{B_{m},(1,2,\dots,s)}_{F_{n},(1,2,\dots,a+1)}(x)$ 
with respect to the $(n+1)$-st column, and 
the determinant $\Delta^{B_{m-1},(1,2,\dots,s)}_{F_{n},(1,2,\dots,a)}(xq^{2}) $
with respect to the $m$-th row. After some calculations 
similar to the ones in the case (ii), 
 we arrive at the right hand side of 
\eqref{bac-pro-iv}.
\\\\
{\it (v) The case} $-s+1 \le m-n \le a-s$ 

Due to \eqref{defe1}, 
\eqref{bac1} is identical to the following relation.
\begin{align}
& \Delta^{B_{m},(n-m-s+a+3,n-m-s+a+4,\dots, n-m+a+2)}_{F_{n},(0,1,\dots, m-n+s-1)}(x)
\Delta^{B_{m-1},(n-m-s+a+2,n-m-s+a+3,\dots, n-m+a+1)}_{F_{n},(1,2,\dots, m-n+s-1)}(x)
 \nonumber \\
& -
\Delta^{B_{m},(n-m-s+a+2,n-m-s+a+3,\dots, n-m+a+1)}_{F_{n},(0,1,\dots, m-n+s-1)}(x)
\Delta^{B_{m-1},(n-m-s+a+3,n-m-s+a+4,\dots, n-m+a+2)}_{F_{n},(1,2,\dots, m-n+s-1)}(x)
\nonumber \\[3pt]
& =
\Delta^{B_{m},(n-m-s+a+3,n-m-s+a+4,\dots, n-m+a+1)}_{F_{n},(1,2,\dots, m-n+s-1)}(x)
\nonumber \\
& 
 \quad \times 
\Delta^{B_{m-1},(n-m-s+a+2,n-m-s+a+3,\dots, n-m+a+2)}_{F_{n},(0,1,\dots, m-n+s-1)}(x)
.
\end{align}
This is nothing but the Pl\"ucker identity \eqref{plucker3}.
\\\\
{\it (vi) The case} $m-n \le -s$

Due to \eqref{defe2}, 
\eqref{bac1} is identical to the following relation.
\begin{align}
\begin{split}
& \Delta^{B_{m},(1,2,\dots,n-m-s,n-m-s+a+2,n-m-s+a+3,\dots, n-m+a+1)}_{F_{n},\emptyset}(x)
\\ 
& \quad \times \Delta^{B_{m-1},(0,1,\dots,n-m-s,n-m-s+a+1,n-m-s+a+2,\dots, n-m+a)}_{F_{n},\emptyset}(x)
\\[3pt]
&  -
\Delta^{B_{m},(1,2,\dots,n-m-s,n-m-s+a+1,n-m-s+a+2,\dots, n-m+a)}_{F_{n},\emptyset}(x)
\\ 
&  \quad \times 
\Delta^{B_{m-1},(0,1,\dots,n-m-s,n-m-s+a+2,n-m-s+a+3,\dots, n-m+a+1)}_{F_{n},\emptyset}(x)
\\[3pt]
&  +
\Delta^{B_{m},(0,1,\dots,n-m-s,n-m-s+a+2,n-m-s+a+3,\dots, n-m+a)}_{F_{n},\emptyset}(x)
\\ 
&  \quad \times 
\Delta^{B_{m-1},(1,2,\dots,n-m-s,n-m-s+a+1,n-m-s+a+2,\dots, n-m+a+1)}_{F_{n},\emptyset}(x)
=0.
\end{split}
\end{align}
This is nothing but the Pl\"ucker identity \eqref{plucker}. 

\app{Conserved quantities and Baxter equations}
\label{Conserved}
In this section, we will present conserved 
quantities and generalized Baxter equations in determinant form 
 based on a similar idea in 
 Theorem 5.1 and Lemma 5.1 in \cite{KLWZ97}. 
\\\\
{\em (i) Formulae from} \eqref{lapQQb1}\\
Here we assume that the function $\Ts_{s}^{(a),B_{m},F_{n}}(x)$ is {\em defined} by 
the right hand side of \eqref{lapQQb1}. 
A condition $a-s \le m-n$ was supposed in the left hand side of \eqref{lapQQb1}, 
but now this condition is relaxed to $s \in {\mathbb C}$ and $a \in {\mathbb Z}_{\ge 0}$. 
Let us introduce $ (\binom{m}{a}+1)\times (\binom{m}{a}+1)$ matrices 
${\mathcal I}_{s}^{(a),B_{m},F_{n}}(x)$
(resp.\ ${\mathcal J}_{s}^{(a),B_{m},F_{n}}(x)$)
 whose $(i,j)$ element 
is given by 
$\Ts_{s+i+j}^{(a),B_{m},F_{n}}(xq^{-s+i-j})$, 
(resp.\ $\Ts_{s-i-j}^{(a),B_{m},F_{n}}(xq^{s+i-j})$)
where $i,j \in \{1,2,\dots, \binom{m}{a}+1 \}$. 
For any square matrix ${\mathcal M}$, we will write a minor determinant whose 
$\alpha $-th row and $\beta$-th column removed from 
${\mathcal M}$ as 
$D\begin{bmatrix}
\alpha \\
\beta 
\end{bmatrix}
({\mathcal M})$. 
Then we obtain conserved quantities as follows. 
\begin{theorem}\label{th-cons1}
For any $a \in {\mathbb Z}_{\ge 0}$ and 
$\alpha,\beta,\gamma \in \{1,2,\dots, \binom{m}{a}+1 \}$, the following quantities 
\begin{align}
\frac{D
\begin{bmatrix} 
\alpha \\
\beta  
\end{bmatrix}
({\mathcal I}_{s}^{(a),B_{m},F_{n}}(x))}{
D
\begin{bmatrix} 
\gamma \\
\beta  
\end{bmatrix}
({\mathcal I}_{s}^{(a),B_{m},F_{n}}(x))}
, \qquad 
\frac{D
\begin{bmatrix} 
\alpha \\
\beta  
\end{bmatrix}
({\mathcal J}_{s}^{(a),B_{m},F_{n}}(x))}{
D
\begin{bmatrix} 
\gamma \\
\beta  
\end{bmatrix}
({\mathcal J}_{s}^{(a),B_{m},F_{n}}(x))}
\end{align}
does not depend on $s$.
\end{theorem}
Let us replace $r$-th column of ${\mathcal I}_{s}^{(a),B_{m},F_{n}}(x)$ 
(resp.\ ${\mathcal J}_{s}^{(a),B_{m},F_{n}}(x)$)
with the column vector whose $i$-th component is $
(\prod_{\gamma \in I}z_{\gamma})^{i}\Qs_{I}(xq^{2i+\frac{m-n}{2}})$ 
(resp.\ $(\prod_{\gamma \in I}z_{\gamma})^{-i}
\Qs_{B_{m}\sqcup F_{n} \setminus I}(xq^{2i-\frac{m-n}{2}})$) and write it as 
${\mathcal I}_{s;r,I}^{(a),B_{m},F_{n}}(x)$ 
(resp.\ ${\mathcal J}_{s;r,I}^{(a),B_{m},F_{n}}(x)$), 
where $i \in \{1,2,\dots,\binom{m}{a}+1 \}$, $I \subset B_{m}$ and 
${\rm Card}(I)=a$. 
Then we obtain the following lemma.
\begin{lemma}
For any $a \in {\mathbb Z}_{\ge 0}$, $s \in {\mathbb C}$ and 
$r \in \{1,2,\dots,\binom{m}{a}+1 \}$, the following relations hold. 
\begin{align}
& {\rm det}({\mathcal I}_{s}^{(a),B_{m},F_{n}}(x))=
{\rm det}({\mathcal J}_{s}^{(a),B_{m},F_{n}}(x))= 0, 
\\[5pt]
& {\rm det}({\mathcal I}_{s;r,I}^{(a),B_{m},F_{n}}(x))=
{\rm det}({\mathcal J}_{s;r,I}^{(a),B_{m},F_{n}}(x))=0 .
\label{baxterEQ-general}
\end{align}
\end{lemma}
(\ref{baxterEQ-general}) for $(m,n)=(M,N)$ corresponds to a $U_{q}(\hat{gl}(M|N))$ 
generalization of the Baxter equation. 
\\\\
{\em (ii) Formulae from} \eqref{lapQQb3} \\
Here we assume that the function $\Ts_{s}^{(a),B_{m},F_{n}}(x)$ is {\em defined} by 
the right hand side of \eqref{lapQQb3}. 
A condition $a-s \ge m-n$ was supposed in the left hand side of \eqref{lapQQb3}, 
but now this condition is relaxed to $a \in {\mathbb C}$ and $s \in {\mathbb Z}_{\ge 0}$. 
Let us introduce $ (\binom{n}{s}+1)\times (\binom{n}{s}+1)$ matrices 
${\mathcal K}_{s}^{(a),B_{m},F_{n}}(x)$
(resp.\ ${\mathcal L}_{s}^{(a),B_{m},F_{n}}(x)$)
 whose $(i,j)$ element 
is given by 
$\Ts_{s}^{(a+i+j),B_{m},F_{n}}(xq^{a-i+j})$, 
(resp.\ $\Ts_{s}^{(a-i-j),B_{m},F_{n}}(xq^{-a-i+j})$)
where $i,j \in \{1,2,\dots, \binom{n}{s}+1 \}$. 
%
%
Then we obtain conserved quantities as follows.
\begin{theorem}\label{th-cons3}
For any $s \in {\mathbb Z}_{\ge 0}$ and 
$\alpha,\beta,\gamma \in \{1,2,\dots, \binom{n}{s}+1 \}$, the following 
quantities 
\begin{align}
\frac{D
\begin{bmatrix} 
\alpha \\
\beta  
\end{bmatrix}
({\mathcal K}_{s}^{(a),B_{m},F_{n}}(x))}{
D
\begin{bmatrix} 
\gamma \\
\beta  
\end{bmatrix}
({\mathcal K}_{s}^{(a),B_{m},F_{n}}(x))}
, \qquad 
\frac{D
\begin{bmatrix} 
\alpha \\
\beta  
\end{bmatrix}
({\mathcal L}_{s}^{(a),B_{m},F_{n}}(x))}{
D
\begin{bmatrix} 
\gamma \\
\beta  
\end{bmatrix}
({\mathcal L}_{s}^{(a),B_{m},F_{n}}(x))}
\end{align}
does not depend on $a$.
\end{theorem}
Let us replace $r$-th column of ${\mathcal K}_{s}^{(a),B_{m},F_{n}}(x)$ 
(resp.\ ${\mathcal L}_{s}^{(a),B_{m},F_{n}}(x)$)
with a column vector whose $i$-th component is $
(\prod_{\gamma \in J}(-z_{\gamma}))^{i}\Qs_{J}(xq^{-2i+\frac{m-n}{2}})$ 
(resp.\ $(\prod_{\gamma \in J}(-z_{\gamma}))^{-i}
\Qs_{B_{m}\sqcup F_{n} \setminus J}(xq^{-2i-\frac{m-n}{2}})$) and write it as 
${\mathcal K}_{s;r,J}^{(a),B_{m},F_{n}}(x)$ 
(resp.\ ${\mathcal L}_{s;r,J}^{(a),B_{m},F_{n}}(x)$), 
where $i \in \{1,2,\dots,\binom{n}{s}+1 \}$, $J \subset F_{n}$ and 
${\rm Card}(J)=s$. 
Then we obtain the following lemma.
\begin{lemma}\label{lem-cons3}
For any $s \in {\mathbb Z}_{\ge 0}$, $a \in {\mathbb C}$ and 
$r \in \{1,2,\dots,\binom{n}{s}+1 \}$, the following relations hold. 
\begin{align}
& {\rm det}({\mathcal K}_{s}^{(a),B_{m},F_{n}}(x))=
{\rm det}({\mathcal L}_{s}^{(a),B_{m},F_{n}}(x))= 0, 
\\[5pt]
& {\rm det}({\mathcal K}_{s;r,J}^{(a),B_{m},F_{n}}(x))=
{\rm det}({\mathcal L}_{s;r,J}^{(a),B_{m},F_{n}}(x))=0 .
\label{baxterEQ-general-2}
\end{align}
\end{lemma}
(\ref{baxterEQ-general-2}) for $(m,n)=(M,N)$ corresponds to a $U_{q}(\widehat{gl}(M|N))$ 
generalization of the Baxter equation. 
Theorem \ref{th-cons3} and Lemma \ref{lem-cons3} become trivial for $m=0$. 
Thus these are peculiar to the superalgebra case $U_{q}(\widehat{gl}(M|N))$ for $N>0$.

We remark that we did {\em not} use concrete function form of $\Qs_{I}(x)$ and  
$\overline{\Qs}_{I}(x)$ in the above theorems and lemmas. 
When one take into account the condition $a-s \le m-n$ 
in \eqref{lapQQb1} or 
 $a-s \ge m-n$ in \eqref{lapQQb3}, 
one have to put
\footnote{This comes from Lemma \ref{vanish-dai}. 
To change this condition may be necessary when one modify the formulae 
in this paper to apply the $T$- and $Y$-system for AdS/CFT \cite{GKV09,BFT09}. 
One will have to connect the $\Ts$-functions for the ``left wing" and the ``right wing". 
}
 some of the matrix elements in 
  ${\mathcal I}_{s}^{(a),B_{m},F_{n}}(x)$, ${\mathcal J}_{s}^{(a),B_{m},F_{n}}(x)$, 
  ${\mathcal K}_{s}^{(a),B_{m},F_{n}}(x)$ and ${\mathcal L}_{s}^{(a),B_{m},F_{n}}(x)$ 
to $0$, and needs some restrictions on $a,s$ in the above theorems and lemmas. 
We also remark that similar theorems also hold based on 
the formulae \eqref{lapQQb2} and \eqref{lapQQb4} 
(they are related to the above theorems by
 $\Qs_{I}(xq^{\mathtt s}) \to \overline{\Qs}_{I}(xq^{-{\mathtt s}})$ 
for any $I \subset {\mathfrak I}$, 
where ${\mathtt s}$ is any shift of $x$ of the $\Qs$-functions). 
 
\end{document}